\documentclass{article}
\usepackage{arxiv}

\usepackage[utf8]{inputenc}
\usepackage[T1]{fontenc}
\usepackage[dvipsnames]{xcolor}
\definecolor{mycolor}{RGB}{14,3,123}
\usepackage[
    colorlinks=true,
    citecolor=blue,
    linkcolor=red,
    urlcolor=magenta,
    filecolor=cyan,
    menucolor=red,
    runcolor=filecolor,
    allcolors=mycolor
]{hyperref}
\usepackage{url}
\usepackage{booktabs}
\usepackage{amsmath}
\usepackage{amsfonts}
\usepackage{amsthm}
\usepackage{bm}
\usepackage{nicefrac}
\usepackage{microtype}
\usepackage{graphicx}
\usepackage{subcaption}
\usepackage[numbers]{natbib}
\usepackage{doi}
\usepackage{enumitem}
\usepackage{algorithm}
\usepackage{algorithmic}
\usepackage{float}
\usepackage{makecell}
\usepackage[capitalize,noabbrev]{cleveref}

\allowdisplaybreaks[4]

\newtheorem{theorem}{Theorem}[section]

\newcommand{\CF}{\mathcal{F}}

\newcommand{\CS}{\mathcal{S}}
\newcommand{\CT}{\mathcal{T}}
\newcommand{\DKL}{D_{\mathrm{KL}}}
\newcommand{\Hist}{I}
\DeclareMathOperator*{\argmax}{arg\,max}

\newcommand{\Param}{\Theta}
\newcommand{\param}{\theta}
\newcommand{\design}{\xi}
\newcommand{\Epsilon}{\mathcal{E}}

\title{Optimal Stopping for Sequential Bayesian\\Experimental Design}

\newif\ifuniqueAffiliation
\uniqueAffiliationtrue

\ifuniqueAffiliation
\author{ \href{https://orcid.org/0000-0001-9773-2993}{\includegraphics[scale=0.06]{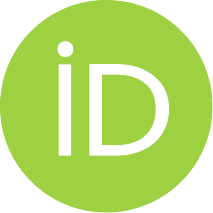}\hspace{1mm}Chen Cheng} \\
    Department of Mechanical Engineering\\
    University of Michigan\\
    Ann Arbor, MI 48109 \\
    \texttt{chech@umich.edu} \\
    \And
    \href{https://orcid.org/0000-0001-6544-2764}{\includegraphics[scale=0.06]{orcid.pdf}\hspace{1mm}Xun Huan} \\
    Department of Mechanical Engineering\\
    University of Michigan\\
    Ann Arbor, MI 48109 \\
    \texttt{xhuan@umich.edu} \\
}
\else
\usepackage{authblk}

\newbox{\orcid}\sbox{\orcid}{\includegraphics[scale=0.06]{orcid.pdf}}
\author[1]{
    \href{https://orcid.org/0000-0001-9773-2993}{\usebox{\orcid}\hspace{1mm}Chen Cheng\thanks{\texttt{chech@umich.edu}}}
}
\author[1]{
    \href{https://orcid.org/0000-0001-6544-2764}{\usebox{\orcid}\hspace{1mm}Xun Huan\thanks{\texttt{xhuan@umich.edu}}}
}
\affil[1]{Department of Mechanical Engineering, University of Michigan, Ann Arbor, MI 48109}
\fi

\renewcommand{\shorttitle}{Optimal Stopping for Sequential Bayesian Experimental Design}

\hypersetup{
pdftitle={Optimal Stopping for Sequential Bayesian Experimental Design},
pdfsubject={Bayesian experimental design; optimal stopping},
pdfauthor={Chen Cheng and Xun Huan},
pdfkeywords={Bayesian experimental design, optimal stopping, Markov decision process, policy gradient, curriculum learning, sequential design},
}

\begin{document}
\date{}
\maketitle

\begin{abstract}
Sequential Bayesian experimental design is often formulated as a fixed-horizon policy optimization problem, in which the number of experiments is specified before data collection begins. In practical campaigns, however, additional measurements may provide diminishing information relative to their cost, making termination an integral part of experimental design. Common threshold-based stopping rules are easy to implement but myopic, because they compare the current state with a fixed criterion rather than the expected value of future experiments. This work develops a Bayesian optimal stopping framework for sequential experimental design by treating design and stopping as coupled decisions in a finite-horizon sequential decision problem. We prove that, for any fixed design policy, the optimal stopping rule terminates when the immediate terminal reward is no smaller than the expected continuation value. We then derive a policy-gradient method for learning continuous design policies with value-based stopping. The resulting optimization is challenging because the design policy, continuation value, and stopping boundary are mutually dependent, and na\"{i}ve training can become trapped in early-stopping local optima. To address this difficulty, we introduce a curriculum strategy that gradually transitions from forced continuation to adaptive stopping during training. Numerical studies on a linear-Gaussian benchmark, a nonlinear test case, and a contaminant source detection problem show that the proposed approach learns stable, resource-aware design--stopping policies, with the largest gains in settings with strong sequential dependence.
\end{abstract}


\section{Introduction}

Sequential experimentation plays a central role across science and engineering, including clinical trials~\cite{Murphy2003}, materials discovery~\cite{Lookman2019}, and environmental monitoring~\cite{Krause2008}. In these settings, experiments are selected adaptively using information gathered from previous observations. Bayesian experimental design (BED)~\cite{Chaloner1995,Ryan2016,Alexanderian2021,Rainforth2024,Huan2024} provides a principled framework for this process by choosing experiments that maximize an expected utility, often based on expected information gain (EIG)~\cite{Lindley1956}. Modern non-myopic, policy-based extensions of sequential BED~\cite{Foster2021,Ivanova2021,Blau2022,Shen2023,Shen2025} learn adaptive design policies that map the current state of knowledge to the next experiment.

A fundamental decision nevertheless remains largely outside the scope of these sequential design frameworks: \emph{when should experimentation stop}? Most non-myopic or policy-based sequential BED formulations assume a fixed experimental horizon determined before data collection begins, while greedy sequential design methods are typically applied until a prescribed budget or external stopping rule is reached. In both cases, the stopping policy is not jointly optimized with the design policy. This omission is consequential in practice, where experimental campaigns often operate under uncertain budgets, limited resources, or diminishing returns. Additional experiments may provide further information, but that information must be weighed against its cost. Determining when further experimentation is no longer worthwhile is therefore an integral part of rational experimental design.

\begin{figure}[htbp]
    \centering
    \includegraphics[width=.5\linewidth]{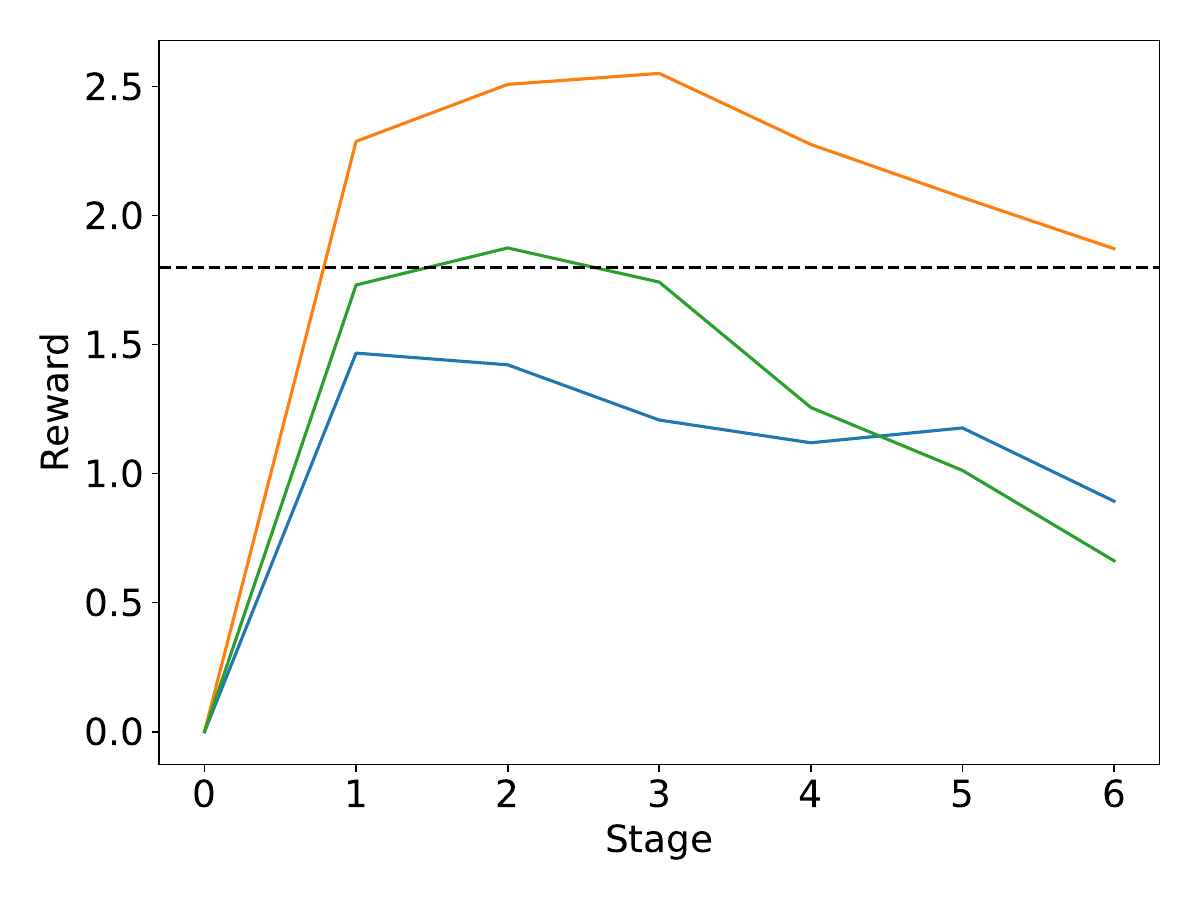}
    \caption{Illustration of threshold-based stopping on accumulated reward trajectories from the linear-Gaussian benchmark. Each curve shows the reward that would be obtained by stopping after a given number of experiments along one realized trajectory. A fixed threshold of 1.8 (dashed line) leads to qualitatively different behavior across trajectories: the orange trajectory stops prematurely after one experiment, missing larger rewards at later stages; the blue trajectory never reaches the threshold and therefore continues past its maximal reward; and the green trajectory stops near its maximum only by chance. The example illustrates that threshold rules are path-dependent and myopic, because they compare the current reward with a fixed cutoff rather than the value of continuing experimentation.}
    \label{fig:threshold_example}
\end{figure}

Existing stopping strategies are commonly based on threshold rules that terminate experimentation once a predefined statistic, such as posterior variance, information gain, or remaining budget, crosses a prescribed cutoff. Similar threshold-based stopping criteria appear throughout active learning~\cite{Zhu2010,Pullar-Strecker2024}, multi-armed bandits~\cite{Audibert2010}, and online A/B testing~\cite{Daskalakis2017}. Although straightforward to implement, such rules are inherently myopic, because they compare the current state against a fixed criterion without explicitly accounting for the expected value of future experiments. Consequently, they may stop too early and miss valuable information or continue unnecessarily after additional measurements cease to justify their cost. \Cref{fig:threshold_example} illustrates this limitation on a simple example.

A natural alternative is to treat stopping as part of the sequential decision problem itself. Importantly, introducing stopping does not require changing the underlying information-theoretic objective; rather, it changes the decision structure by making the experimental horizon an optimization variable. This viewpoint leads to a Bayesian optimal stopping formulation in which the decision to terminate is determined by comparing the value of stopping immediately with the expected value of continuing experimentation. 
Optimal stopping theory has a long history in stochastic processes, dynamic programming, and sequential decision-making~\cite{Haggstrom1966,Peskir2006,Shiryaev2008,Bertsekas2012,Puterman2014}.

Related ideas have appeared in several neighboring settings. Berry et al.~\cite{Berry2002} introduced a ``terminator'' for adaptive Bayesian dose-ranging drug trials, deciding whether to abandon the drug, continue dose finding, or shift to a confirmatory phase. Their approach uses Bayesian decision analysis with constrained backward induction and forward simulation, reducing the stopping rule to a table indexed by low-dimensional posterior summaries of treatment advantage. This work demonstrates the value of incorporating termination into adaptive experimentation, but is tailored to dose-ranging clinical trials and regulatory decision-making rather than providing a general information-theoretic formulation for sequential BED. In Bayesian optimization, Knowledge Gradient methods~\cite{Frazier2008,Ryzhov2012} select measurements by maximizing the expected one-step increase in terminal decision value, and related cost-aware stopping rules continue sampling only when the anticipated value of additional evaluations justifies their cost~\cite{Garnett2023,Xie2025}. These methods provide principled stopping-like criteria for optimization objectives, but they are tied to terminal optimization value rather than posterior information gain. Other learning-based optimal stopping formulations typically cast the decision as a binary action problem, with actions such as \{\text{continue}, \text{stop}\}, and focus on learning the stopping policy for an exogenously evolving process~\cite{Fathan2021,Li2023}. In contrast, sequential BED with stopping requires jointly optimizing the stopping rule and the experimental design selected upon continuation, which may lie in a continuous design space. A general framework that jointly optimizes experimental design and stopping decisions in information-theoretic sequential BED is therefore still lacking.

Simply augmenting an existing sequential design method with a stopping action is insufficient. Introducing stopping fundamentally changes the structure of the decision problem by creating an endogenous horizon, terminal rewards, and a state-dependent stopping boundary. These changes raise both theoretical and computational challenges. From a theoretical perspective, the optimal stopping rule depends on the continuation value induced by future design decisions. From a computational perspective, design and stopping become tightly coupled because the design policy determines future information, future information determines continuation value, and continuation value determines the stopping boundary. Na\"{i}ve joint optimization can therefore create a self-reinforcing feedback loop that favors premature stopping and poor exploration.

In this work, we develop a Bayesian optimal stopping framework for sequential experimental design. Our contributions are:
\begin{itemize}
\item We formulate sequential BED with stopping as a finite-horizon sequential decision problem, equivalently a Markov decision process, in which experimental design and stopping are jointly optimized.
\item We derive a value-based characterization of the optimal stopping policy and prove that, for any fixed design policy, experimentation should terminate exactly when the terminal reward exceeds the expected continuation value.
\item We develop a computational framework based on policy-gradient optimization and value-function approximation for learning optimal stopping and experimental design policies in continuous design spaces.
\item We identify a stopping-induced training instability arising from the coupling between design and continuation value, and mitigate it through a curriculum-learning strategy that improves robustness in problems with strong sequential dependence.
\end{itemize}
We demonstrate the proposed framework on a linear-Gaussian benchmark, a nonlinear inference test case, and a contaminant source detection problem. The results show that the learned policies automatically balance information gain against experimental cost and can substantially outperform fixed-horizon and threshold-based stopping strategies, particularly when future experimental opportunities depend strongly on previous design decisions.

The remainder of the paper is organized as follows.
\Cref{sec:formulation} formulates sequential BED with stopping.
\Cref{sec:theory} develops the theoretical framework, including the optimal stopping characterization and policy-gradient theorem.
\Cref{sec:computation} presents the computational methodology, including value-function approximation and curriculum learning.
\Cref{sec:results} presents numerical experiments.
\Cref{sec:conclusion} concludes.

\section{Problem Formulation}\label{sec:formulation}

We formulate sequential BED with optimal stopping as a finite-horizon Markov decision process (MDP), building on MDP formulations of sequential design such as Shen et al.~\cite{Shen2023,Shen2025}. The key extension is that the experimental horizon is endogenous, allowing the decision maker at each stage to either stop and receive a terminal reward or continue by selecting a new experiment.

\subsection{Bayesian updating and notation}
\label{sec:bayesian_update}

We consider at most $N$ experiments indexed by $k=0,1,\ldots,N-1$. Let $\Param \in \mathbb{R}^p$ denote the unknown model parameter, $\design_k \in \Xi_k \subseteq \mathbb{R}^d$ the design for experiment $k$, and $Y_k \in \mathbb{R}^n$ the corresponding observation. We use uppercase letters for random variables and lowercase letters for their realizations; for example, $\Param$ denotes the random parameter and $\param$ denotes a realization. The information history available before experiment $k$ is
\begin{align}
\Hist_k = \{ \design_0,y_0,\ldots,\design_{k-1},y_{k-1} \},
\qquad \Hist_0 = \emptyset .
\end{align}
The stage-$k$ belief is represented by the density $p(\param | \Hist_k)$; this belief serves as the prior for experiment $k$ and is the posterior induced by the previous experiments.

Given design $\design_k$ and history $\Hist_k$, observations are assumed to follow
\begin{align}
Y_k = G_k(\Param,\design_k;\Hist_k) + \Epsilon_k ,
\label{eq:observation_model}
\end{align}
where $G_k$ is the forward map, which may depend on the current history, and $\Epsilon_k \in \mathbb{R}^n$ is additive observation noise with known density $p_{\epsilon}$. We assume the observation noises are conditionally independent across stages given $(\param,\design_k,\Hist_k)$. After observing $y_k$, Bayes' rule gives
\begin{align}
p(\param | y_k,\design_k,\Hist_k)
=
\frac{
p(y_k | \param,\design_k,\Hist_k)p(\param | \Hist_k)
}{
p(y_k | \design_k,\Hist_k)
},
\label{eq:bayes_rule}
\end{align}
where $p(y_k | \param,\design_k,\Hist_k)$ is the likelihood induced by \cref{eq:observation_model}, and
\begin{align}
p(y_k | \design_k,\Hist_k)
=
\int
p(y_k | \param,\design_k,\Hist_k)
p(\param | \Hist_k)
\,\mathrm{d}\param
\label{eq:marginal_likelihood}
\end{align}
is the marginal likelihood. The updated density $p(\param | y_k,\design_k,\Hist_k)$ becomes the belief for the next stage, denoted $p(\param | \Hist_{k+1})$.

\subsection{Sequential design with stopping}
\label{sec:sequential_design_stopping}

\begin{figure}[htbp]
\centering
\includegraphics[width=0.95\linewidth]{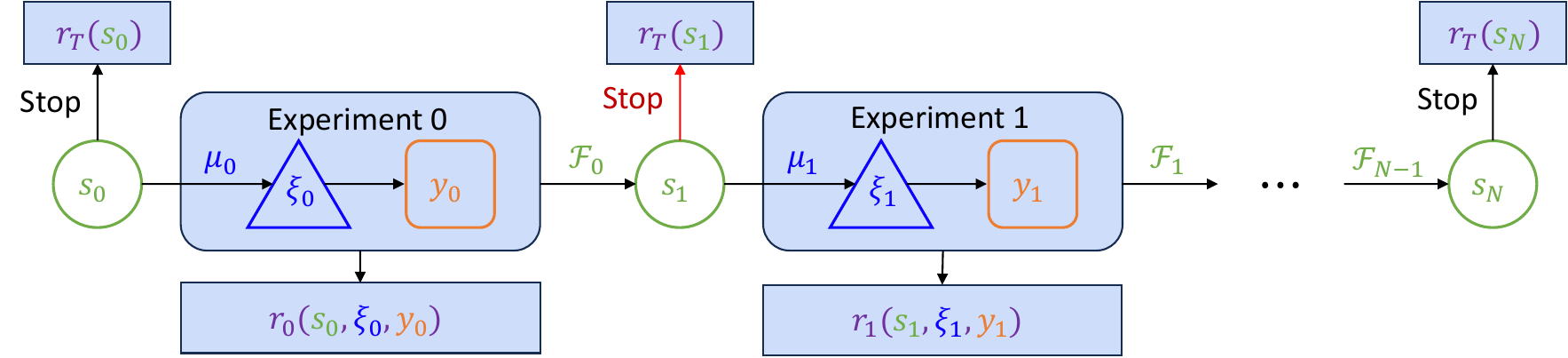}
\caption{Sequential BED with optimal stopping. At each stage, the decision maker either stops and receives a terminal reward, or continues by selecting a design. If continuing, an observation is collected and the belief state is updated by Bayes' rule. The process repeats until stopping or until the maximum number of experiments is reached.}
\label{fig:flowchart}
\end{figure}

At each stage, the decision maker first decides whether to stop or continue. If the process continues, a design is selected, an observation is collected, and the belief is updated. If the process stops, no further experiments are performed. \Cref{fig:flowchart} summarizes this sequential decision process.

We represent the \emph{state} at stage $k$ by $s_k \in \CS$. The state contains the current belief about $\Param$ and any additional deterministic variables needed for future design decisions. In the simplest case, the history $\Hist_k$ itself is a Markov state, since it determines the current belief $p(\param | \Hist_k)$ and all information needed for subsequent Bayesian updating. In applications with additional physical variables, such as a mobile sensor location, we write the state abstractly as
\begin{align}
s_k = (s_k^b,s_k^p),
\end{align}
where $s_k^b$ denotes the belief component and $s_k^p$ denotes the physical component. We also introduce a terminal state $T$, which is reached once experimentation stops.

The \emph{stopping policy} is a collection of decision rules
\begin{align}
\psi = \{\varphi_k:\CS \to \{0,1\},\; k=0,\ldots,N-1\},
\end{align}
where $\varphi_k(s_k)=1$ denotes stopping and $\varphi_k(s_k)=0$ denotes continuation. The \emph{design policy} is a collection of mappings
\begin{align}
\pi = \{\mu_k:\CS \to \Xi_k,\; k=0,\ldots,N-1\},
\end{align}
so that, upon continuation at stage $k$, the design is
\begin{align}
\design_k = \mu_k(s_k).
\end{align}

When an experiment is performed, the state evolves according to
\begin{align}
s_{k+1} = \CF_k(s_k,\design_k,y_k),
\end{align}
where $\CF_k$ encodes both the Bayesian update in \cref{eq:bayes_rule} and any deterministic evolution of the physical state. If stopping is selected, the process enters the terminal state. If no stopping occurs before stage $N$, the process terminates after evaluating the final state $s_N$. Thus,
\begin{align}
s_{k+1}
=
\begin{cases}
\CF_k(s_k,\design_k,y_k),
& s_k \neq T \text{ and } \varphi_k(s_k)=0,\\
T,
& \text{otherwise}.
\end{cases}
\label{eq:state_transition}
\end{align}
This transition structure induces a state-dependent stopping time and therefore an endogenous experimental horizon.

\subsection{Information rewards and experimental costs}
\label{sec:reward_formulations}

The reward balances information gain against experimental cost. We distinguish between a stage reward $r_k(s_k,\design_k,y_k)$, collected when experiment $k$ is performed, and a terminal reward $r_T(s_k)$, collected when experimentation stops. Let $c_k(\design_k) \leq 0$ denote the cost of experiment $k$. Information gain is measured by Kullback--Leibler (KL) divergence. We consider two reward formulations.

The first is the \emph{terminal formulation}, in which information gain and accumulated cost are evaluated only when experimentation stops:
\begin{align}
r_k(s_k,\design_k,y_k) &= 0,
\label{eq:terminal_reward_stage}\\
r_T(s_k)
&=
\begin{cases}
\DKL\left(
p_{\Param|\Hist_k}\,||\,p_{\Param|\Hist_0}
\right)
+
\sum\limits_{i=0}^{k-1} c_i(\design_i),
& s_k \neq T \text{ and } \varphi_k(s_k)=1,\\
0,
& \text{otherwise}.
\end{cases}
\label{eq:terminal_reward}
\end{align}
The terminal reward in \cref{eq:terminal_reward} applies to stages $k<N$ at which stopping is selected. If no stopping decision occurs before stage $N$, the same expression is evaluated at $s_N$ with $k=N$.
Here $p_{\Param|\Hist_k}$ denotes the posterior distribution of $\Param$ given $\Hist_k$. The terminal formulation directly measures the information accumulated from the prior to the belief at stopping.

The second is the \emph{incremental formulation}, in which information gain is accumulated stage by stage:
\begin{align}
r_k(s_k,\design_k,y_k)
&=
\begin{cases}
\DKL\left(
p_{\Param|\Hist_{k+1}}\,||\,p_{\Param|\Hist_k}
\right)
+
c_k(\design_k),
& s_k \neq T \text{ and } \varphi_k(s_k)=0,\\
0,
& \text{otherwise},
\end{cases}
\label{eq:incremental_reward_stage}\\
r_T(s_k) &= 0 .
\label{eq:incremental_reward_terminal}
\end{align}
The terminal and incremental formulations assign rewards differently along a trajectory, but they yield the same expected utility under any fixed design and stopping policies. Their equivalence for policy-dependent stopping times is established in \Cref{sec:theory}.

\subsection{Optimization problem}
\label{sec:optimization_problem}

The goal is to jointly optimize the design policy $\pi$ and stopping policy $\psi$. For any pair $(\pi,\psi)$, define the expected utility
\begin{align}
U(\pi,\psi)
=
\mathbb{E}_{Y_{0 : N-1} | \pi, \psi, s_0}\left[
\sum_{k=0}^{N-1}
\left\{
r_k(s_k,\design_k,Y_k) + r_T(s_k)
\right\}
+ r_T(s_N)
\right],
\label{eq:objective_soed}
\end{align}
where the expectation is taken over the parameter, observations, and state trajectory induced by the initial belief, observation model, and fixed policies $(\pi,\psi)$. The additional term $r_T(s_N)$ accounts for forced termination when no stopping decision occurs before stage $N$. The optimal policies solve
\begin{align}
(\pi^{\ast},\psi^{\ast})
\in
\argmax_{\pi,\psi}
\;
U(\pi,\psi)
\label{eq:joint_optimization_problem}
\end{align}
subject to, for $k=0,\ldots,N-1$,
\begin{align}
\varphi_k(s_k) &\in \{0,1\},\\
\design_k &= \mu_k(s_k) \in \Xi_k,\\
s_{k+1}
&=
\begin{cases}
\CF_k(s_k,\design_k,Y_k),
& s_k \neq T \text{ and } \varphi_k(s_k)=0,\\
T,
& \text{otherwise}.
\end{cases}
\end{align}
This formulation unifies experimental design and stopping as coupled decision variables. The stopping policy determines the experimental horizon, while the design policy determines the information available for future stopping decisions. The next section analyzes this coupling and derives the value-based structure of the optimal stopping rule.

\section{Theoretical Framework}
\label{sec:theory}

This section analyzes the structure of sequential BED with stopping. We first characterize the optimal stopping rule induced by a fixed design policy. We then show that the terminal and incremental reward formulations in \cref{sec:reward_formulations} yield the same expected utility, even when the stopping time is policy dependent. Finally, we derive a policy-gradient expression for optimizing parameterized continuous design policies under value-based stopping.

\subsection{Optimal stopping characterization}
\label{sec:optimal_stopping_characterization}

Fix a design policy $\pi=\{\mu_k\}_{k=0}^{N-1}$ and a stopping policy $\psi=\{\varphi_k\}_{k=0}^{N-1}$. For stages $k=0,\ldots,N$, let $V_k^{\pi,\psi}(s_k)$ denote the \emph{value function}, defined as the expected cumulative reward from state $s_k$ onward when following design policy $\pi$ and stopping policy $\psi$:
\begin{align}
V_k^{\pi,\psi}(s_k)
=
\mathbb{E}_{Y_{k:N-1}|\pi,\psi,s_k}
\left[
\sum_{\ell=k}^{N-1}
\left\{
r_\ell(s_\ell,\mu_\ell(s_\ell),Y_\ell)
+
r_T(s_\ell)
\right\}
+
r_T(s_N)
\right].
\label{eq:value_function}
\end{align}
The terminal condition is
\begin{align}
V_N^{\pi,\psi}(s_N) = r_T(s_N).
\label{eq:value_terminal_condition}
\end{align}

Equivalently, the value function satisfies the recursion
\begin{align}
V_k^{\pi,\psi}(s_k)
=
\begin{cases}
r_T^S(s_k),
& \varphi_k(s_k)=1,\\
\mathbb{E}_{Y_k|s_k,\mu_k(s_k)}
\left[
r_k^C(s_k,\mu_k(s_k),Y_k)
+
V_{k+1}^{\pi,\psi}(s_{k+1})
\right],
& \varphi_k(s_k)=0,
\end{cases}
\label{eq:value_function_recursion}
\end{align}
for $k=0,\ldots,N-1$, with $s_{k+1}=\CF_k(s_k,\mu_k(s_k),Y_k)$ under continuation.
Here $r_T^S(s_k)$ denotes the terminal reward that would be collected if stopping were selected at $s_k$, and $r_k^C(s_k,\design_k,y_k)$ denotes the stage reward collected if continuation were selected, design $\design_k$ were performed, and observation $y_k$ were realized. Under the terminal reward formulation, $r_k^C=0$; under the incremental formulation, $r_k^C$ contains the stagewise information gain and experimental cost.

For a fixed design policy $\pi$, the \emph{continuation value} at stage $k$ is
\begin{align}
Q_k^{\pi,\psi}(s_k,\design_k)
=
\mathbb{E}_{Y_k|s_k,\design_k}
\left[
r_k^C(s_k,\design_k,Y_k)
+
V_{k+1}^{\pi,\psi}(s_{k+1})
\right],
\label{eq:continuation_value}
\end{align}
where $s_{k+1}=\CF_k(s_k,\design_k,Y_k)$. 
After this continuation step, future decisions follow $(\pi,\psi)$ through $V_{k+1}^{\pi,\psi}$.
Thus, for any fixed continuation policy, stopping can be expressed as a comparison between the immediate terminal reward $r_T^S(s_k)$ and the expected value of performing one more experiment.

\begin{theorem}[Optimal stopping rule]
\label{thm:optimal_stopping_rule}
For a fixed design policy $\pi=\{\mu_k\}_{k=0}^{N-1}$, an optimal stopping policy $\psi^\ast=\{\varphi_k^\ast\}_{k=0}^{N-1}$ satisfies
\begin{align}
\varphi_k^\ast(s_k)
=
\mathbf{1}_{s_k \in \CT_k},
\qquad
k=0,\ldots,N-1,
\label{eq:stopping_policy_indicator}
\end{align}
where the stopping set $\CT_k \subseteq \CS$ is
\begin{align}
\CT_k
=
\left\{
s_k \in \CS:
r_T^S(s_k)
\geq
Q_k^{\pi,\psi^\ast}
\left(
s_k,\mu_k(s_k)
\right)
\right\}.
\label{eq:stopping_set_q}
\end{align}
Equivalently, experimentation stops at stage $k$ exactly when the terminal reward is no smaller than the expected continuation value,
\begin{align}
r_T^S(s_k)
\geq
\mathbb{E}_{Y_k|s_k,\mu_k(s_k)}
\left[
r_k^C(s_k,\mu_k(s_k),Y_k)
+
V_{k+1}^{\pi,\psi^\ast}
\left(
\CF_k(s_k,\mu_k(s_k),Y_k)
\right)
\right].
\label{eq:stopping_set_v}
\end{align}
\end{theorem}

The proof is given in \cref{app:optimal_stopping_rule_proof}. Because the continuation value depends on the stopping policy followed after the next observation, \cref{eq:stopping_set_q} gives a dynamic-programming fixed-point characterization of the stopping boundary. \Cref{thm:optimal_stopping_rule} shows that, for any fixed design policy, stopping is determined by a comparison between the value of terminating immediately and the value of continuing. The resulting boundary is therefore not a fixed threshold on a posterior statistic, but a value-based boundary that accounts for future experimental opportunities.

\subsection{Terminal--incremental reward equivalence}
\label{sec:terminal_incremental_equivalence}\
The reward formulations in \cref{sec:reward_formulations} differ in how information gain is assigned along a trajectory. The terminal formulation accumulates all information gain at stopping, whereas the incremental formulation distributes information gain across stages. The following result shows that these two formulations yield the same expected utility, even when the stopping time is determined by the policy.
Related terminal--incremental equivalences are used in fixed-horizon sequential design by Foster et al.~\cite{Foster2021} and Shen et al.~\cite{Shen2023}; here the result is stated for policy-dependent stopping times.

\begin{theorem}[Terminal--incremental equivalence with stopping]
\label{thm:equiv_terminal_incremental}
Let $U_T(\pi,\psi)$ denote the expected utility under the terminal formulation in \cref{eq:terminal_reward_stage}--\cref{eq:terminal_reward}, and let $U_I(\pi,\psi)$ denote the expected utility under the incremental formulation in \cref{eq:incremental_reward_stage}--\cref{eq:incremental_reward_terminal}. Then, for any design policy $\pi$ and stopping policy $\psi$,
\begin{align}
U_T(\pi,\psi)=U_I(\pi,\psi).
\label{eq:terminal_incremental_equivalence}
\end{align}
\end{theorem}

The proof is given in \cref{app:equiv_terminal_incremental}. This equivalence is useful both conceptually and computationally. Conceptually, it shows that introducing stopping changes the decision structure, but not the underlying information-theoretic objective. Computationally, it permits either reward representation to be used depending on whether one prefers sparse terminal rewards or denser stagewise rewards.

\subsection{Policy-gradient theorem}
\label{sec:policy_gradient_theorem}

\Cref{thm:optimal_stopping_rule} shows that, for a fixed design policy, the optimal stopping rule is determined by the continuation value. We now parameterize the design policy and derive a gradient expression for optimizing it.

Let each design map $\mu_k$ be parameterized by $w_k$, written $\mu_{k,w_k}$, and let
\begin{align}
w=\{w_k\}_{k=0}^{N-1}
\end{align}
collect all design-policy parameters. The corresponding design policy is denoted $\pi_w$. For each $w$, the stopping policy induced by the value-based stopping rule is denoted $\psi_w$, with
\begin{align}
\varphi_{k,w}(s_k)
=
\mathbf{1}_{s_k \in \CT_{k,w}},
\label{eq:parameterized_stopping_policy}
\end{align}
where
\begin{align}
\CT_{k,w}
=
\left\{
s_k \in \CS:
r_T^S(s_k)
\geq
Q_k^{\pi_w,\psi_w}
\left(
s_k,\mu_{k,w_k}(s_k)
\right)
\right\}.
\label{eq:parameterized_stopping_set}
\end{align}
The expected utility becomes
\begin{align}
U(w)
=
\mathbb{E}_{Y_{0:N-1}|\pi_w,\psi_w,s_0}
\left[
\sum_{k=0}^{N-1}
\left\{
r_k(s_k,\mu_{k,w_k}(s_k),Y_k)
+
r_T(s_k)
\right\}
+
r_T(s_N)
\right].
\label{eq:expected_utility_w}
\end{align}

Let
\begin{align}
\tau_w
=
\inf\{k \in \{0,\ldots,N-1\}: \varphi_{k,w}(s_k)=1\}
\wedge N
\label{eq:parameterized_stopping_time}
\end{align}
be the stopping time induced by $(\pi_w,\psi_w)$. 

The following result gives the design-policy gradient under standard regularity conditions, assuming that the design policy is differentiable in $w$, the continuation value $Q_k^{\pi_w,\psi_w}(s_k,\design_k)$ is differentiable in its current-design argument $\design_k$, the event that a trajectory lies exactly on the stopping boundary has probability zero under the trajectory distribution induced by $(\pi_w,\psi_w)$, and differentiation may be interchanged with the relevant expectations.

\begin{theorem}[Policy gradient]
\label{thm:policy_gradient}
Let $\nabla_{\design_k}Q_k^{\pi_w,\psi_w}(s_k,\design_k)$ denote the partial derivative of the continuation value with respect to the current design argument, holding the future policies $(\pi_w,\psi_w)$ fixed. Then the gradient of the expected utility with respect to $w$ is
\begin{align}
\nabla_w U(w)
=
\sum_{k=0}^{N-1}
\mathbb{E}_{s_k|\pi_w,\psi_w,s_0}
\left[
\mathbf{1}_{k<\tau_w}
\nabla_w \mu_{k,w_k}(s_k)
\cdot
\nabla_{\design_k}
Q_k^{\pi_w,\psi_w}(s_k,\design_k)
\big|_{\design_k=\mu_{k,w_k}(s_k)}
\right].
\label{eq:policy_gradient}
\end{align}
\end{theorem}

The proof is given in \cref{app:policy_gradient_expression}. The indicator $\mathbf{1}_{k<\tau_w}$ reflects that only experiments performed before stopping contribute to the design-policy gradient. The theorem reduces joint design--stopping optimization to gradient-based optimization of the design policy, with stopping determined implicitly by the continuation value. The resulting computational challenge is to estimate the state expectations and continuation-value gradients in \cref{eq:policy_gradient}; this is addressed in \cref{sec:computation}.

\section{Computational Methodology}
\label{sec:computation}\ 
The policy-gradient expression in\cref{eq:policy_gradient} reduces the joint optimization to two computational tasks: estimating expectations over stopped trajectories and evaluating gradients of the continuation value with respect to the current design. This section presents a practical approximation strategy. We use Monte Carlo trajectory simulation to approximate the state expectations, a differentiable design policy to parameterize continuous designs, and a learned continuation-value approximation to provide gradients for policy optimization. We then introduce a curriculum strategy that stabilizes training by delaying the influence of stopping decisions early in optimization.

\subsection{Monte Carlo policy-gradient approximation}
\label{sec:mc_policy_gradient}

Let $M$ denote the number of simulated trajectories used at each policy update. For a fixed parameter vector $w$, trajectory $m$ is generated by following the current design policy $\pi_w$ and the induced stopping policy $\psi_w$ until the stopping time $\tau_w^{(m)}$. The policy-gradient theorem gives the Monte Carlo approximation
\begin{align}
\nabla_w U(w)
\approx
\frac{1}{M}
\sum_{m=1}^{M}
\sum_{k=0}^{N-1}
\mathbf{1}_{k<\tau_w^{(m)}}
\nabla_w \mu_{k,w_k}\left(s_k^{(m)}\right)
\cdot
\nabla_{\design_k}
Q_k^{\pi_w,\psi_w}
\left(s_k^{(m)},\design_k\right)
\bigg|_{\design_k=\mu_{k,w_k}\left(s_k^{(m)}\right)} .
\label{eq:policy_grad_mc}
\end{align}
The main computational difficulty is that the continuation value $Q_k^{\pi_w,\psi_w}$ is generally unavailable in closed form. Direct nested Monte Carlo estimation of its design gradient would be expensive and high variance. We therefore approximate the continuation value with a differentiable surrogate.

\subsection{Policy and continuation-value parameterizations}
\label{sec:policy_value_parameterization}

The design policy is parameterized by a differentiable function $\mu_w(k,s_k)$, implemented as a neural network with parameters $w$. Rather than using separate networks for each stage, we use a single network that takes the stage index and state representation as inputs and returns a design in $\Xi_k$:
\begin{align}
\design_k = \mu_w(k,s_k).
\end{align}
When the admissible design set is constrained, the network output is mapped to $\Xi_k$ by an appropriate transformation or projection.

In the numerical experiments, the state is represented by the experimental history $\Hist_k$. Since the history grows with $k$, we use a fixed-length zero-padded representation. The stage index is encoded as a one-hot vector
\begin{align}
e_{k+1}
=
[0,\ldots,0,\underbrace{1}_{(k+1)\text{th}},0,\ldots,0]^\top
\in \mathbb{R}^{N}.
\end{align}
The history is represented by zero-padding unobserved future entries:
\begin{align}
s_k
\quad \longrightarrow \quad
\left[
e_{k+1}^\top,
\design_0^\top,\ldots,\design_{k-1}^\top,
0,\ldots,0,
y_0^\top,\ldots,y_{k-1}^\top,
0,\ldots,0
\right]^\top .
\label{eq:state_representation}
\end{align}
For design dimension $d$ and observation dimension $n$, this representation has dimension $N+(N-1)(d+n)$. This construction follows the fixed-size history representation used in related sequential design methods such as Shen et al.~\cite{Shen2023}.

The continuation value is approximated by a differentiable function
\begin{align}
Q_{\eta}(k,s_k,\design_k)
\approx
Q_k^{\pi_w,\psi_w}(s_k,\design_k),
\end{align}
with parameters $\eta$. The continuation-value network takes the same state representation as the design policy, augmented with the candidate design $\design_k$, and outputs a scalar continuation value. Its differentiability with respect to $\design_k$ provides the gradient
\begin{align}
\nabla_{\design_k} Q_{\eta}(k,s_k,\design_k),
\end{align}
which is used in the Monte Carlo policy-gradient estimator.

Computationally, the method can be viewed as an actor-critic approximation---the design-policy network $\pi_{w}$ acts as the actor, while the continuation-value network $Q_{\eta}$ acts as the critic supplying gradients with respect to the current design. We use this terminology only to describe the numerical implementation; the stopping rule itself is determined by the optimal stopping characterization in \cref{thm:optimal_stopping_rule}.

\subsection{Continuation-value learning}
\label{sec:critic_training}

The continuation-value approximation is trained to satisfy the Bellman relation implied by \cref{eq:continuation_value}. For a simulated trajectory, define the one-step target
\begin{align}
z_k^{(m)}
=
r_k^C\left(s_k^{(m)},\design_k^{(m)},y_k^{(m)}\right)
+
\widehat{V}_{k+1,\eta}
\left(s_{k+1}^{(m)}\right),
\label{eq:critic_target}
\end{align}
where
\begin{align}
\widehat{V}_{k+1,\eta}(s_{k+1})
=
\begin{cases}
r_T(s_N),
& k+1=N,\\
\max\left\{
r_T^S(s_{k+1}),
Q_{\eta}\left(k+1,s_{k+1},\mu_w(k+1,s_{k+1})\right)
\right\},
& k+1<N.
\end{cases}
\label{eq:value_target}
\end{align}
The maximum in \cref{eq:value_target} implements the value-based stopping rule by assigning, after the next state is reached, the larger of the predicted values for stopping immediately and continuing with the current design policy.

The continuation-value parameters are updated by minimizing the empirical squared Bellman residual over states at which continuation was selected:
\begin{align}
\mathcal{L}(\eta)
=
\frac{1}{M}
\sum_{m=1}^{M}
\sum_{k=0}^{N-1}
\mathbf{1}_{k<\tau_w^{(m)}}
\left[
Q_{\eta}
\left(k,s_k^{(m)},\design_k^{(m)}\right)
-
z_k^{(m)}
\right]^2 .
\label{eq:value_loss}
\end{align}
In practice, the target in \cref{eq:critic_target} is treated as fixed when updating $\eta$, as in standard temporal-difference regression. The resulting continuation-value approximation supplies differentiable, low-variance gradients for updating the design policy.

\subsection{Trajectory simulation and exploration}
\label{sec:trajectory_simulation}

Each training iteration generates trajectories under the current policy. A parameter realization is first sampled from the prior,
\begin{align}
\param^{(m)} \sim p(\param),
\end{align}
and observations are then simulated sequentially from the likelihood. The design policy itself is deterministic, but during training we perturb the selected design to encourage exploration:
\begin{align}
\design_k^{(m)}
=
\mu_w\left(k,s_k^{(m)}\right)
+
\Epsilon_{\mathrm{explore}},
\qquad
\Epsilon_{\mathrm{explore}}
\sim
\mathcal{N}\left(0,\sigma_{\mathrm{explore}}^2 I_d\right).
\label{eq:exploration_noise}
\end{align}
The exploration perturbation is used only to broaden the training distribution for trajectory generation and critic fitting; the actor update evaluates the design-gradient at the deterministic policy output.
The exploration scale $\sigma_{\mathrm{explore}}$ can be held fixed or decreased during training. At deployment, no exploration perturbation is added.

After collecting $y_k^{(m)}$, the state is updated by Bayes' rule through
\begin{align}
s_{k+1}^{(m)}
=
\CF_k\left(s_k^{(m)},\design_k^{(m)},y_k^{(m)}\right).
\end{align}
Stopping is then determined by comparing the terminal reward with the learned continuation value:
\begin{align}
r_T^S\left(s_{k+1}^{(m)}\right)
\geq
Q_{\eta}
\left(
k+1,
s_{k+1}^{(m)},
\mu_w(k+1,s_{k+1}^{(m)})
\right),
\label{eq:learned_stopping_rule}
\end{align}
with forced termination at stage $N$.

\subsection{Training instability and curriculum learning}
\label{sec:curriculum_learning}\
Directly training the design and continuation-value approximations can be unstable because stopping and continuation values are mutually dependent. The stopping set depends on the continuation value, the continuation value depends on the design policy and future stopping decisions, and the design policy is trained only on trajectories that have not yet stopped. Early in training, inaccurate designs can produce low information gain, which leads the continuation-value approximation to underestimate the benefit of future experiments. This can trigger premature stopping, truncating trajectories before the policy has explored informative later-stage designs.

To mitigate this feedback, we use a curriculum strategy inspired by curriculum learning \cite{Bengio2009}. Let $p_{\mathrm{stop}}(\ell)\in[0,1]$ denote the probability of enforcing the learned stopping rule at training iteration $\ell$. When the stopping condition in \cref{eq:learned_stopping_rule} is satisfied, training stops the trajectory with probability $p_{\mathrm{stop}}(\ell)$ and overrides the stopping decision with probability $1-p_{\mathrm{stop}}(\ell)$. Thus, early iterations can force longer trajectories, while later iterations increasingly follow the learned value-based stopping rule.

A simple schedule is
\begin{align}
p_{\mathrm{stop}}(\ell)
=
\min\left\{1,\frac{\ell}{\ell_{\mathrm{warm}}}\right\},
\label{eq:stopping_curriculum}
\end{align}
where $\ell_{\mathrm{warm}}$ controls the length of the curriculum phase. This strategy is used during training only. At deployment, stopping is deterministic and follows \cref{eq:learned_stopping_rule}. The curriculum breaks the self-reinforcing cycle of early stopping by ensuring that the design policy and continuation-value approximation are exposed to later-stage trajectories before stopping decisions dominate the training distribution.

Additional comparisons with a generic hybrid-action reinforcement learning baseline are provided in \cref{app:additional_results}; these results further illustrate the difficulty of learning stopping and continuous design decisions through a single stochastic action parameterization.

\subsection{Information-gain computation}
\label{sec:information_gain_computation}

Reward evaluation requires computing KL divergences between posterior distributions. In the numerical experiments, the parameter dimension is small enough that we use grid-based posterior approximation. The parameter space is discretized, posterior densities are evaluated pointwise up to normalization, and KL divergences are approximated by quadrature on the grid. For example,
\begin{align}
\DKL\left(
p_{\Param|\Hist_k}\,||\,p_{\Param|\Hist_0}
\right)
\approx
\sum_{j}
p(\param_j|\Hist_k)
\log
\frac{
p(\param_j|\Hist_k)
}{
p(\param_j|\Hist_0)
}
\Delta \param_j .
\label{eq:grid_kl}
\end{align}
For higher-dimensional parameter spaces, the same framework can be combined with approximate posterior representations, including variational inference \cite{Foster2019}, MCMC-based density estimation, or amortized posterior inference methods such as conditional diffusion models \cite{Baldassari2023}. These alternatives affect reward evaluation but do not change the decision-theoretic structure or policy-gradient expression.

\subsection{Algorithm summary}
\label{sec:algorithm_summary}

\Cref{alg:pg_soed} summarizes the training procedure. The algorithm alternates between generating stopped trajectories, fitting the continuation-value approximation, and updating the design policy using the Monte Carlo policy-gradient estimator.
\begin{algorithm}[htbp]
\caption{Policy-gradient training for sequential BED with optimal stopping.}
\label{alg:pg_soed}
\begin{algorithmic}[1]
\STATE Set initial state $s_0$, maximum horizon $N$, number of training iterations $L$, number of trajectories $M$, learning rates $\alpha_w$ and $\alpha_{\eta}$, exploration scale $\sigma_{\mathrm{explore}}$, and stopping curriculum $p_{\mathrm{stop}}(\ell)$.
\STATE Initialize design-policy parameters $w$ and continuation-value parameters $\eta$.
\FOR{$\ell=1,\ldots,L$}
    \FOR{$m=1,\ldots,M$}
        \STATE Set $s_0^{(m)}=s_0$ and sample $\param^{(m)}\sim p(\param)$.
        \FOR{$k=0,\ldots,N-1$}
            \STATE Evaluate stopping condition
                $r_T^S\left(s_{k}^{(m)}\right)
                \geq
                Q_{\eta}
                \left(
                k,
                s_{k}^{(m)},
                \mu_w(k,s_{k}^{(m)})
                \right).$
            \IF{stopping condition holds}
                \STATE Stop with probability $p_{\mathrm{stop}}(\ell)$; otherwise continue.
                \IF{trajectory is stopped}
                    \STATE Set $\tau_w^{(m)}=k$ and terminate trajectory.
                    \STATE \textbf{break}
                \ENDIF
            \ENDIF
            \STATE Select exploratory design
            $
            \design_k^{(m)}
            =
            \mu_w\left(k,s_k^{(m)}\right)
            +
            \Epsilon_{\mathrm{explore}},
            \Epsilon_{\mathrm{explore}}\sim
            \mathcal{N}\left(0,\sigma_{\mathrm{explore}}^2 I_d\right).
            $
            \STATE Simulate observation
            $
            y_k^{(m)}
            \sim
            p\left(\cdot|\param^{(m)},\design_k^{(m)},\Hist_k^{(m)}\right).
            $
            \STATE Update state
            $
            s_{k+1}^{(m)}
            =
            \CF_k\left(s_k^{(m)},\design_k^{(m)},y_k^{(m)}\right).
            $
            \IF{$k+1=N$}
                \STATE Set $\tau_w^{(m)}=N$ and terminate trajectory.
                \STATE \textbf{break}
            \ENDIF
        \ENDFOR
    \ENDFOR
    \STATE Compute rewards and targets using \cref{eq:critic_target}--\cref{eq:value_target}.
    \STATE Update $\eta$ by stochastic gradient descent on \cref{eq:value_loss}: $\eta \leftarrow \eta - \alpha_{\eta}\nabla_{\eta}\mathcal{L}(\eta)$.
    \STATE Estimate $\nabla_w U(w)$ using \cref{eq:policy_grad_mc}, with $Q_k^{\pi_w,\psi_w}$ replaced by $Q_{\eta}$.
    \STATE Update $w \leftarrow w+\alpha_w \nabla_w U(w)$.
\ENDFOR
\STATE Return the design policy $\pi_w$ and stopping rule induced by $Q_{\eta}$.
\end{algorithmic}
\end{algorithm}

\subsection{Computational considerations}
\label{sec:computational_considerations}

Training requires repeated forward-model evaluations, posterior updates, reward evaluations, and continuation-value regression. The dominant cost depends on the problem: in inexpensive low-dimensional examples, grid-based posterior updates and KL computations are often the main cost, whereas in complex physical models the forward map may dominate. These costs are incurred offline during training.

Once trained, deployment is inexpensive. At a new state, the next design is obtained by one forward pass through the design policy, and the stopping decision is obtained by comparing $r_T^S(s_k)$ with one continuation-value evaluation. Thus the online cost is a posterior update, a policy evaluation, and a continuation-value evaluation, without nested online optimization or lookahead simulation. This makes the approach suitable for autonomous sequential experimentation when offline training is feasible and online decisions must be made rapidly.

\section{Numerical Experiments}
\label{sec:results}

We evaluate the proposed framework on three numerical examples that test progressively more challenging aspects of sequential BED with stopping. The linear-Gaussian benchmark validates the policy-gradient method against analytical solutions. The nonlinear case tests non-conjugate Bayesian updating while retaining transparent posterior computation. The contaminant source detection problem introduces constrained sensor movement and stronger sequential dependence, making it the main stress test for the curriculum strategy.

Training hyperparameters are reported in \cref{tab:hyperparameters}. These settings largely follow Shen et al.~\cite{Shen2023}. Additional hyperparameter tuning may further improve performance, but is not pursued here.

\begin{table}[htbp]
    \centering
    \caption{Hyperparameter settings for the numerical experiments.}
    \label{tab:hyperparameters}
    \begin{tabular}{|l|c|c|c|}
    \hline
    & \textbf{Linear-Gaussian} & \textbf{Nonlinear} & \textbf{Source detection} \\
    \hline
    Policy network architecture
    & \multicolumn{3}{c|}{$N + (N-1)(d+n) \to 80 \to 80 \to d$} \\
    \hline
    \makecell[l]{Continuation-value network\\ architecture}
    & \multicolumn{3}{c|}{$N + (N-1)(d+n) + d \to 80 \to 80 \to 1$} \\
    \hline
    Policy learning rate $\alpha_w$
    & \multicolumn{3}{c|}{$1.5 \times 10^{-1}$} \\
    \hline
    \makecell[l]{Continuation-value learning rate\\ $\alpha_{\eta}$}
    & \multicolumn{3}{c|}{$1.0 \times 10^{-3}$} \\
    \hline
    Continuation-value batch size
    & \multicolumn{3}{c|}{$500$} \\
    \hline
    Exploration scale $\sigma_{\mathrm{explore}}$
    & \multicolumn{2}{c|}{$1.0$} & $0.05$ \\
    \hline
    Exploration-scale decay
    & \multicolumn{3}{c|}{$0.99$} \\
    \hline
    Monte Carlo size $M$
    & \multicolumn{3}{c|}{$1000$} \\
    \hline
    \end{tabular}
\end{table}

For experiments using curriculum learning, we control the probability of enforcing the learned stopping rule using a shifted sigmoid schedule,
\begin{align}
p_{\mathrm{stop}}(\ell)
=
\frac{1}{1+\exp\{-a(\ell-\ell_0)\}},
\label{eq:sigmoid_stopping_schedule}
\end{align}
where $\ell$ is the training iteration, $a$ controls the transition steepness, and $\ell_0$ determines the midpoint. The parameters are chosen so that $p_{\mathrm{stop}}(\ell)$ is near zero early in training, increases smoothly during mid-training, and exceeds $0.999$ for the final 30 iterations. Thus, early training encourages longer trajectories, while late training follows the value-based stopping rule almost deterministically.

\subsection{Linear-Gaussian benchmark}
\label{sec:case_linear_gaussian}

We first consider a linear-Gaussian benchmark for which analytical solutions are available. The observation model is
\begin{align}
Y_k
=
G(\Param,\design_k)+\Epsilon_k
=
\Param \design_k+\Epsilon_k,
\qquad
\Epsilon_k \sim \mathcal{N}(0,1^2),
\label{eq:linear_gaussian_model}
\end{align}
with prior $\Param\sim\mathcal{N}(0,3^2)$. Designs are constrained to $\design_k\in[0.1,3]$.

This benchmark has a useful simplifying structure. Because the model is conjugate, the posterior variance update is deterministic conditional on the chosen design and does not depend on the realized observation. Consequently, the expected information gain and continuation value are stage-dependent but observation-independent. The optimal adaptive stopping policy therefore reduces to choosing a fixed stopping stage. Details of the analytical derivation are provided in \cref{app:linear_gaussian_analytic_solution}. \Cref{tab:analytic_result} reports the analytical utility obtained by stopping after exactly $N_{\mathrm{fixed}}$ experiments, and the bold entry in each row identifies the optimal stopping stage among $N_{\mathrm{fixed}}=1,2,3,4$ for this benchmark.

\begin{table}[htbp]
\centering
\caption{Analytical utility for the linear-Gaussian benchmark when stopping after exactly $N_{\mathrm{fixed}}$ experiments. Because the posterior variance update is deterministic in this conjugate case, the optimal adaptive stopping rule reduces to a fixed stopping stage; the bold entry in each row identifies this optimal stage among $N_{\mathrm{fixed}}=1,2,3,4$.}
\label{tab:analytic_result}
\begin{tabular}{|c|cccc|}
\hline
Cost & $N_{\mathrm{fixed}}=1$ & $N_{\mathrm{fixed}}=2$ & $N_{\mathrm{fixed}}=3$ & $N_{\mathrm{fixed}}=4$ \\
\hline
$c_k=0$     & 2.203 & 2.547 & 2.749 & \textbf{2.892} \\
$c_k=-0.5$  & \textbf{1.703} & 1.547 & 1.249 & 0.892 \\
$c_k=-0.25$ & 1.953 & \textbf{2.047} & 1.999 & 1.892 \\
\hline
\end{tabular}
\end{table}

\begin{figure}[htbp]
  \centering
    \includegraphics[width=0.32\linewidth]{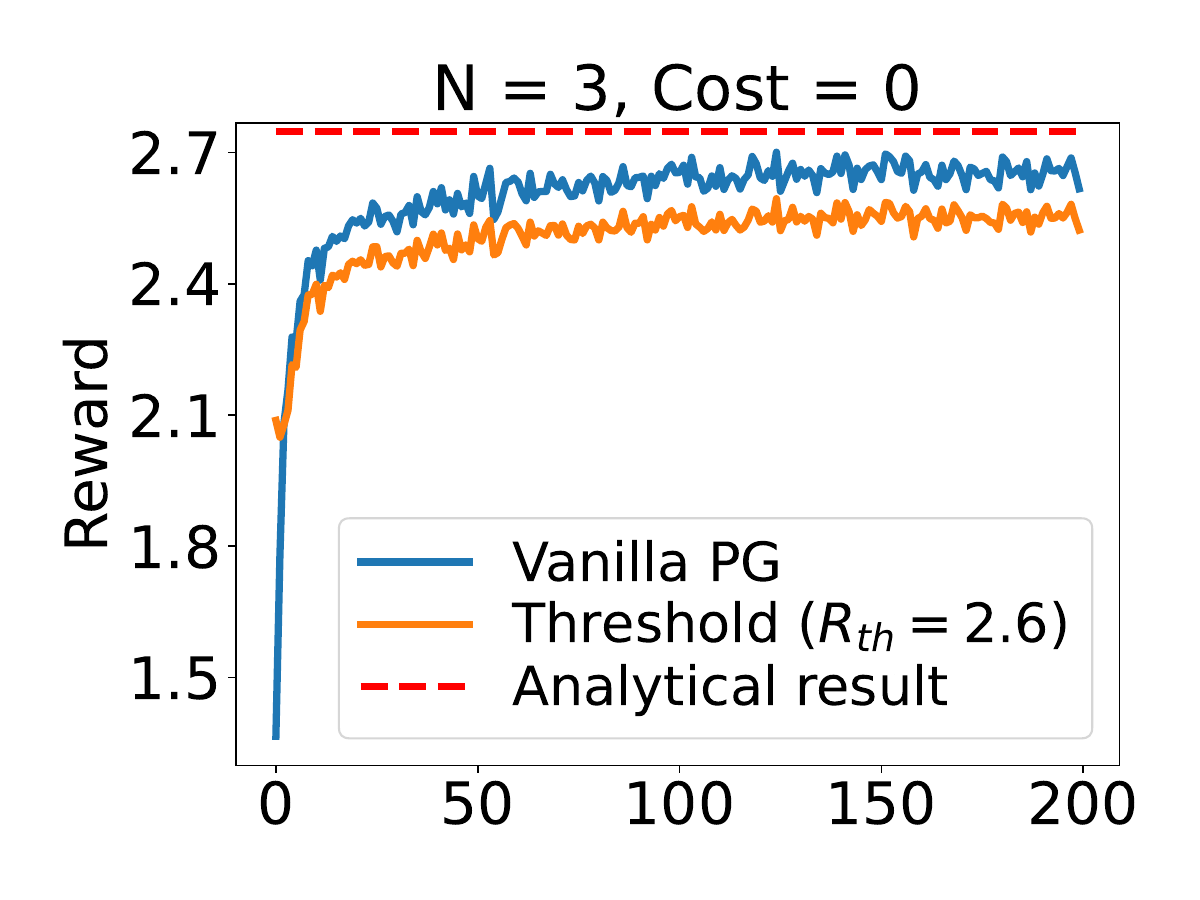}
    \includegraphics[width=0.32\linewidth]{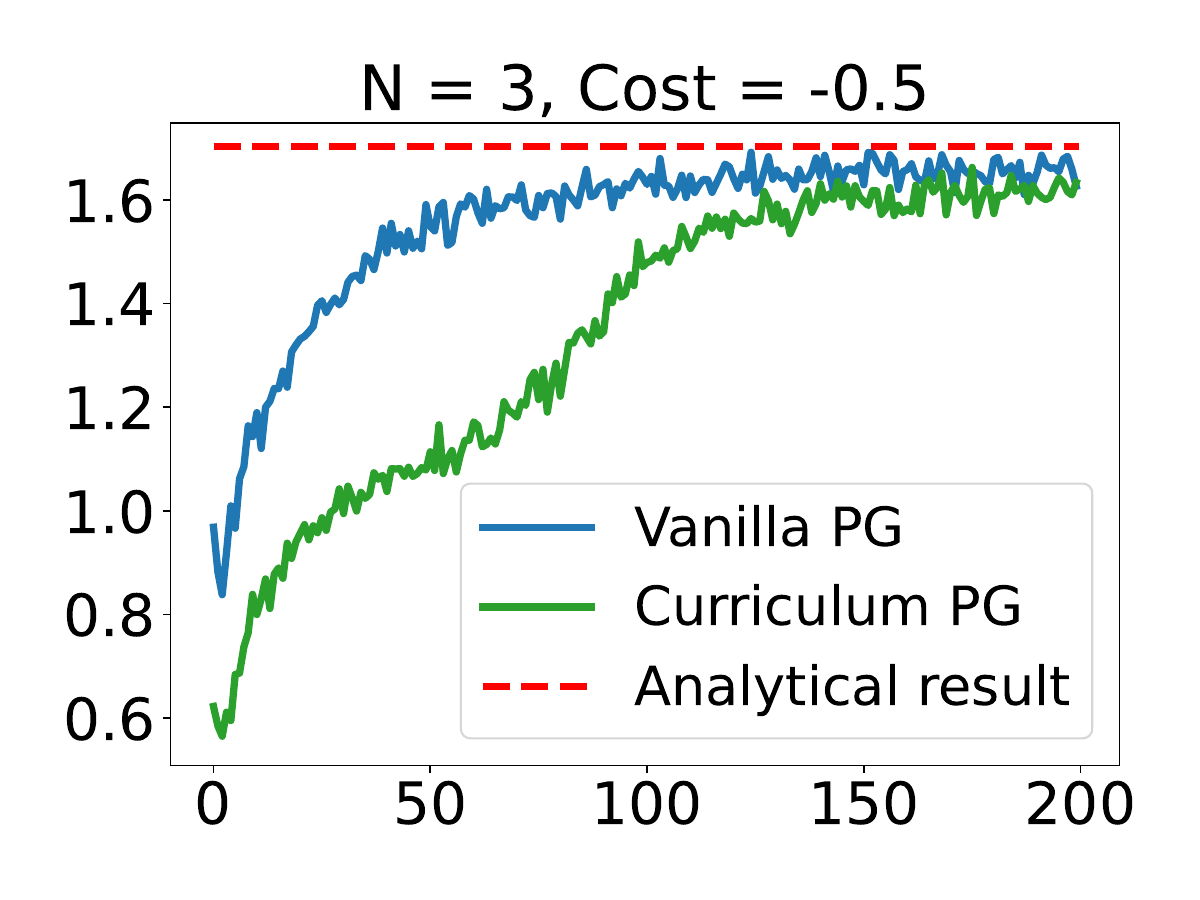}
    \includegraphics[width=0.32\linewidth]{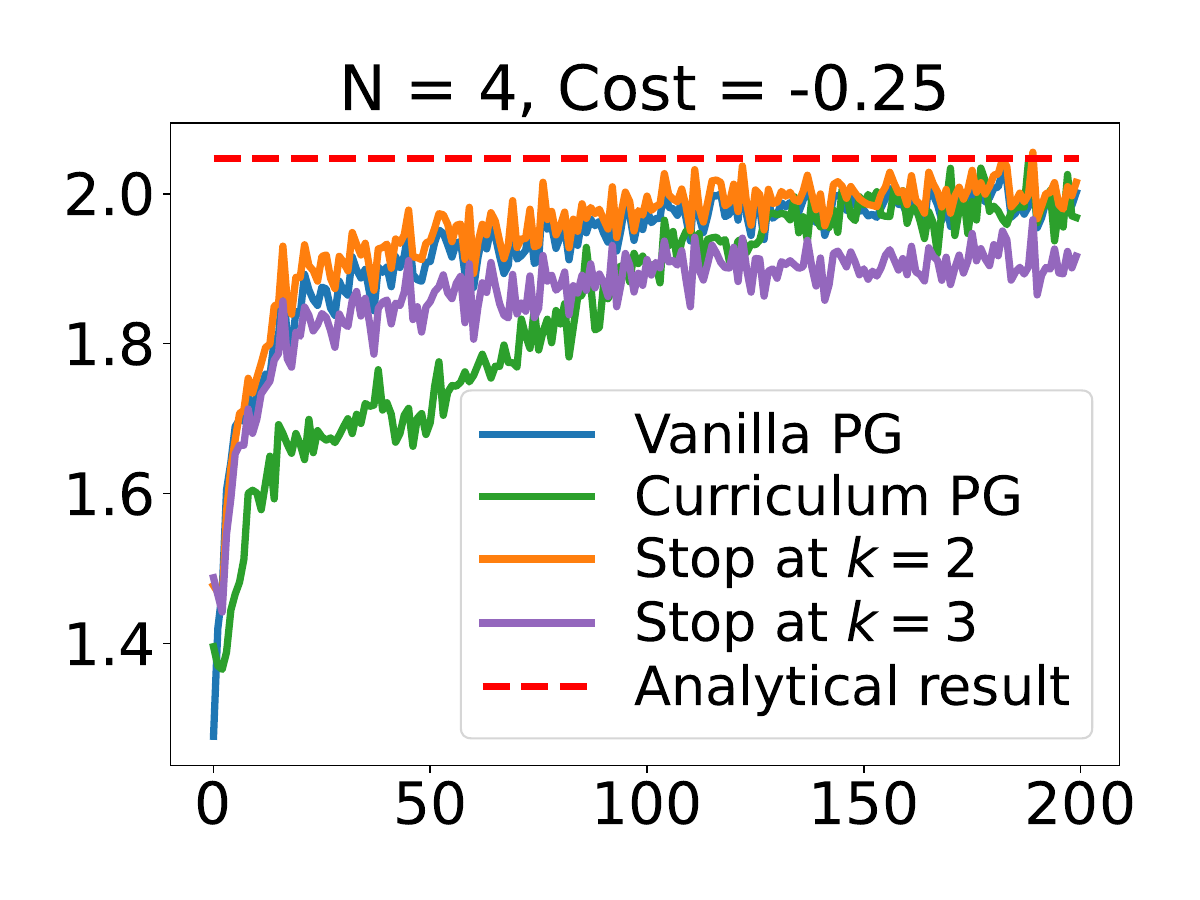}\\
    \includegraphics[width=0.32\linewidth]{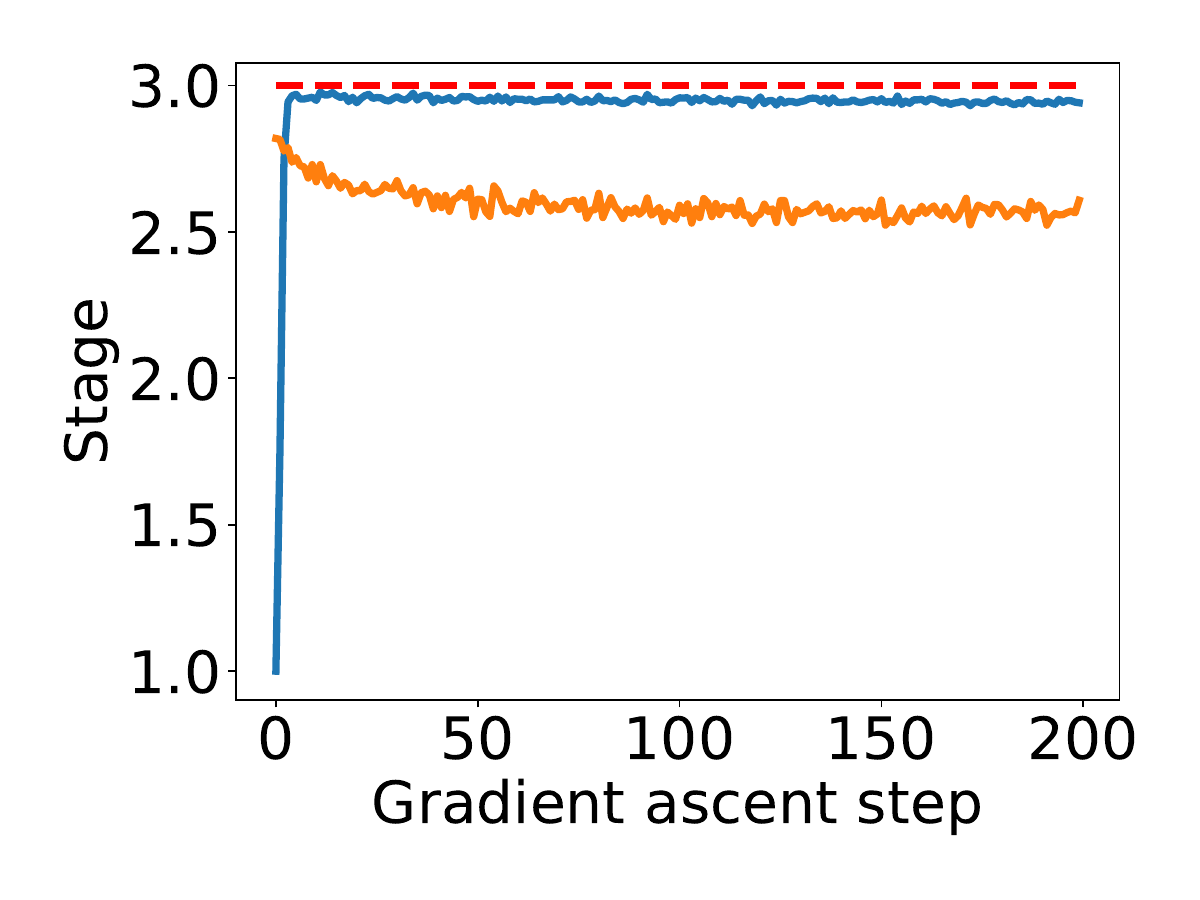}
    \includegraphics[width=0.32\linewidth]{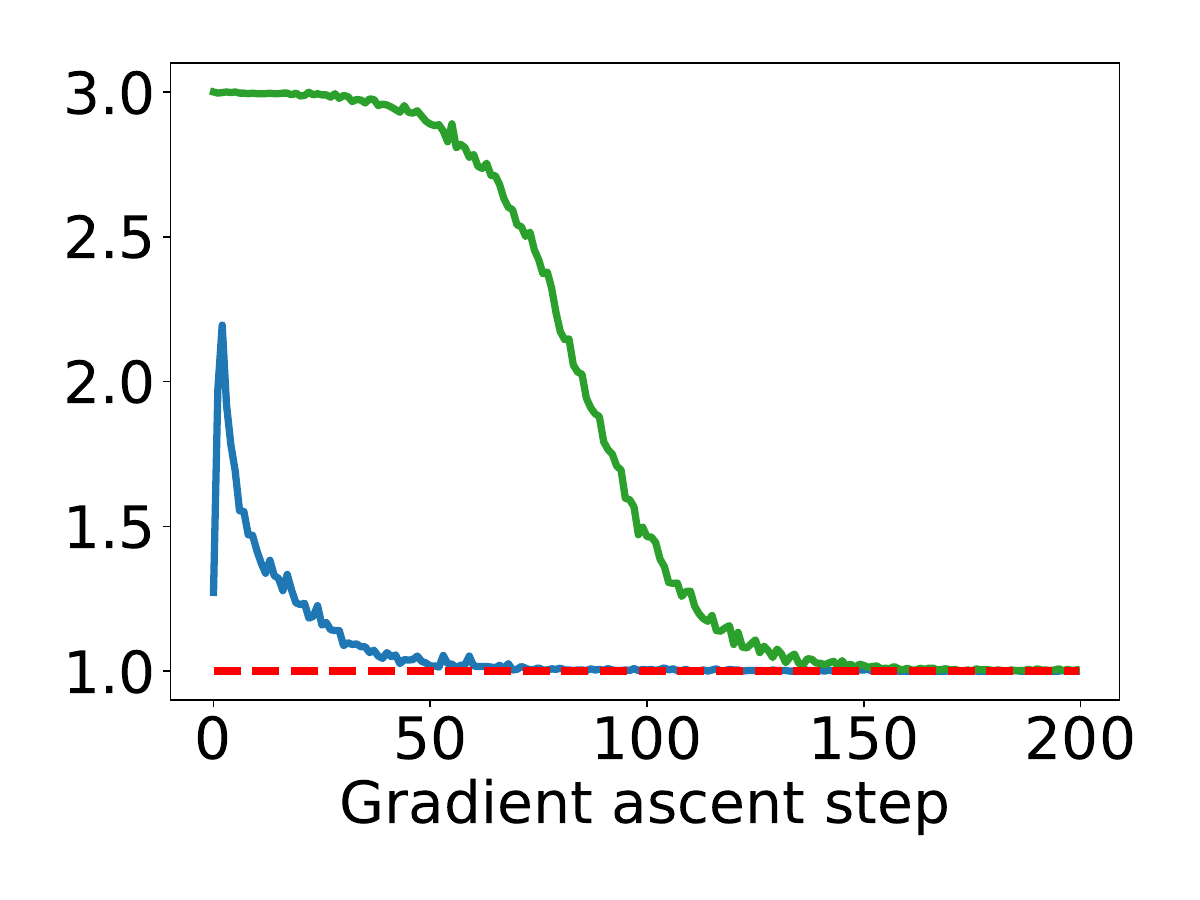}
    \includegraphics[width=0.32\linewidth]{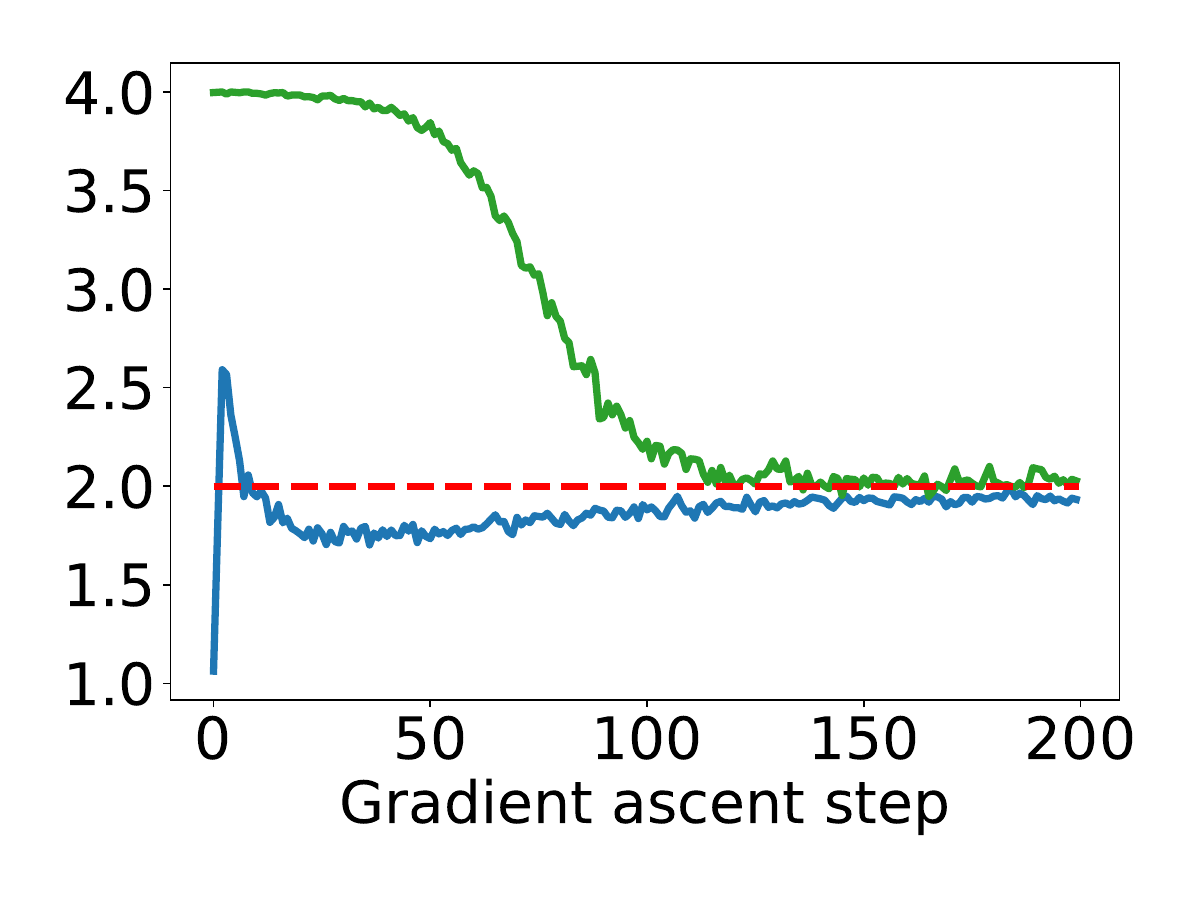}
  \caption{Training convergence for the linear-Gaussian benchmark. Top: average reward. Bottom: average stopping stage. Columns show representative combinations of maximum horizon $N$ and experimental cost.}
  \label{fig:linear_gaussian_convergence}
\end{figure}

The zero-cost case provides a basic validation of the stopping rule. When $c_k=0$, additional experiments carry no penalty. Since each additional experiment provides a nonnegative EIG contribution and the posterior variance decreases deterministically, it is optimal to continue until the maximum allowable stage. Consistent with this analytical structure, the learned policy in the left column of \cref{fig:linear_gaussian_convergence}, trained under maximum allowable horizon $N=3$ as an example, converges to stopping at stage $3$ and achieves the analytical optimum; the average values in the plots are averaged over the $M$ training trajectories. In contrast, threshold-based stopping can terminate prematurely and underperform because it does not account for the continuation value.

The negative-cost cases test whether the method learns to trade information gain against experimental cost. In the $N=3$, $c_k=-0.5$ setting, the analytical solution stops after one experiment, and both vanilla training (i.e., training without curriculum learning) and curriculum training converge to this early-stopping policy. In the $N=4$, $c_k=-0.25$ setting, the two methods achieve similar final rewards but exhibit different convergence behavior. Vanilla training tends to underestimate the stopping stage and can become trapped in shorter-horizon behavior, whereas the curriculum strategy more reliably explores longer trajectories before settling near the analytical optimum. The benefit of curriculum learning is modest in this benchmark, but the example illustrates an early-stopping bias that becomes more consequential in problems with stronger sequential dependence.

In the right column of \cref{fig:linear_gaussian_convergence}, the curves labeled ``Stop at $k=2$'' and ``Stop at $k=3$'' are fixed-stage reference policies that force termination after two and three experiments, respectively. These references correspond conceptually to $N_{\mathrm{fixed}}=2$ and $N_{\mathrm{fixed}}=3$ in \cref{tab:analytic_result}, although their numerical values may differ slightly because they are evaluated within the same numerical setup as the $N=4$ training experiment.

The remaining discrepancies from the analytical values are small and are primarily attributable to function approximation and optimization error. In particular, the $c_k=-0.25$ setting has a relatively flat reward landscape, where stopping at nearby stages yields similar utilities and makes the stopping stage sensitive to approximation error. Even in this case, the learned policies capture the correct information-cost tradeoff.

\begin{figure}[htbp]
  \centering
  \includegraphics[width=0.45\linewidth]{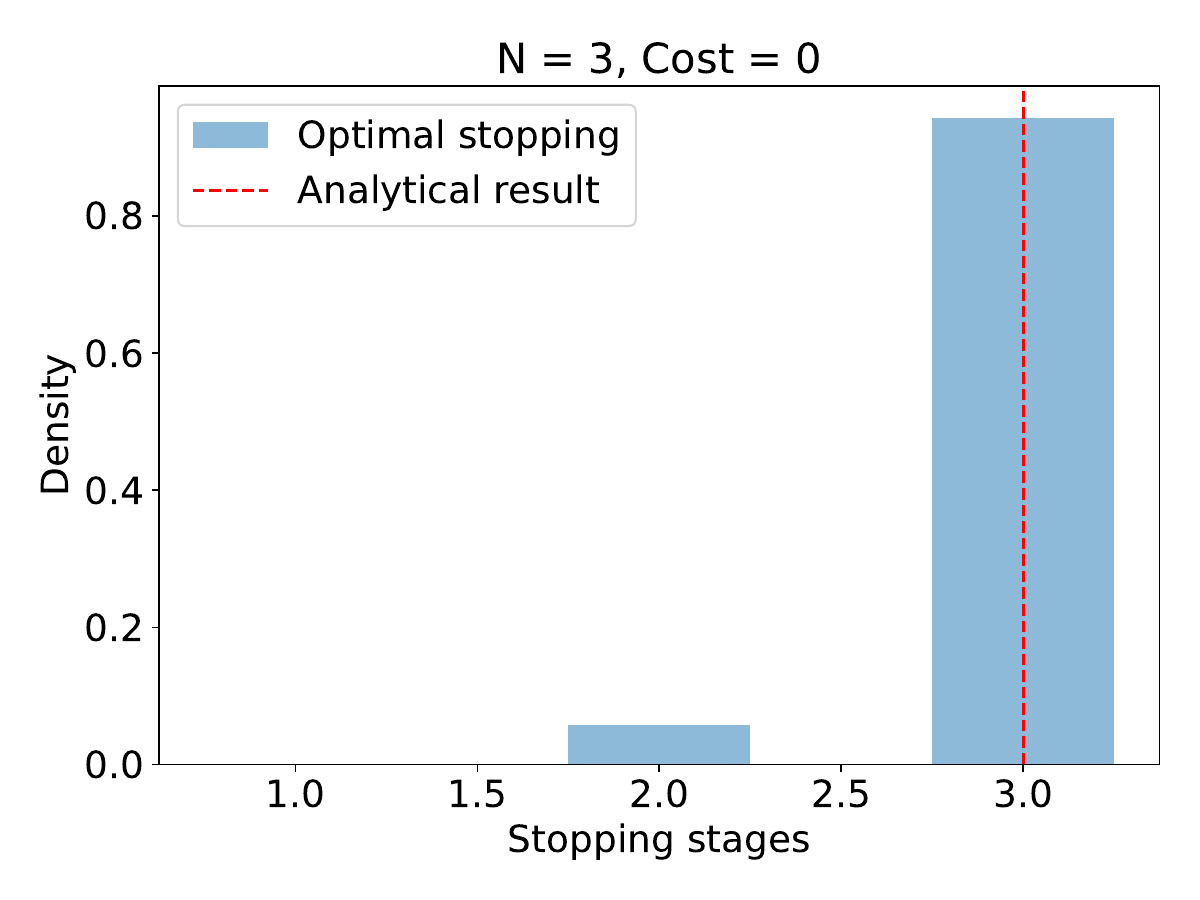}
  \includegraphics[width=0.45\linewidth]{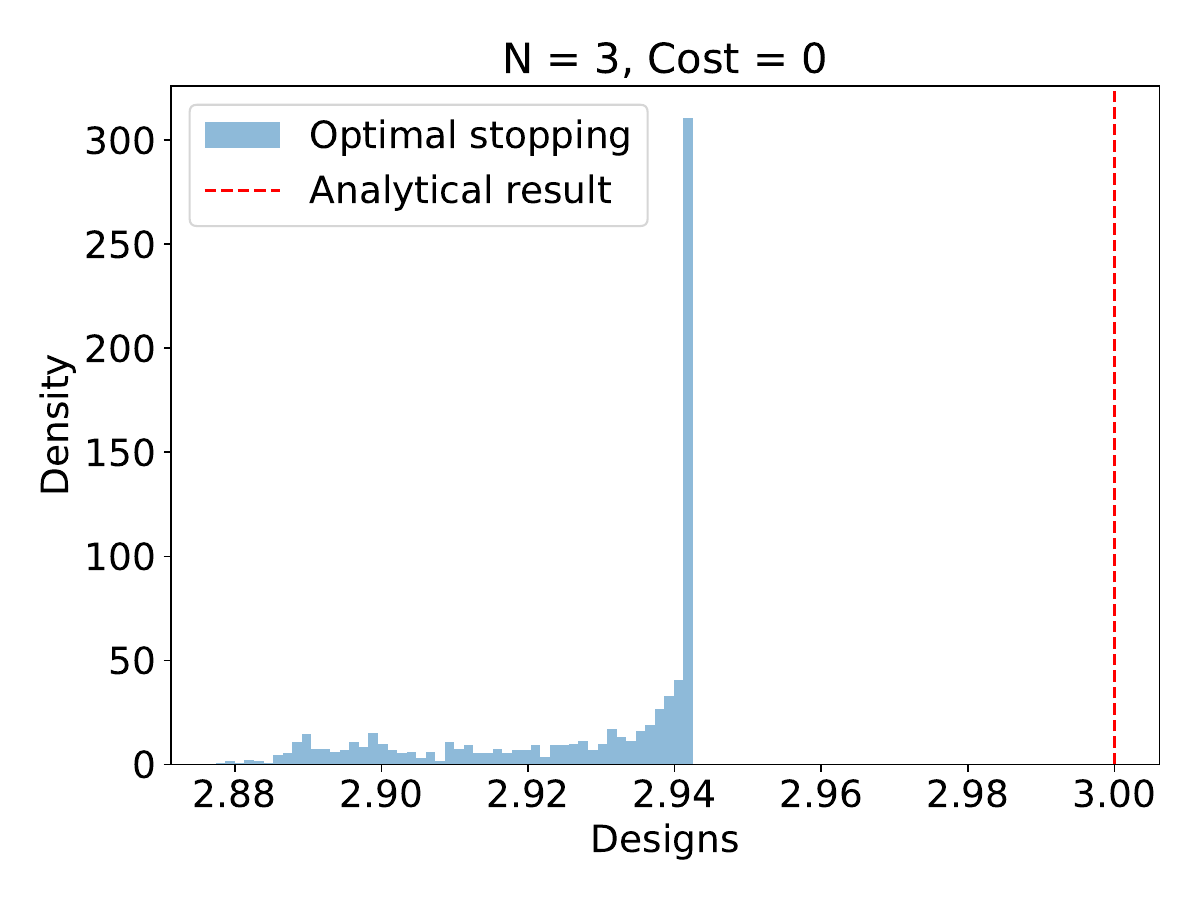}
  \caption{Distributional analysis for the linear-Gaussian benchmark with zero experimental cost ($c_k=0$). Left: stopping-stage distribution. Right: design distribution.}
  \label{fig:linear_gaussian_histograms}
\end{figure}

\Cref{fig:linear_gaussian_histograms} further confirms that the trained policy recovers the analytical structure of the problem. With zero experimental cost, the stopping-stage distribution concentrates near the maximum allowable stage, while the design distribution concentrates near the upper design bound $\design=3$, where the signal-to-noise ratio is largest. The sharp peak near $\design=2.94$ is computed from the deterministic policy outputs recorded during training, rather than from the noise-perturbed designs used to generate exploratory training trajectories. Since the linear-Gaussian optimum is observation-independent, the deterministic policy outputs are nearly identical across trajectories once the network has saturated near the upper design bound. The small gap from the analytical optimum $\design^\ast=3$ is expected for the standard bounded-output neural parameterization used here: unconstrained network outputs are mapped to the feasible design interval through a sigmoid transformation, so finite pre-activation values can approach but do not exactly attain the upper bound. Thus, the learned policy recovers both the expected stopping behavior and the expected design behavior in this benchmark.

\subsection{One-dimensional nonlinear test case}
\label{sec:case_nonlinear}

We next consider a one-dimensional nonlinear case to assess the method beyond the conjugate setting. The observation model is
\begin{align}
Y_k
=
G(\Param,\design_k)+\Epsilon_k
=
\Param^3 \design_k^2
+
\Param \exp\left(-|0.2-\design_k|\right)
+
\Epsilon_k,
\qquad
\Epsilon_k \sim \mathcal{N}(0,0.03^2),
\label{eq:nonlinear_model}
\end{align}
with prior $\Param\sim\mathcal{U}(0,1)$ and design constraint $\design_k\in[0,1]$. In contrast to the linear-Gaussian benchmark, this model is non-conjugate: the posterior is not Gaussian, and the continuation value depends on realized observations through the evolving posterior shape. Consequently, the stopping rule is adaptive. The stopping stage is now a trajectory-dependent random variable rather than a fixed deterministic stage, and its average need not be an integer. At the same time, the scalar parameter allows accurate grid-based posterior inference, so this example isolates the effect of nonlinearity without introducing high-dimensional posterior approximation error.

\begin{figure}[htbp]
  \centering
    \includegraphics[width=0.32\linewidth]{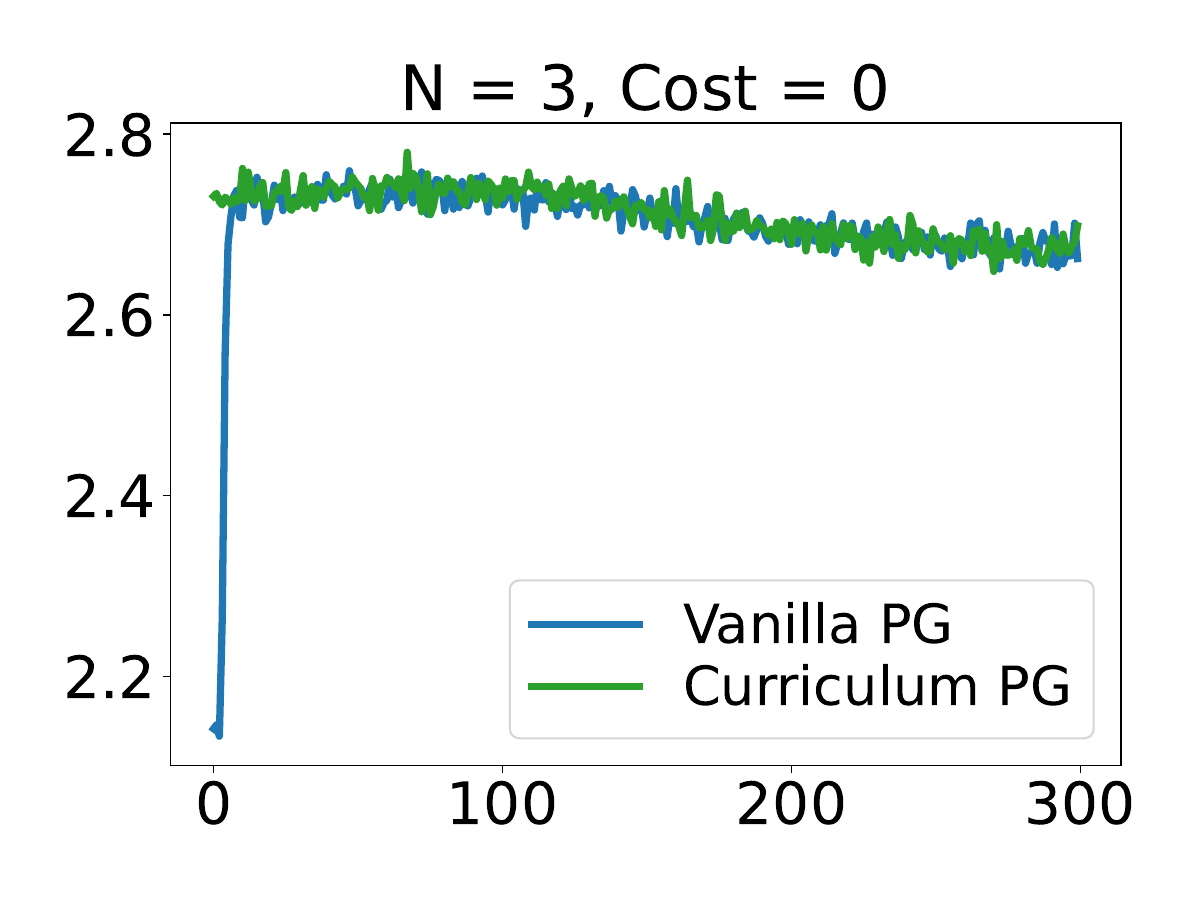}
    \includegraphics[width=0.32\linewidth]{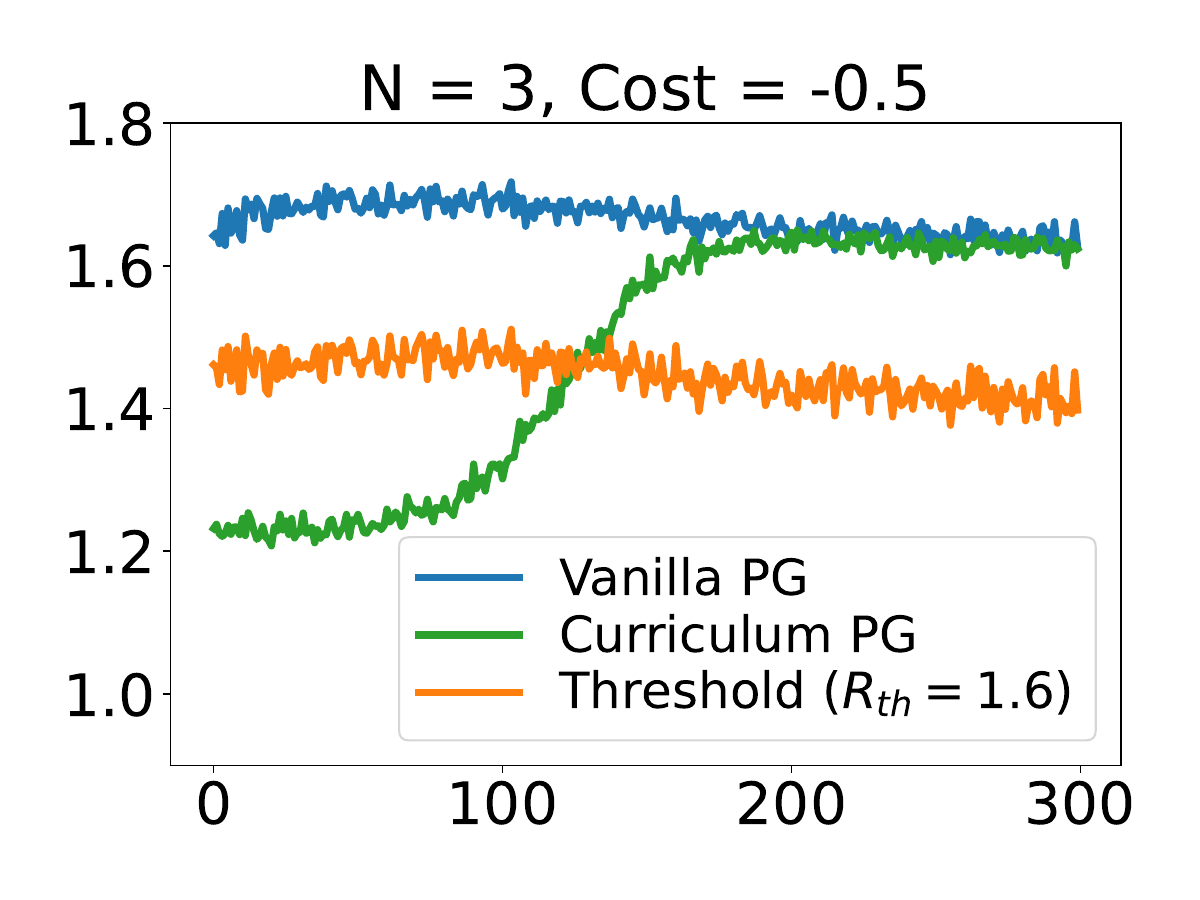}
    \includegraphics[width=0.32\linewidth]{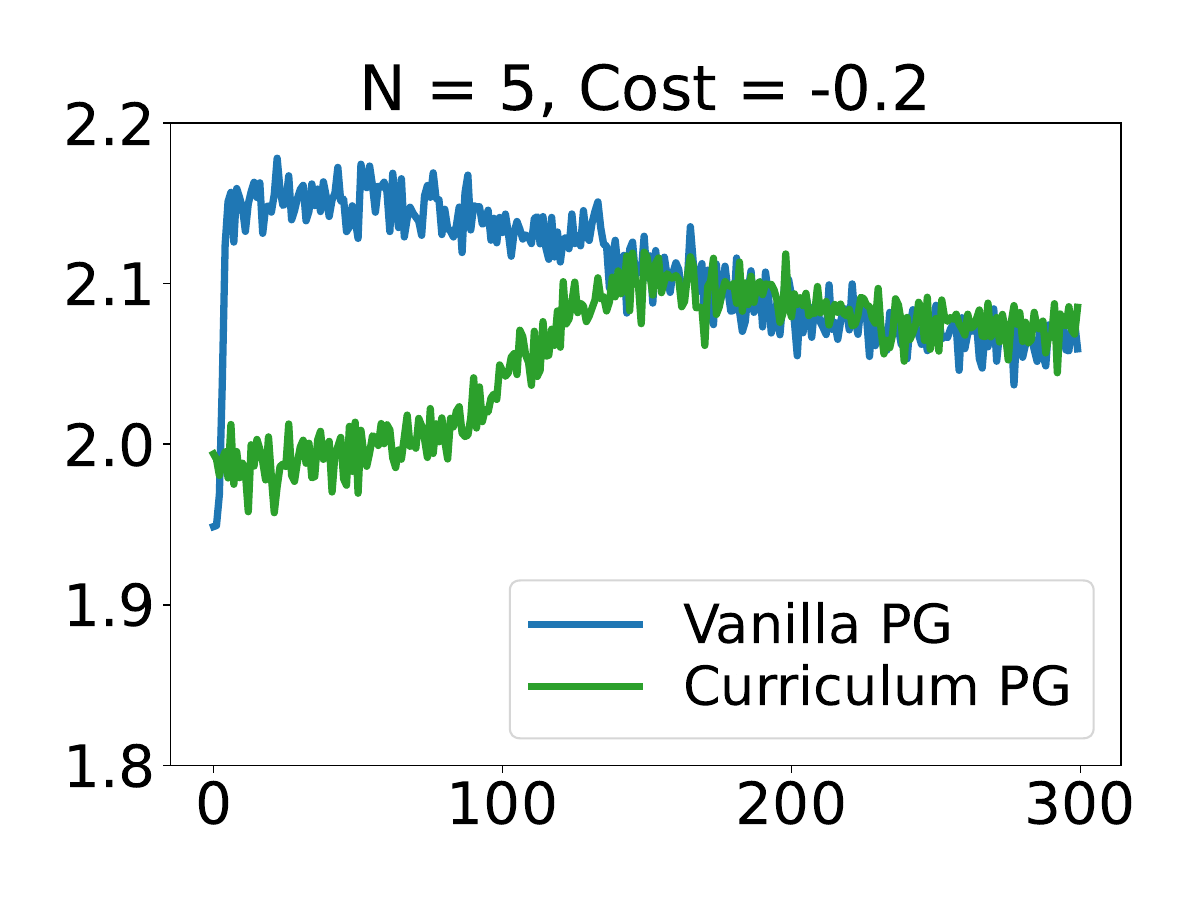}\\
    \includegraphics[width=0.32\linewidth]{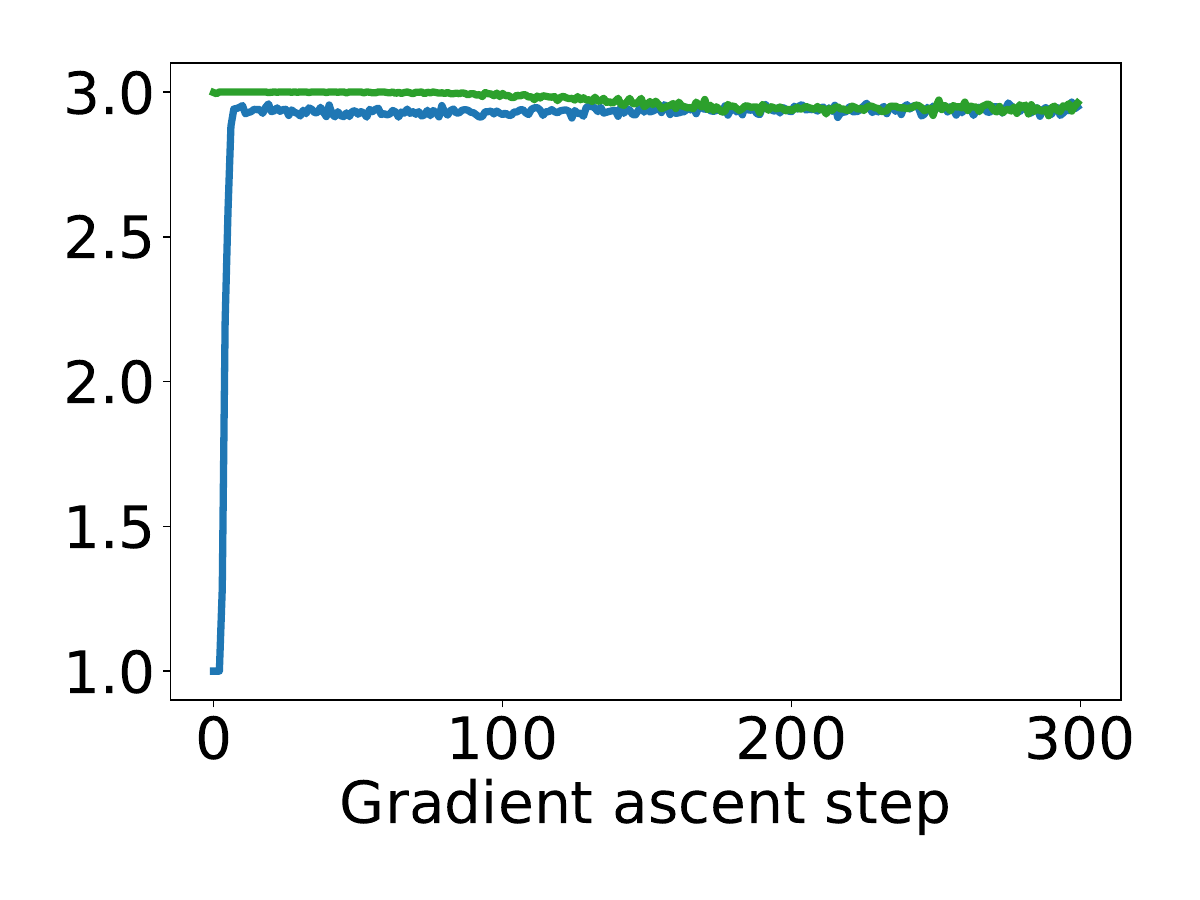}
    \includegraphics[width=0.32\linewidth]{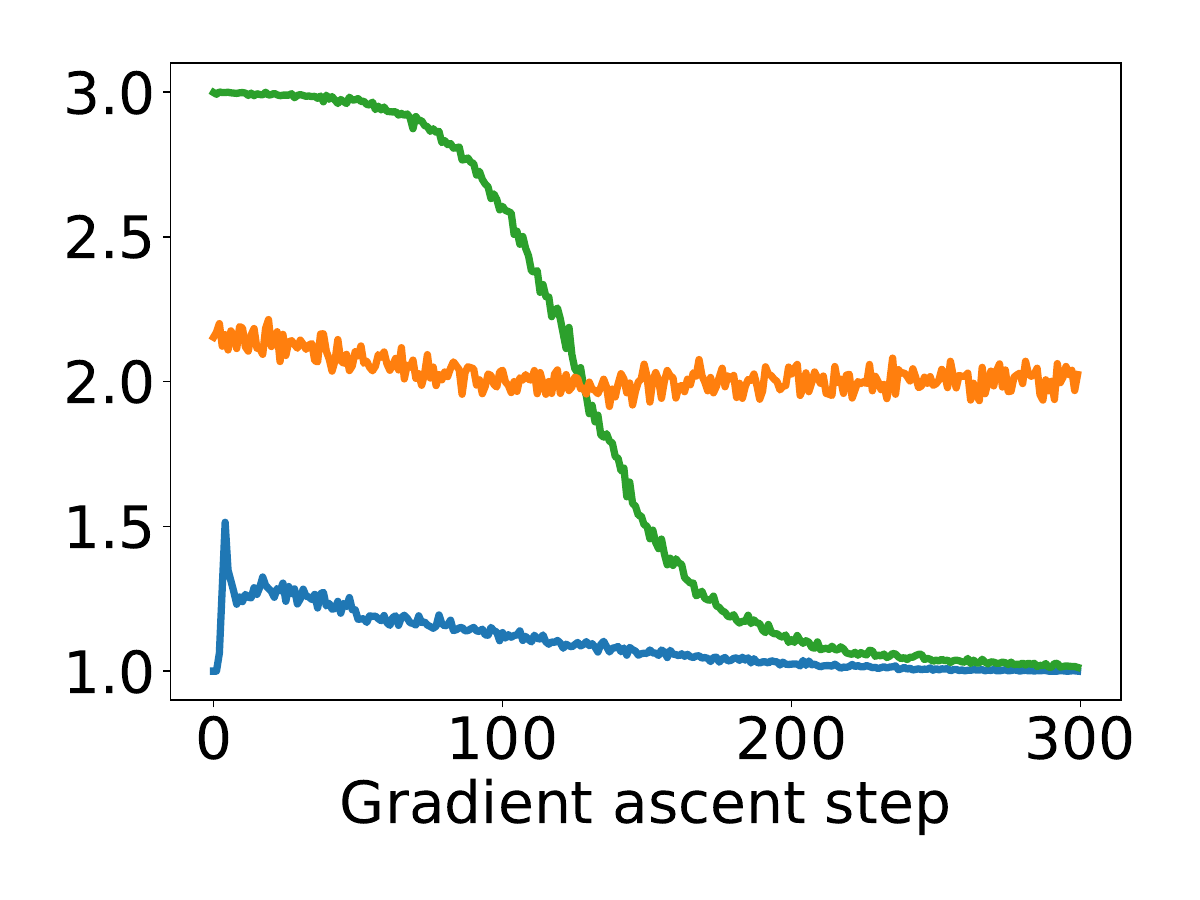}
    \includegraphics[width=0.32\linewidth]{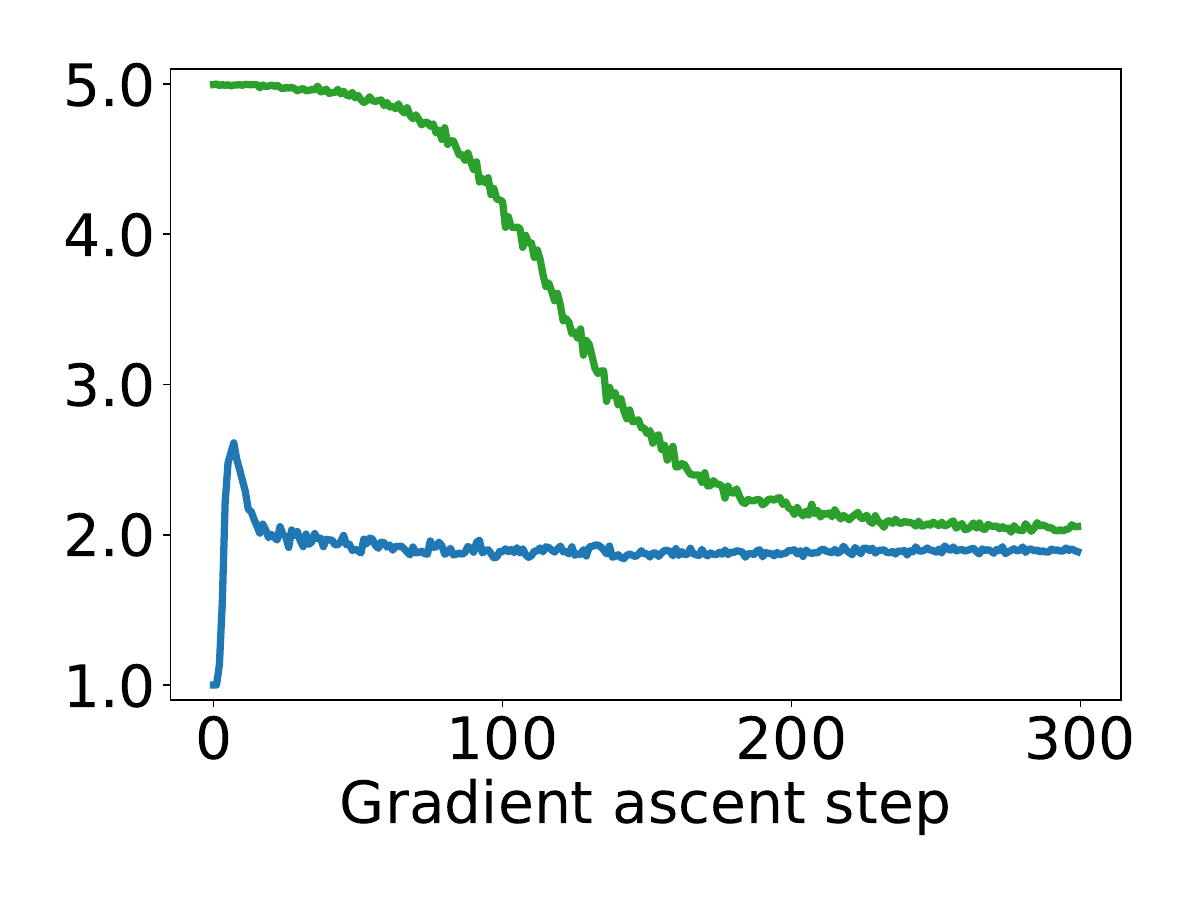}
  \caption{Training convergence for the one-dimensional nonlinear test problem. Top: average reward. Bottom: average stopping stage. Columns show representative combinations of maximum horizon $N$ and experimental cost.}
  \label{fig:nonlinear_convergence}
\end{figure}

\Cref{fig:nonlinear_convergence} summarizes three representative settings. With zero experimental cost and maximum horizon $N=3$, both vanilla and curriculum training rapidly reach comparable rewards, and the average stopping stage remains near the maximum allowable stage. This is consistent with the nonnegative expected information-gain contribution of additional measurements in the absence of experimental cost.

When a constant cost $c_k=-0.5$ is imposed for the same maximum horizon, the learned behavior shifts toward early termination. Both methods converge to stopping after approximately one experiment, indicating that the first nonlinear observation provides enough information that further measurements are not worth their cost. The curriculum method initially explores longer trajectories, as intended, and then transitions to the same early-stopping regime as the stopping rule is enforced more frequently during training. In contrast, the threshold-based rule with threshold 1.6 tends to terminate later, around stage 2, and consequently achieves a lower reward.

For the longer-horizon case $N=5$ with moderate cost $c_k=-0.2$, both methods converge to similar final rewards and stopping stages, with termination occurring around stage two. The training curves exhibit larger fluctuations than in the previous two settings. This behavior is consistent with a flatter information-cost tradeoff, where adjacent stopping stages yield similar expected utilities. In the nonlinear problem, this tradeoff is further complicated by observation-dependent posterior updates, where each measurement changes the posterior shape and hence the value of subsequent designs. Consequently, small changes in the learned policy can shift trajectories among several nearly equivalent stopping behaviors, producing visible fluctuations even when the final performance is comparable.

\subsection{Contaminant source detection in a convection-diffusion field}
\label{sec:case_source_detection}

We next consider a mobile-sensor problem for contaminant source detection. This example exhibits stronger sequential dependence than the preceding two cases. In the linear-Gaussian benchmark, the optimal design and stopping decisions are observation-independent. In the one-dimensional nonlinear example, observations affect future decisions through the updated posterior. Here, the dependence is stronger still because the design variable is a sensor displacement, so the current sensor location depends on the cumulative history of previous designs, while the posterior depends on the cumulative history of previous observations. Effective experimentation therefore requires jointly learning where to move the sensor and when further movement is no longer worth the cost.

Contaminant transport is modeled by a convection-diffusion equation for the concentration field $G(z,t;\Param)$ on a two-dimensional square domain:
\begin{align}
\frac{\partial G(z,t;\Param)}{\partial t}
=
\nabla^2 G(z,t;\Param)
-
u(t)\cdot \nabla G(z,t;\Param)
+
S(z,t;\Param),
\qquad
z\in [z_L,z_R]^2,\quad t>0 .
\label{eq:convection_diffusion_pde}
\end{align}
The source term is Gaussian,
\begin{align}
S(z,t;\Param)
=
\frac{q}{2\pi h^2}
\exp\left(
-
\frac{
(\Param_x-z_x)^2+(\Param_y-z_y)^2
}{
2h^2
}
\right),
\label{eq:source_term}
\end{align}
where $\Param=(\Param_x,\Param_y)$ is the unknown source location, $h$ is the known source width, and $q$ is the known source strength. The velocity field is $u(t)=[u_x(t),u_y(t)]^\top$. In the numerical experiment, we infer the source location with prior
\begin{align}
\Param_x,\Param_y \sim \mathcal{U}(0,1),
\end{align}
and fix $h=0.05$, $q=2$, and $u_x(t)=u_y(t)=10t/0.2$.

A mobile sensor measures the concentration field at discrete times $t_k=k\Delta t$, $k=0,\ldots,N-1$, with $\Delta t=5.0\times 10^{-2}$. The observation model is
\begin{align}
Y_k
=
G(s_{k+1}^p,t_k;\Param)
+
\Epsilon_k,
\qquad
\Epsilon_k\sim\mathcal{N}(0,\sigma_{\epsilon}^2),
\label{eq:source_detection_observation}
\end{align}
where $s_k^p$ denotes the physical sensor location. The design $\design_k$ is the sensor displacement, so the physical state evolves as
\begin{align}
s_{k+1}^p
=
s_k^p+\design_k .
\label{eq:sensor_motion}
\end{align}
Thus, unlike the previous examples, future admissible sensor locations depend directly on previous design choices.

\begin{figure}[htbp]
  \centering
  \includegraphics[width=\linewidth]{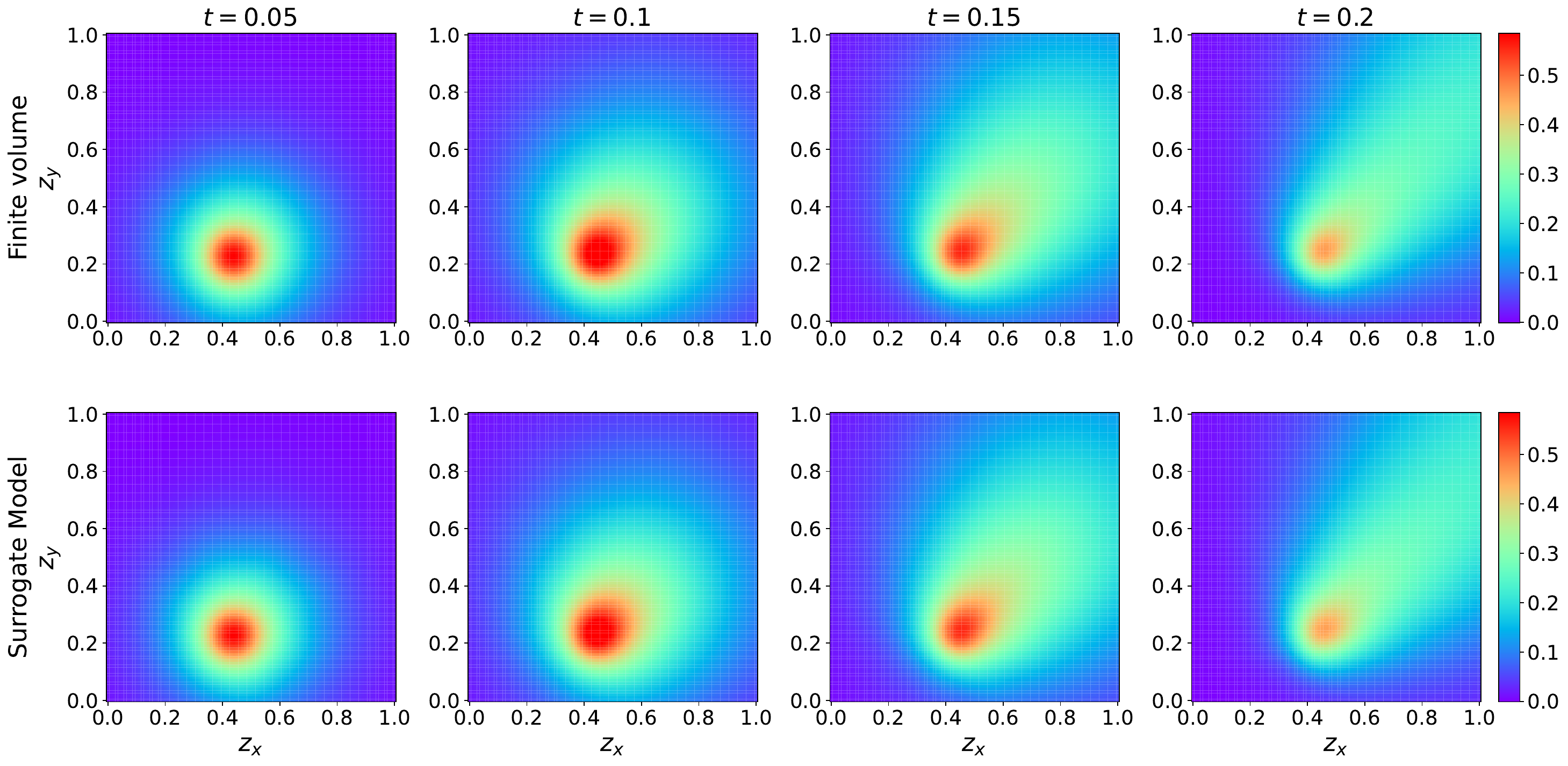}
  \caption{Comparison of the concentration field $G$ at four time points for $\param_x=0.43$ and $\param_y=0.22$, computed using the finite-volume solver (upper panel) and the neural-network surrogate (lower panel). The finite-volume solution is computed on the wider domain $[-1,2]^2$ and displayed on the region of interest $[0,1]^2$.}
  \label{fig:surrogate_comparison}
\end{figure}

The forward model in \cref{eq:convection_diffusion_pde} is solved numerically using a second-order finite-volume method on a uniform grid with spatial resolution $\Delta z_x=\Delta z_y=0.01$, together with a second-order fractional-step time integrator with time step $\Delta t_{\mathrm{PDE}}=5.0\times 10^{-4}$. Because repeated forward solves are expensive during policy training, we construct neural-network surrogate models for $G(z,t_k;\param)$ at each measurement time $t_k$. Each surrogate takes the input $(z,\param)$, where $\param=(\param_x,\param_y)$ is a source-location realization, and maps it through five hidden layers with 40, 80, 40, 20, and 10 neurons, respectively, to a scalar concentration output. The training set is generated from 2000 prior samples $\param^{(i)}$, restricted to the reachable region, and split 80\%/20\% into training and test sets. \Cref{fig:surrogate_comparison} compares concentration contours from the finite-volume solver and the neural-network surrogate. The contours are visually nearly indistinguishable, and the test mean-squared errors are on the order of $10^{-6}$. The surrogate also provides a speedup of approximately $10^5$ relative to the finite-volume solver, making repeated trajectory simulation feasible during training.

\begin{figure}[htbp]
  \centering
  \includegraphics[width=0.32\linewidth]{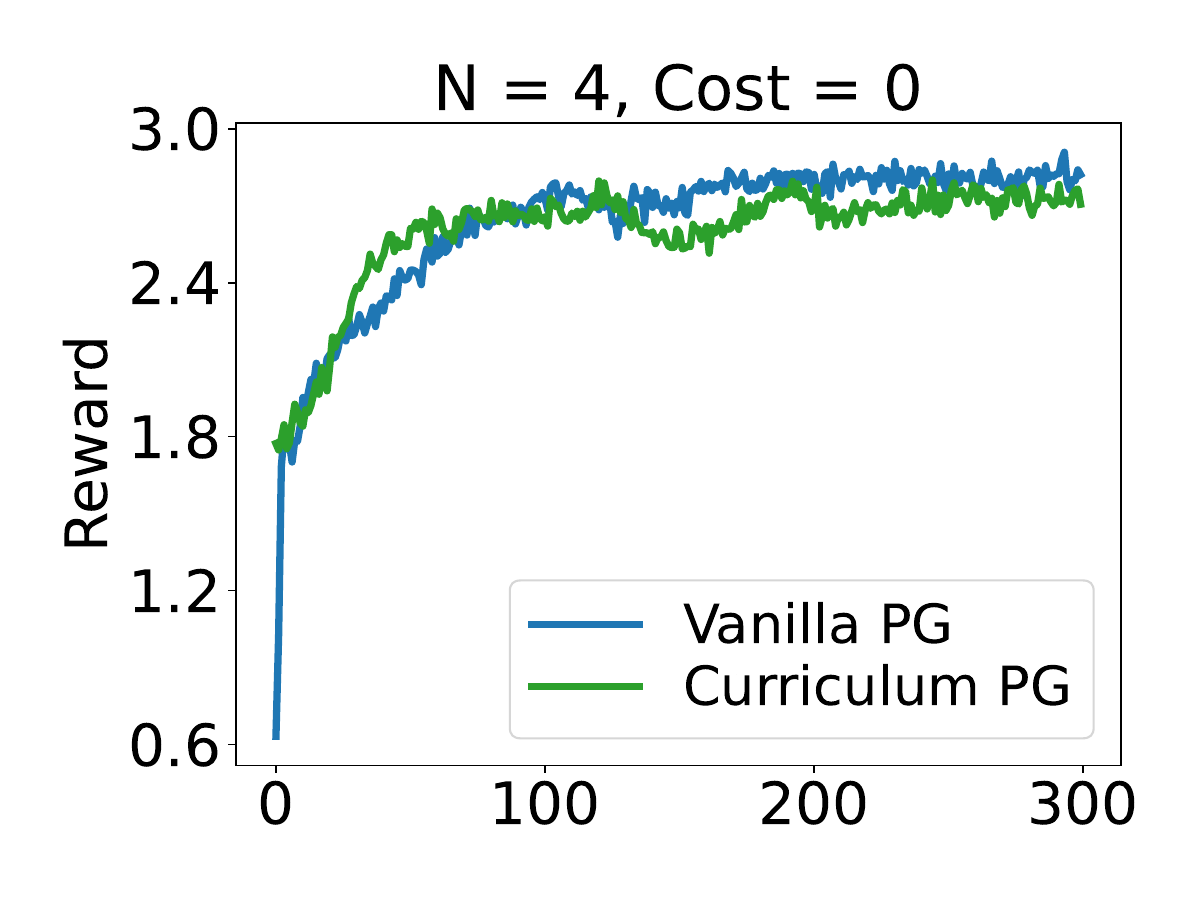}
  \includegraphics[width=0.32\linewidth]{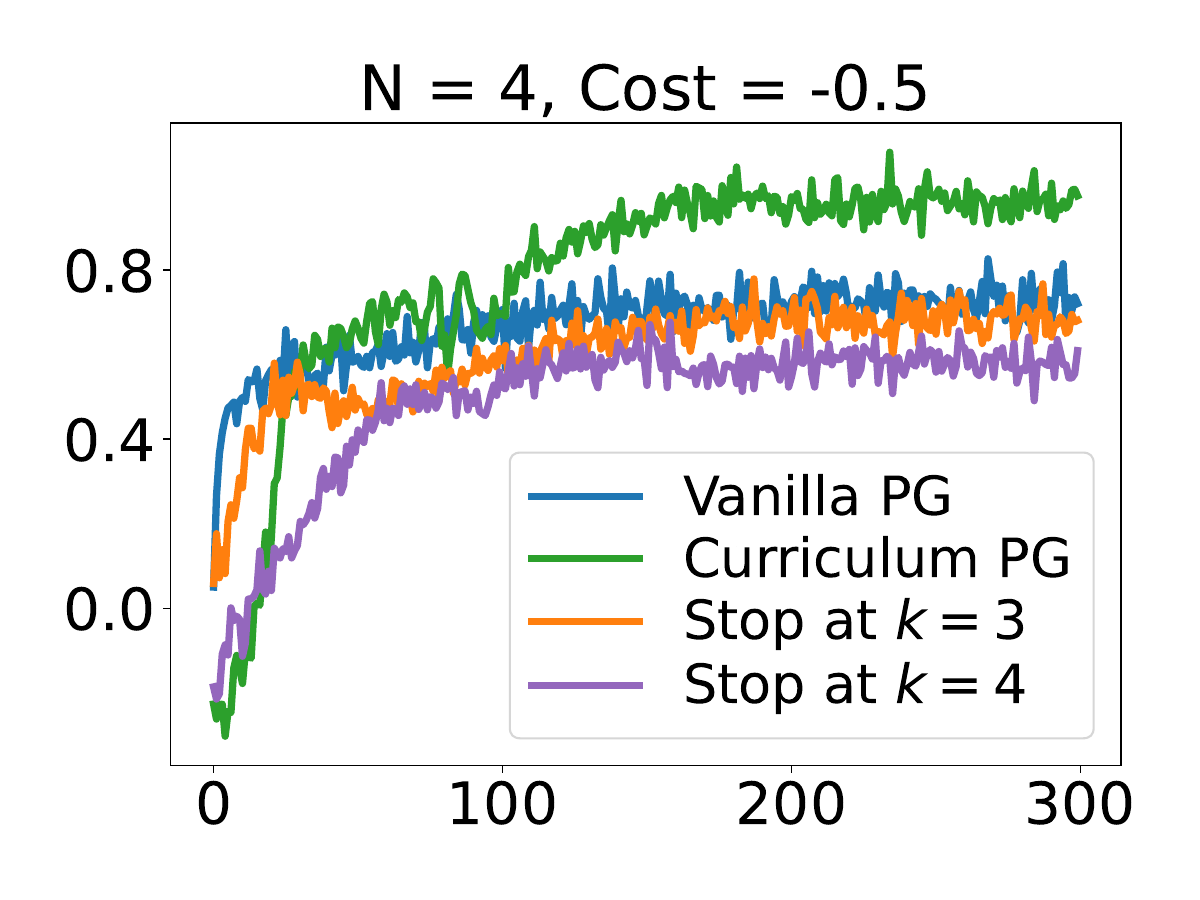}
  \includegraphics[width=0.32\linewidth]{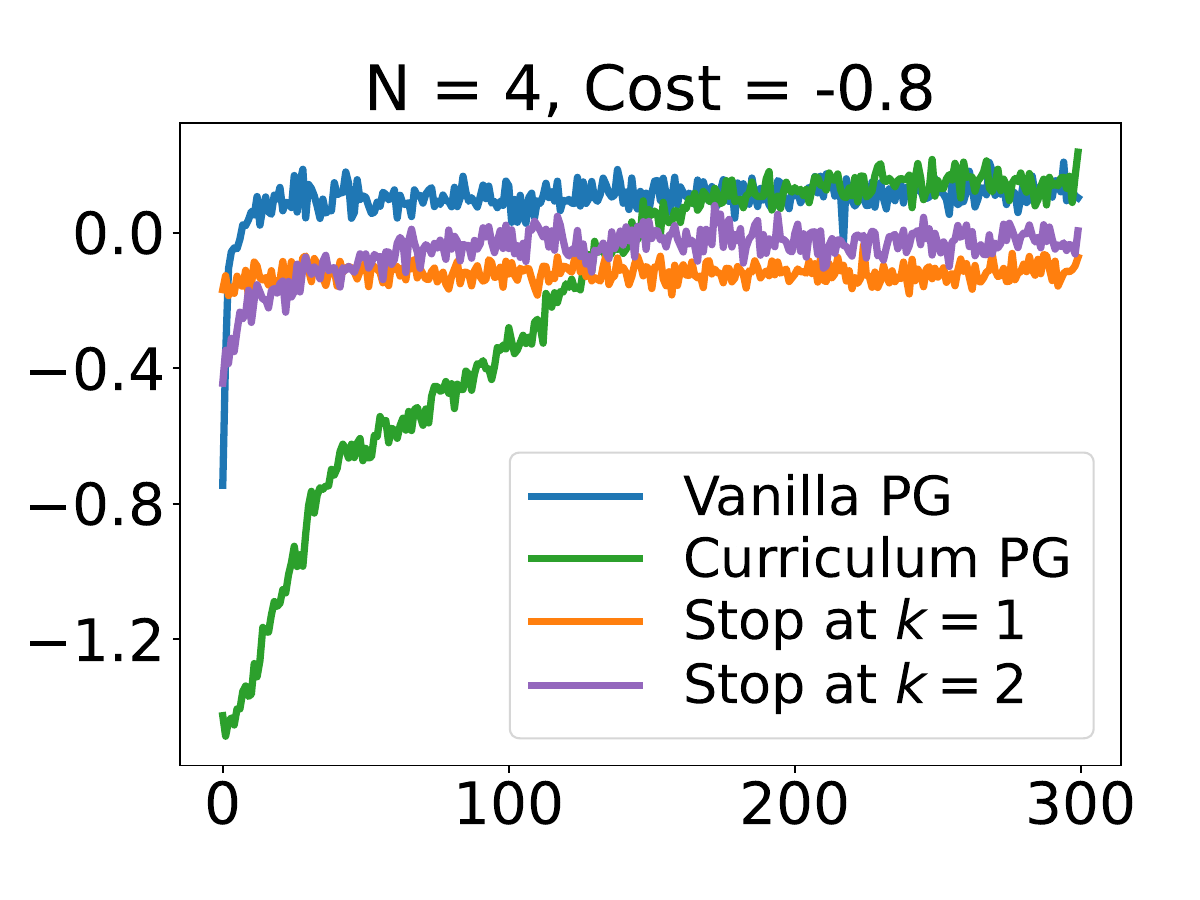}\\
  \includegraphics[width=0.32\linewidth]{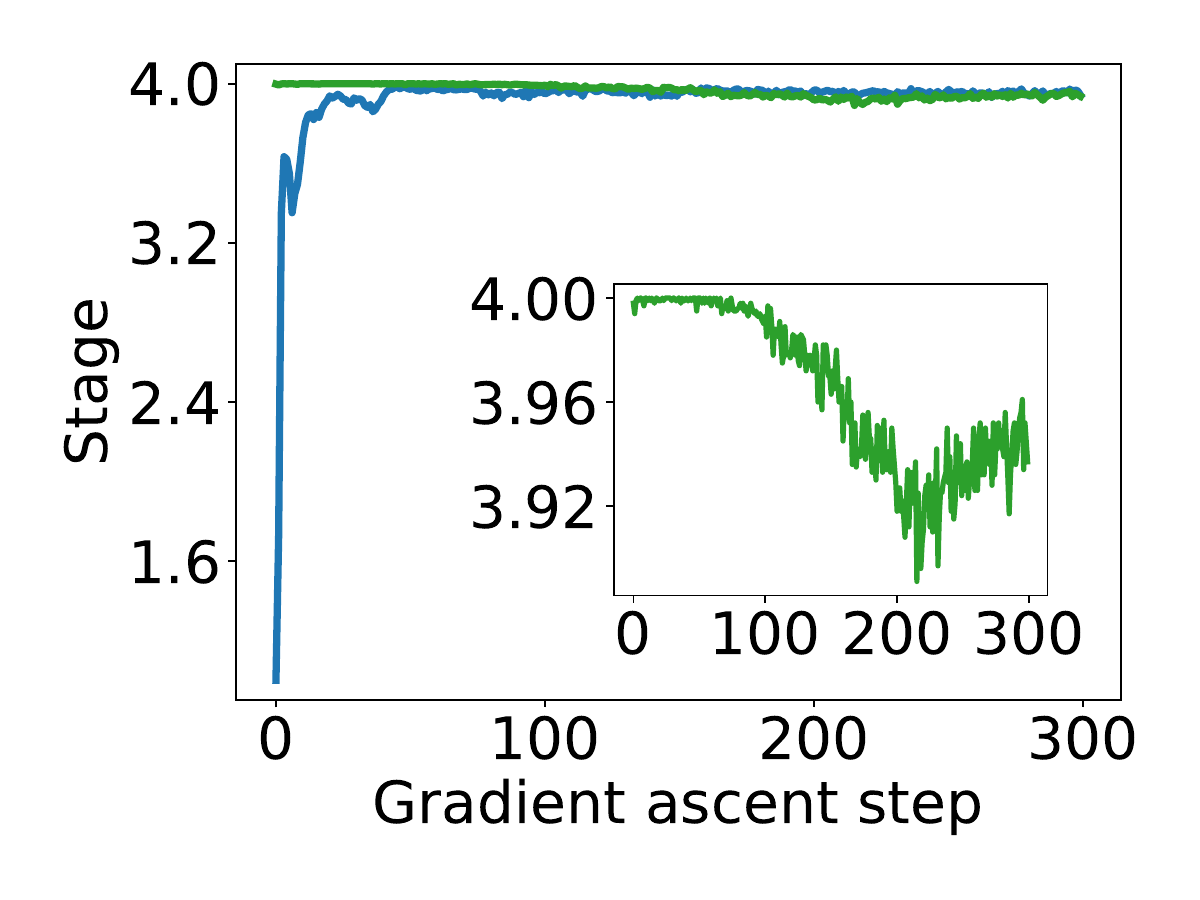}
  \includegraphics[width=0.32\linewidth]{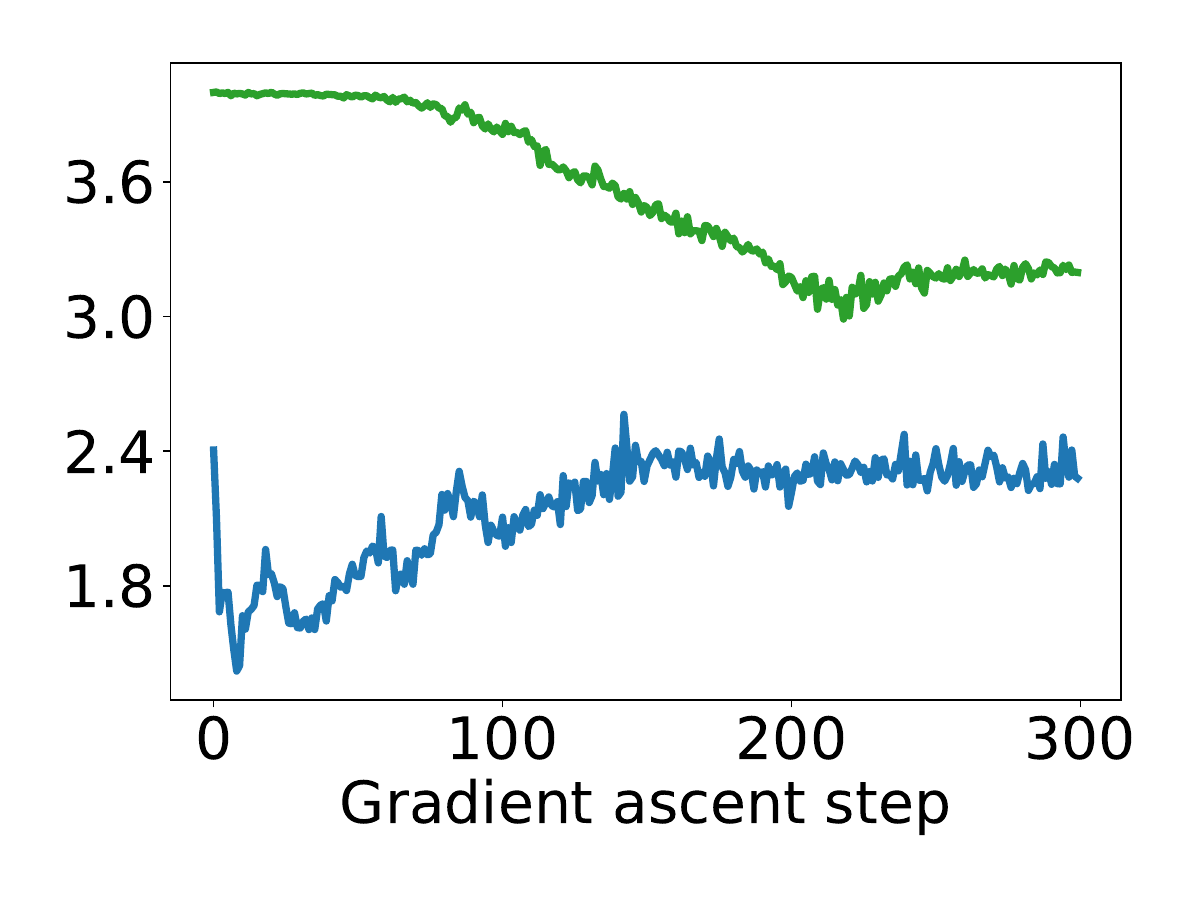}
  \includegraphics[width=0.32\linewidth]{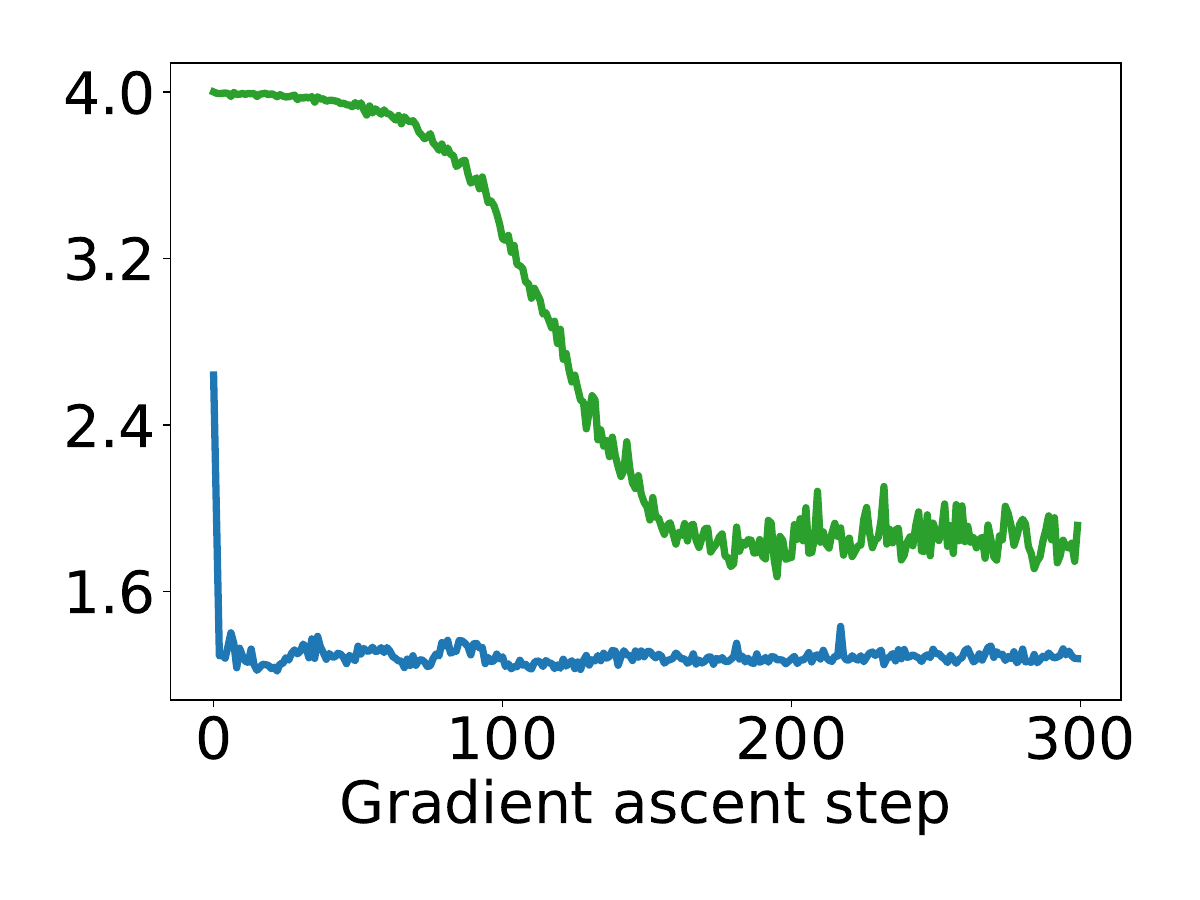}
  \caption{Training convergence for the convection-diffusion source detection problem with maximum horizon $N=4$. Top: average reward. Bottom: average stopping stage. Columns show different experimental costs.}
  \label{fig:convection_diffusion_convergence}
\end{figure}

\Cref{fig:convection_diffusion_convergence} shows training behavior for maximum horizon $N=4$ under three constant-cost settings. When $c_k=0$, both vanilla and curriculum training converge to policies that continue to the maximum allowable stage. This is expected since, without experimental cost, continuation is favored because additional measurements provide nonnegative expected information gain. Vanilla training reaches the high-reward regime rapidly, while curriculum training initially follows the forced-continuation schedule before approaching comparable performance.

The negative-cost cases reveal the main training challenge. Because the design variables are sensor movements rather than independent sensor locations, early movement decisions affect which regions can be sampled later. If the continuation-value approximation underestimates the value of future measurements early in training, the induced stopping rule truncates trajectories before the policy can experience informative later-stage movements. Those missing trajectories then reinforce the underestimated continuation value, creating a self-reinforcing early-stopping local optimum.

For moderate cost, $c_k=-0.5$, the two learning outcomes are clearly separated. Vanilla training converges to an earlier-stopping, lower-reward policy, with average stopping stage around 2.4. Curriculum training, by forcing longer trajectories early in training, allows the design policy and continuation-value approximation to encounter more informative later-stage sensor movements. It converges to a later-stopping policy, with average stopping stage slightly above 3 and substantially higher reward. This is the clearest example in our experiments in which the circular design--stopping dependency produces a practical failure mode, with training settling into a lower-reward early-stopping basin without curriculum learning.

For the larger cost, $c_k=-0.8$, vanilla and curriculum training again converge to different stopping behaviors, but the final rewards are similar. Vanilla training favors the earlier-stopping policy, while curriculum training favors a later-stopping policy because it maintains longer trajectories during early optimization. The comparable final rewards suggest that, in this setting, the reward landscape contains multiple near-optimal design--stopping strategies. Thus, curriculum learning does not always improve the final objective value, but it changes which basin of attraction the training procedure reaches.

\begin{figure}[htbp]
  \centering
  \begin{subfigure}[b]{\linewidth}
    \includegraphics[width=\linewidth]{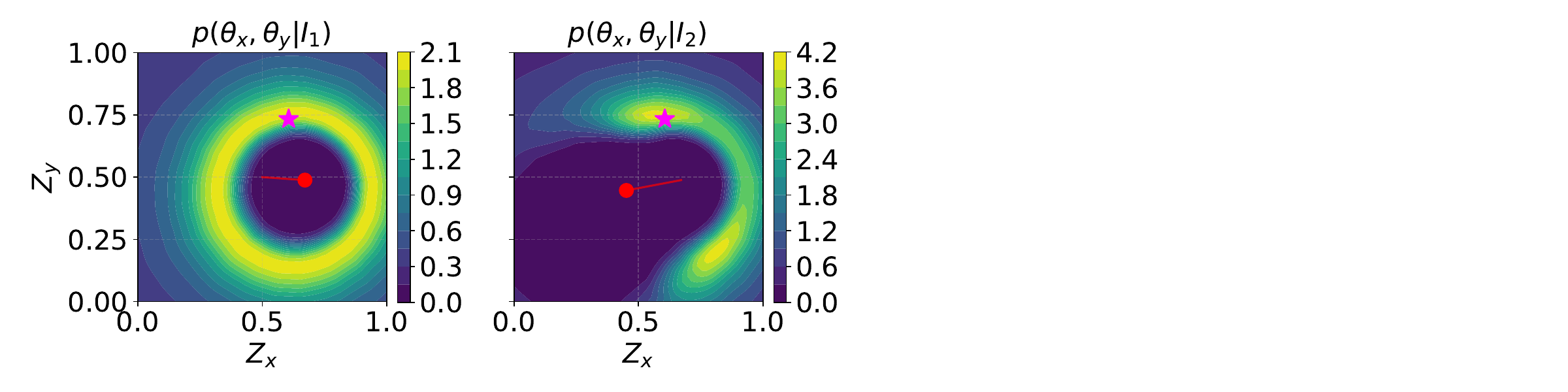}
    \caption{Vanilla policy, $\param=(0.61,0.73)$, total reward $=-0.564$ (information gain $=0.436$, experimental cost $=-1.0$).}
  \end{subfigure}

  \begin{subfigure}[b]{\linewidth}
    \includegraphics[width=\linewidth]{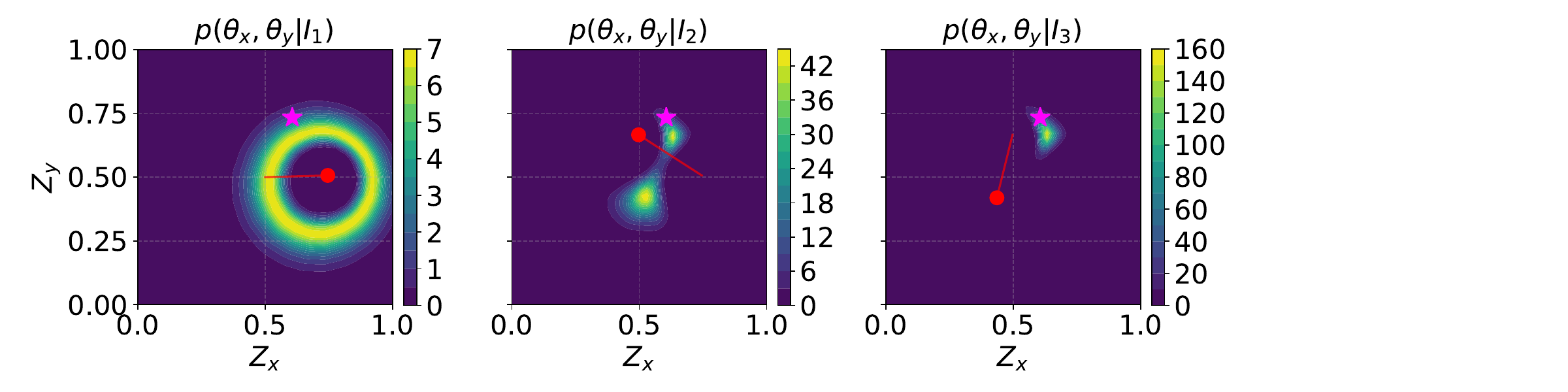}
    \caption{Curriculum policy, $\param=(0.61,0.73)$, total reward $=2.198$ (information gain $=3.698$, experimental cost $=-1.5$).}
  \end{subfigure}
  
  \caption{Representative source-detection episodes for the vanilla policy (upper panel) and curriculum policy (lower panel) in the $c_k=-0.5$ setting. Purple star: true source location. Red dots: sensor positions. Red lines: sensor movements. Contours: posterior density.}
  \label{fig:convection_diffusion_posterior}
\end{figure}

\Cref{fig:convection_diffusion_posterior} illustrates the difference between the learned policies for $c_k=-0.5$ on a representative source realization. Starting from the same initial condition, the two policies make different early movement decisions and therefore collect different subsequent measurements. The vanilla policy stops after two experiments and obtains total reward $-0.564$. The curriculum-trained policy continues for three experiments and obtains total reward $2.198$. This example shows that premature stopping affects not only the stopping rule but also the learned design policy itself, because the policy is trained only on the trajectories that survive the stopping decision.

\begin{figure}[htbp]
  \centering
  \includegraphics[width=0.49\linewidth]{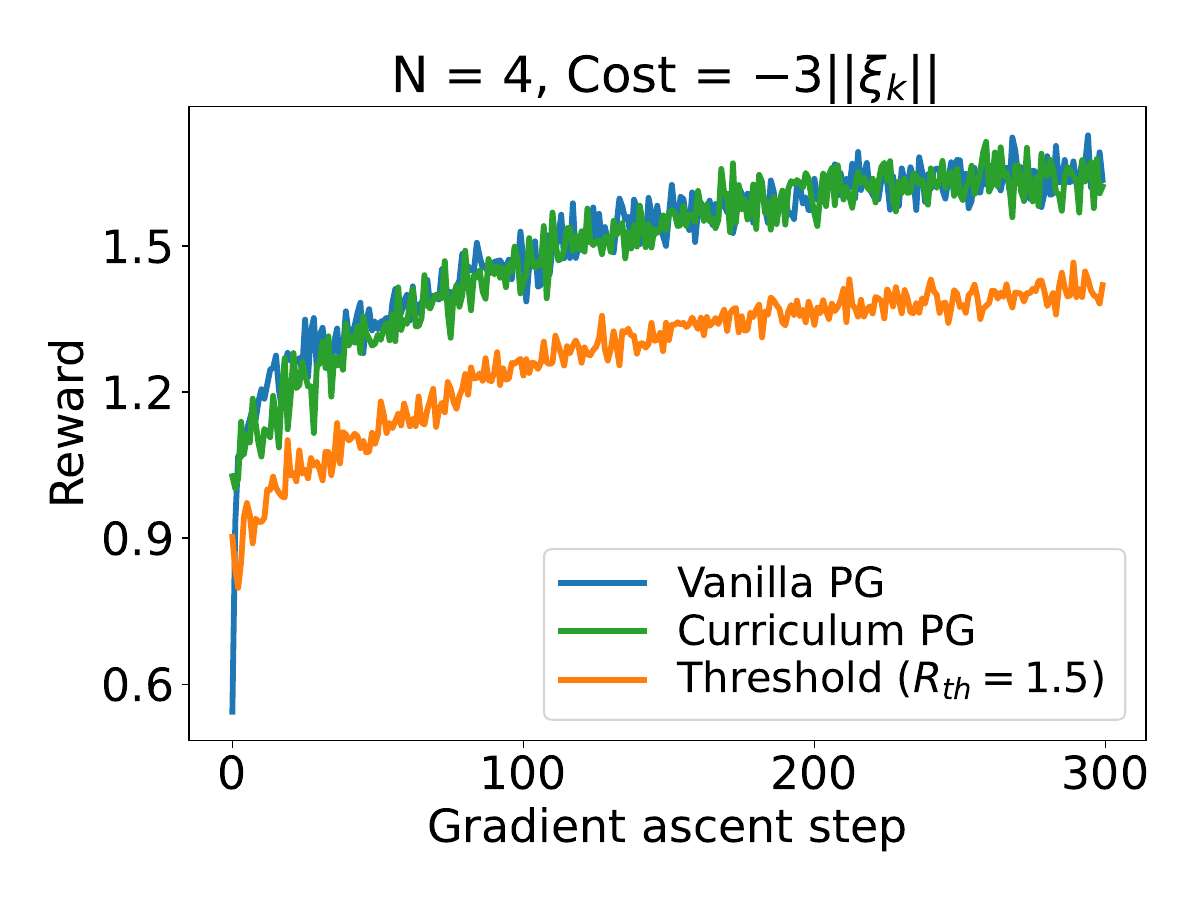}
  \includegraphics[width=0.49\linewidth]{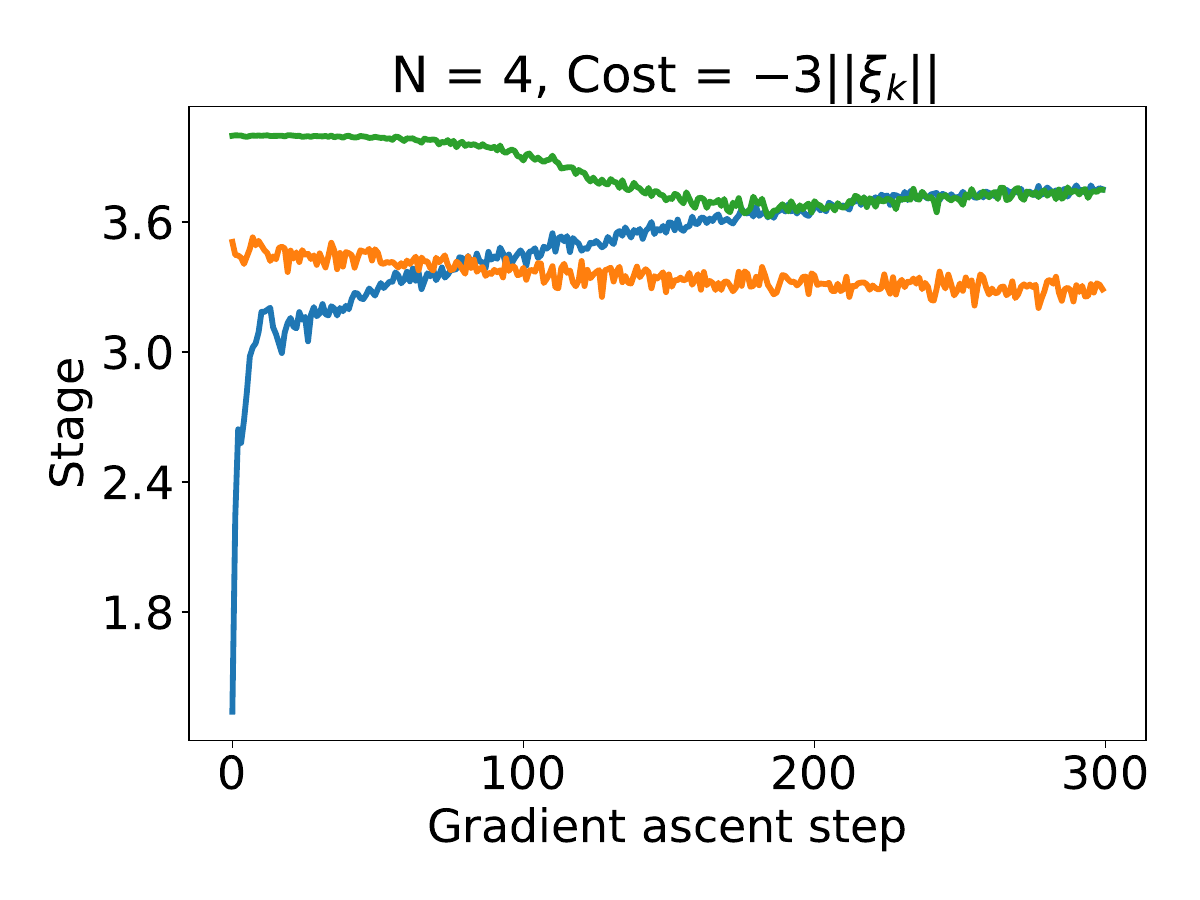}
  \caption{Training convergence for the convection-diffusion source detection problem with design-dependent cost $c_k(\design_k)=-3\|\design_k\|$. Left: average reward. Right: average stopping stage.}
  \label{fig:convection_diffusion_design_dependent_convergence}
\end{figure}

We also consider a design-dependent cost,
\begin{align}
c_k(\design_k)
=
-3\|\design_k\|,
\label{eq:source_detection_design_dependent_cost}
\end{align}
under which longer sensor movements incur larger penalties. This setting tests whether the framework can incorporate spatial movement costs into the joint design--stopping problem. \Cref{fig:convection_diffusion_design_dependent_convergence} shows that vanilla and curriculum training converge to similar rewards and stopping stages under the design-dependent cost, with average stopping stage approaching approximately 3.6--3.7. In this case, the movement penalty regularizes the design policy by discouraging large displacements, and the early-stopping failure mode is less pronounced than in the constant-cost $c_k=-0.5$ case. The threshold-based rule again underperforms the learned stopping policies.

\begin{figure}[htbp]
  \centering
  \includegraphics[width=\linewidth]{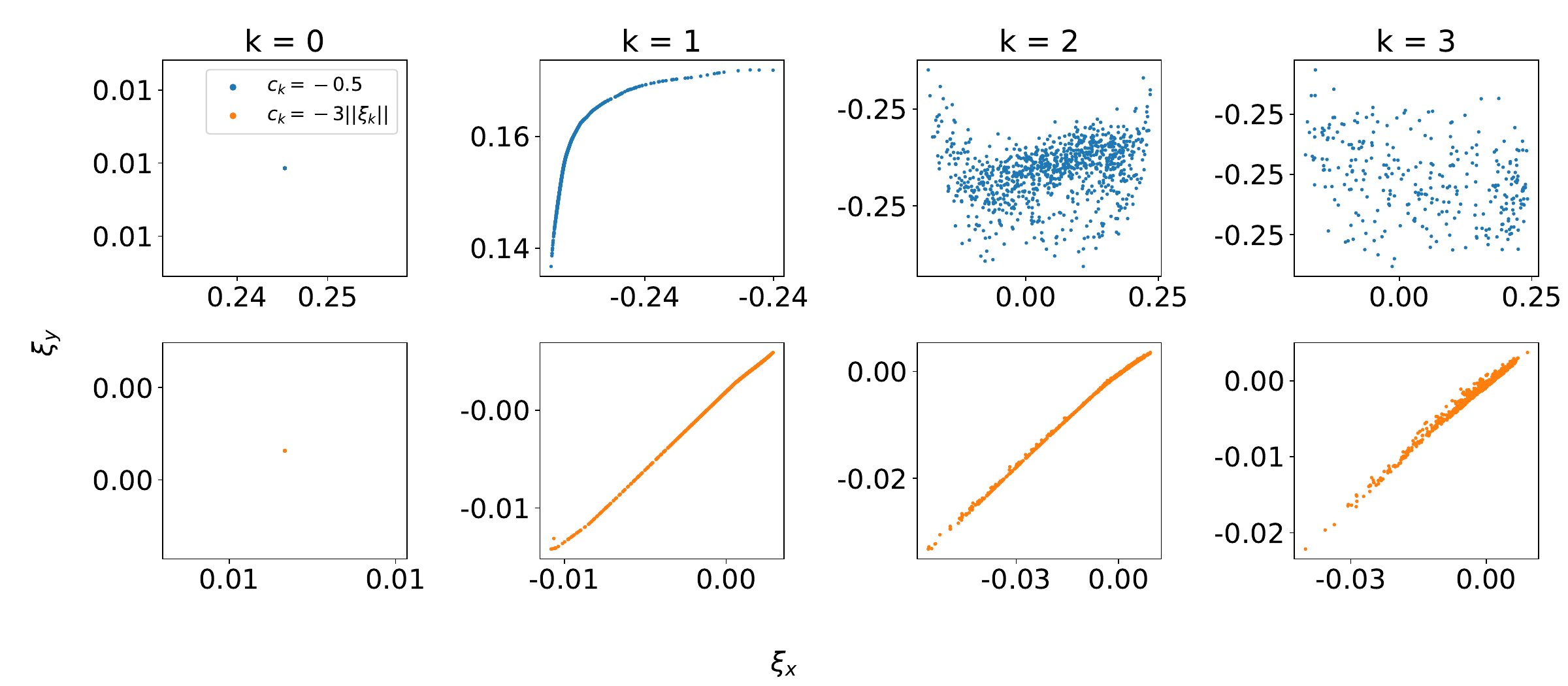}
  \caption{Comparison of learned sensor movement policies under constant and design-dependent costs in the convection-diffusion source detection problem. Upper panels: constant cost $c_k=-0.5$. Lower panels: design-dependent cost $c_k(\design_k)=-3\|\design_k\|$.}
  \label{fig:design_comparison}
\end{figure}

The design-dependent cost also changes the learned movement strategy. \Cref{fig:design_comparison} compares the spatial distribution of learned sensor movements under the constant and design-dependent costs across experimental stages. At stage $k=0$, the fixed initial state makes the learned movement deterministic within each cost setting, although the selected initial movement differs between the two cost structures. Under the constant cost, later-stage movements become more dispersed, especially for $k=2$ and $k=3$, because movement length is not directly penalized. Under the design-dependent cost, movements concentrate along a narrower trajectory with smaller magnitudes, reflecting a policy that balances information acquisition against movement penalty. This behavior demonstrates that the proposed framework can incorporate design-dependent costs without changing the decision-theoretic formulation.

\subsection{Curriculum schedule sensitivity}
\label{sec:curriculum_schedule_sensitivity}

\begin{figure}[htbp]
  \centering
  \includegraphics[width=.45\linewidth]{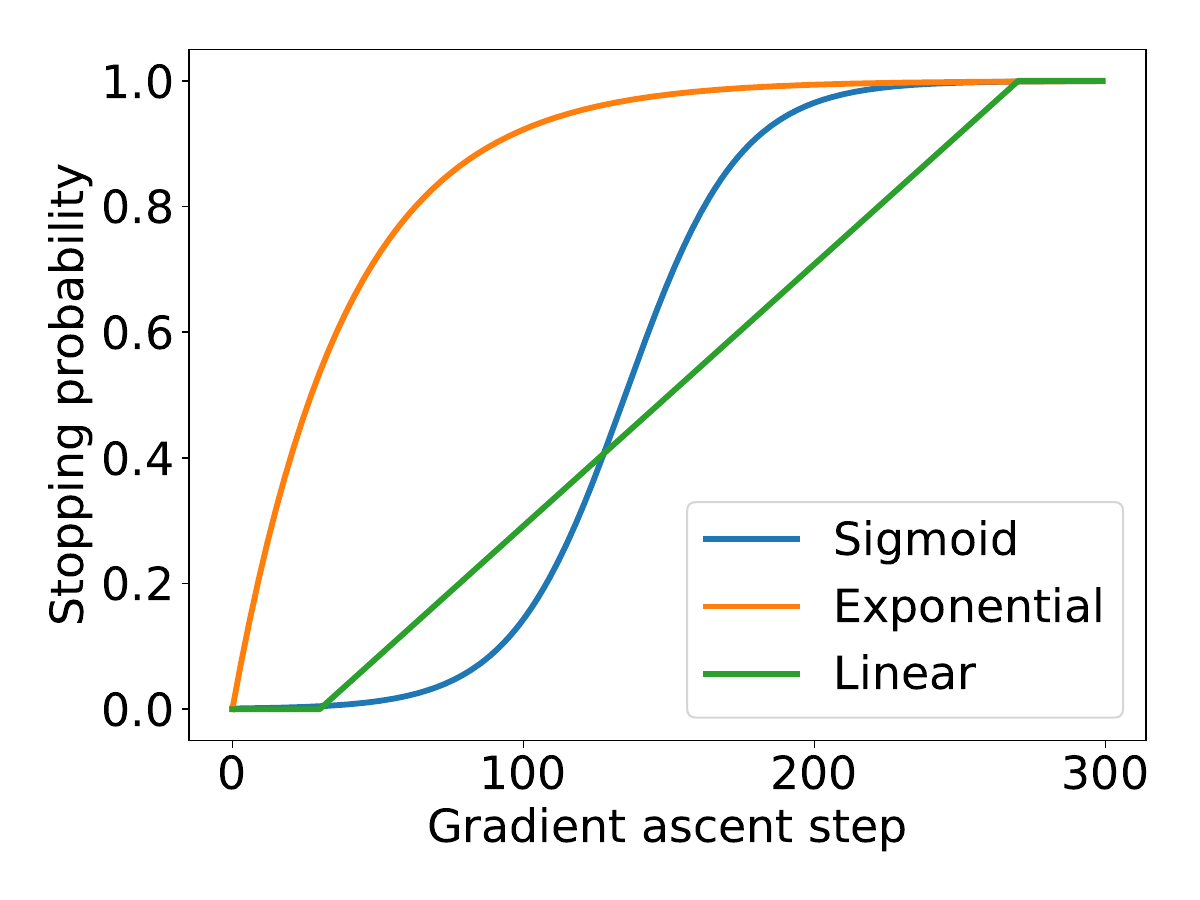}
  \caption{Representative stopping-probability schedules $p_{\mathrm{stop}}(\ell)$ over 300 training iterations.}
  \label{fig:stopping_probability}
\end{figure}

\begin{figure}[htbp]
  \centering
  \includegraphics[width=0.45\linewidth]{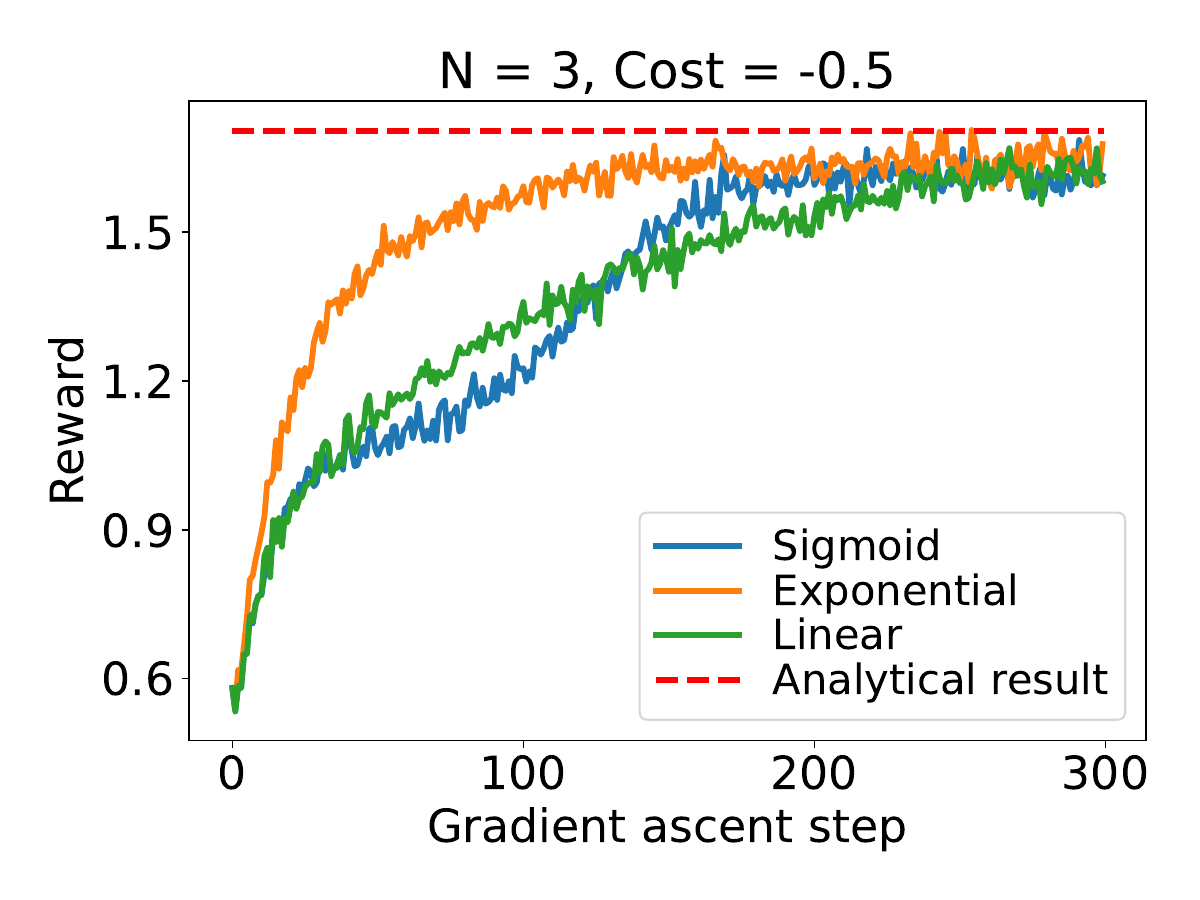}
  \includegraphics[width=0.45\linewidth]{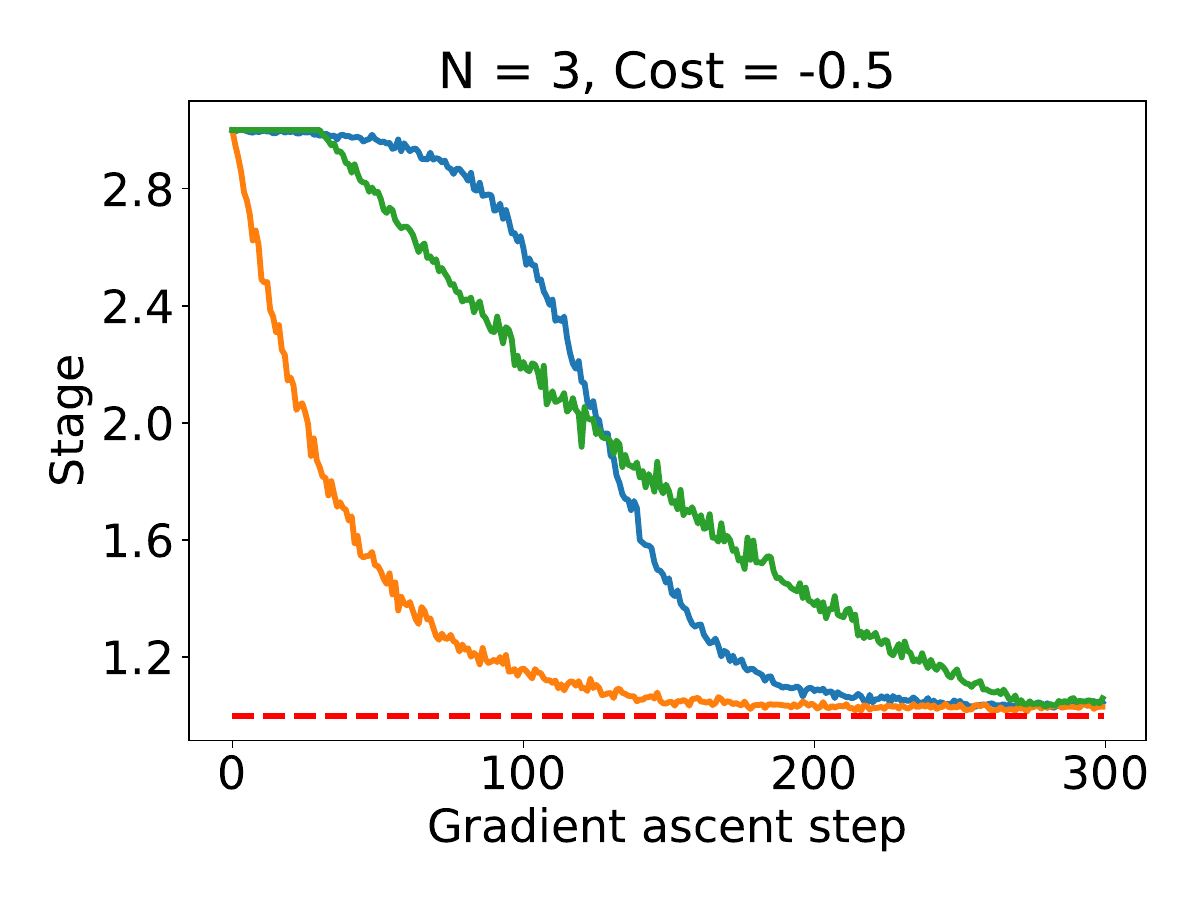}
  \caption{Comparison of curriculum schedules on the linear-Gaussian benchmark. Left: average reward. Right: average stopping stage.}
  \label{fig:linear_gaussian_schedule_comparison}
\end{figure}

Finally, we examine the role of the curriculum schedule. \Cref{fig:stopping_probability} shows representative schedules for the stopping probability $p_{\mathrm{stop}}(\ell)$, including linear, exponential, and sigmoid increases. The schedule comparison shows that simple problems are relatively insensitive to the precise curriculum shape. In the linear-Gaussian benchmark, shown in \cref{fig:linear_gaussian_schedule_comparison}, all schedules converge to the optimal policy, differing mainly in convergence speed and transient behavior.

\begin{figure}[htbp]
  \centering
  \includegraphics[width=0.45\linewidth]{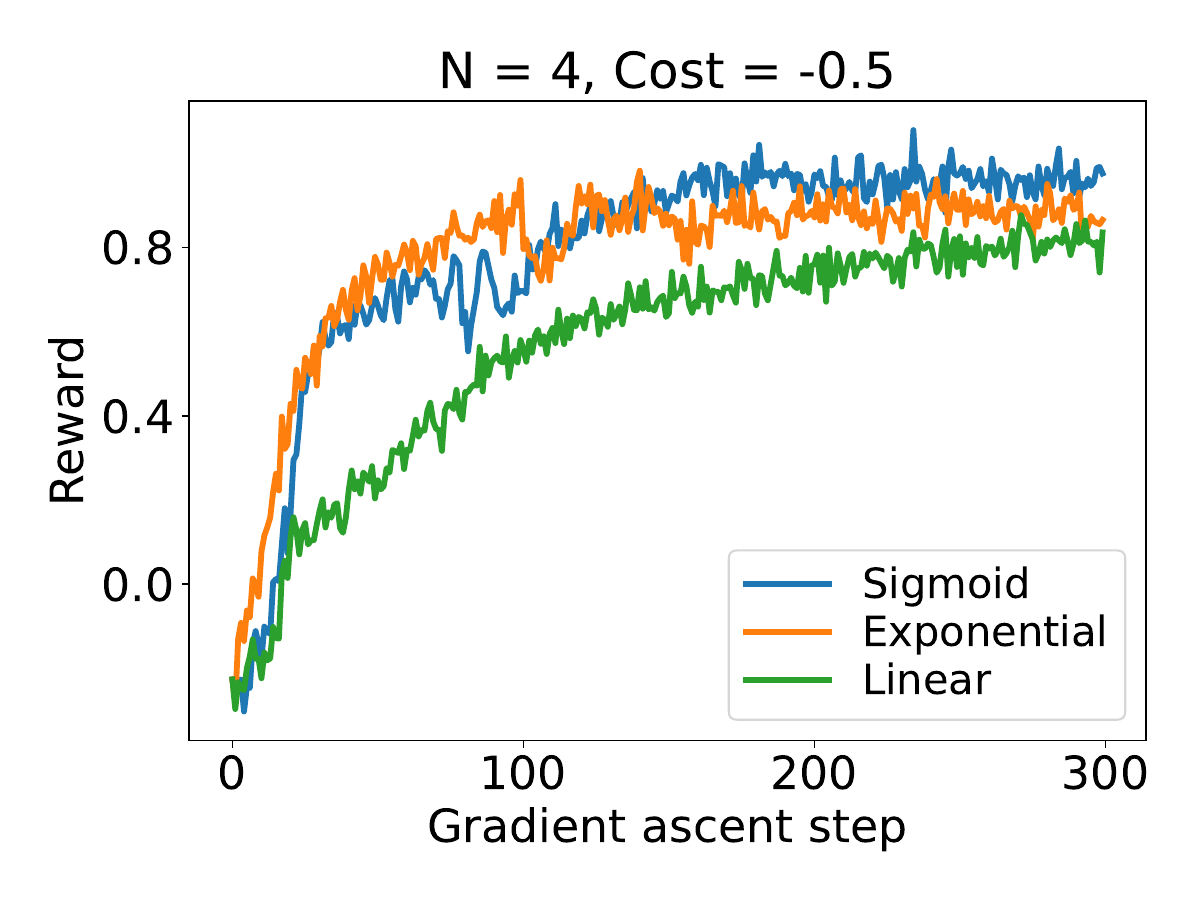}
  \includegraphics[width=0.45\linewidth]{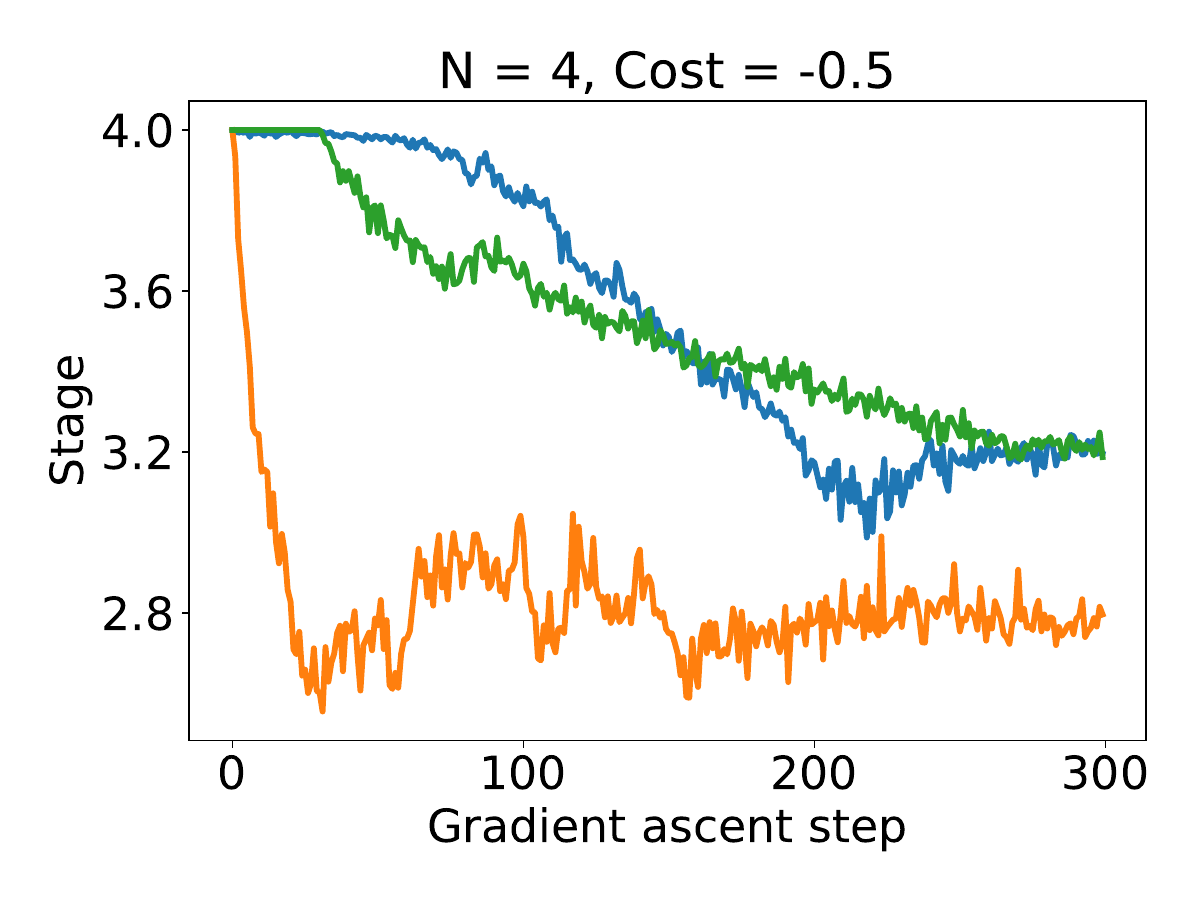}
  \caption{Comparison of curriculum schedules on the convection-diffusion source detection problem. Left: average reward. Right: average stopping stage.}
  \label{fig:convection_diffusion_schedule_comparison}
\end{figure}

In the source-detection problem, shown in \cref{fig:convection_diffusion_schedule_comparison}, the schedule matters more. A fast exponential schedule activates stopping too quickly, reducing later-stage exploration and leading to unstable or suboptimal training. Slower linear and sigmoid schedules maintain longer trajectories early in training and produce more stable final performance. These results suggest that curriculum schedules should be chosen based on the strength of sequential dependence---simple problems tolerate aggressive schedules, while problems with movement constraints and strong path dependence benefit from gradual activation of stopping.

Overall, the source-detection example demonstrates why optimal stopping is not simply an add-on to sequential design. When designs influence future feasible measurements, premature stopping can change the data distribution used for training and prevent the policy from discovering informative later-stage actions. Curriculum learning mitigates this effect by preserving longer trajectories during early optimization, with the clearest gains appearing when early stopping would otherwise suppress later-stage exploration.

\section{Conclusions}
\label{sec:conclusion}

This paper developed a framework for sequential BED with optimal stopping. We formulated stopping as part of the sequential decision problem, rather than prescribing a fixed experimental horizon or relying on threshold-based termination rules. This formulation yields a value-based stopping criterion: experimentation should terminate when the terminal reward is no smaller than the expected continuation value.

We showed that sequential BED with stopping can be formulated as a unified decision problem in which experimental designs and stopping decisions are jointly optimized. For a fixed design policy, we derived the optimal stopping rule and showed that the stopping boundary is determined by a comparison between terminal reward and continuation value. We also established the equivalence between terminal and incremental information-reward formulations under policy-dependent stopping times, and derived a policy-gradient expression for optimizing continuous design policies with value-based stopping.

The resulting computational problem is challenging because the design policy, continuation value, and stopping boundary are mutually dependent. To address this issue, we developed a continuation-value-based policy-gradient method and introduced a curriculum strategy that gradually transitions from forced continuation to adaptive stopping during training. Computationally, the method can be viewed as an actor-critic solution approach for the finite-horizon MDP formulation, with the continuation-value approximation serving as the critic and the design policy serving as the actor. This implementation retains the value-based stopping structure implied by the optimal stopping theorem. Across a linear-Gaussian benchmark, a nonlinear non-conjugate inference case, and a convection-diffusion source detection problem, the method learned resource-aware design--stopping policies. The results show that curriculum learning is most useful when premature stopping prevents the policy from observing informative later-stage trajectories, as in the movement-constrained source detection problem with moderate experimental cost.

Several limitations motivate future work. First, reward evaluation can be expensive when posterior updates or KL-divergence computations are costly, especially in high-dimensional parameter spaces. Future work could combine the proposed stopping framework with scalable posterior approximations, amortized inference, or surrogate models for information gain. Second, although the value-based stopping rule avoids direct optimization over discrete stopping actions, it introduces a circular dependence between the continuation-value approximation and the training data distribution. Alternative training schemes, including direct stopping-policy parameterizations or offline pretraining of continuation values, may provide complementary ways to improve robustness. Finally, this paper focuses on information gain for parameter inference, but the same decision-theoretic framework can accommodate goal-oriented utilities, such as predictive performance or task-specific scientific objectives.

Overall, this work establishes a principled framework for resource-aware stopping in sequential BED. The framework connects optimal stopping with information-theoretic design, yields an interpretable value-based stopping rule, and provides a practical learning algorithm for continuous design spaces. It enables autonomous experimental systems to jointly determine the next experiment and decide when additional experimentation no longer justifies its cost.

\appendix
\crefalias{section}{appendix}

\section{Proof of \cref{thm:optimal_stopping_rule}}
\label{app:optimal_stopping_rule_proof}

\begin{proof}
Fix a design policy $\pi=\{\mu_k\}_{k=0}^{N-1}$. For any stopping policy $\psi$, the value function satisfies the recursion
\begin{align}
V_k^{\pi,\psi}(s_k)
=
\begin{cases}
r_T^S(s_k),
& \varphi_k(s_k)=1,\\
\mathbb{E}_{Y_k|s_k,\mu_k(s_k)}
\left[
r_k^C(s_k,\mu_k(s_k),Y_k)
+
V_{k+1}^{\pi,\psi}(s_{k+1})
\right],
& \varphi_k(s_k)=0,
\end{cases}
\label{eq:proof_value_recursion}
\end{align}
for $k=0,\ldots,N-1$, with terminal condition
\begin{align}
V_N^{\pi,\psi}(s_N)=r_T(s_N).
\label{eq:proof_terminal_condition}
\end{align}
Here $s_{k+1}=\CF_k(s_k,\mu_k(s_k),Y_k)$ whenever continuation is selected.

Let $V_k^{\pi,\ast}$ denote the optimal value over stopping policies from stage $k$ onward, with the design policy $\pi$ fixed. We prove by backward induction that
\begin{align}
V_k^{\pi,\ast}(s_k)
=
\max
\left\{
r_T^S(s_k),
Q_k^{\pi,\psi^\ast}(s_k,\mu_k(s_k))
\right\},
\label{eq:proof_bellman_optimality}
\end{align}
where the first argument corresponds to stopping, and the second term
\begin{align}
Q_k^{\pi,\psi^\ast}(s_k,\mu_k(s_k))
=
\mathbb{E}_{Y_k|s_k,\mu_k(s_k)}
\left[
r_k^C(s_k,\mu_k(s_k),Y_k)
+
V_{k+1}^{\pi,\ast}(s_{k+1})
\right]
\label{eq:proof_continuation_value}
\end{align}
corresponds to continuation. Here $Q_k^{\pi,\psi^\ast}$ denotes the continuation value when optimal stopping is followed after the next state, so that $V_{k+1}^{\pi,\ast}=V_{k+1}^{\pi,\psi^\ast}$.

At the terminal stage, $V_N^{\pi,\ast}(s_N)=r_T(s_N)$, since no further decisions remain. Now suppose that $V_{k+1}^{\pi,\ast}$ is the optimal value from stage $k+1$ onward. At stage $k$, there are only two admissible decisions. If stopping is selected, the reward is $r_T^S(s_k)$. If continuation is selected, the design is fixed by $\pi$ as $\mu_k(s_k)$, the observation $Y_k$ is generated from the predictive distribution conditional on $(s_k,\mu_k(s_k))$, a stage reward $r_k^C(s_k,\mu_k(s_k),Y_k)$ is collected, and the optimal value thereafter is $V_{k+1}^{\pi,\ast}(s_{k+1})$; the resulting quantity is taken in expectation over the predictive distribution of the unknown observation $Y_k$. Therefore the optimal value at stage $k$ is the maximum of these two quantities, which gives \cref{eq:proof_bellman_optimality}. This completes the backward induction.

With the convention of stopping in the case of a tie, stopping is optimal at stage $k$ exactly when
\begin{align}
r_T^S(s_k)
\geq
Q_k^{\pi,\psi^\ast}(s_k,\mu_k(s_k)).
\end{align}
Thus the optimal stopping set is
\begin{align}
\CT_k
=
\left\{
s_k \in \CS:
r_T^S(s_k)
\geq
Q_k^{\pi,\psi^\ast}(s_k,\mu_k(s_k))
\right\},
\end{align}
and the corresponding optimal stopping policy is
\begin{align}
\varphi_k^\ast(s_k)=\mathbf{1}_{s_k\in\CT_k}.
\end{align}
This proves \cref{thm:optimal_stopping_rule}.
\end{proof}

\section{Proof of \cref{thm:equiv_terminal_incremental}}
\label{app:equiv_terminal_incremental}

\begin{proof}
Fix design and stopping policies $(\pi,\psi)$. For each history $\Hist_k$, define the accumulated terminal quantity
\begin{align}
A_k
=
\DKL\left(
p_{\Param|\Hist_k}\,||\,p_{\Param|\Hist_0}
\right)
+
\sum_{i=0}^{k-1} c_i(\design_i),
\qquad k=0,\ldots,N,
\label{eq:proof_accumulated_terminal_quantity}
\end{align}
with the convention that the empty sum is zero. Thus $A_0=0$.

We first record a conditional identity for the KL terms. For any fixed history $\Hist_k$ and design $\design_k$, Bayes' rule implies
\begin{align}
\mathbb{E}_{Y_k|\Hist_k,\design_k}
\left[
p(\param | \Hist_{k+1})
\right]
=
p(\param | \Hist_k).
\label{eq:proof_posterior_mean_identity}
\end{align}
Therefore,
\begin{align}
&\mathbb{E}_{Y_k|\Hist_k,\design_k}
\left[
\DKL\left(
p_{\Param|\Hist_{k+1}}\,||\,p_{\Param|\Hist_0}
\right)
\right]
\nonumber \\
&\quad =
\mathbb{E}_{Y_k|\Hist_k,\design_k}
\left[
\int
p(\param | \Hist_{k+1})
\log
\frac{p(\param | \Hist_{k+1})}{p(\param | \Hist_0)}
\,\mathrm{d}\param
\right]
\nonumber \\
&\quad =
\mathbb{E}_{Y_k|\Hist_k,\design_k}
\left[
\int
p(\param | \Hist_{k+1})
\log
\frac{p(\param | \Hist_{k+1})}{p(\param | \Hist_k)}
\,\mathrm{d}\param
\right]
+
\mathbb{E}_{Y_k|\Hist_k,\design_k}
\left[
\int
p(\param | \Hist_{k+1})
\log
\frac{p(\param | \Hist_k)}{p(\param | \Hist_0)}
\,\mathrm{d}\param
\right]
\nonumber \\
&\quad =
\mathbb{E}_{Y_k|\Hist_k,\design_k}
\left[
\DKL\left(
p_{\Param|\Hist_{k+1}}\,||\,p_{\Param|\Hist_k}
\right)
\right]
+
\DKL\left(
p_{\Param|\Hist_k}\,||\,p_{\Param|\Hist_0}
\right),
\label{eq:proof_kl_conditional_identity}
\end{align}
where the last equality follows because, in the second term, only $p(\param|\Hist_{k+1})$ depends on $Y_k$, and its expectation over the posterior predictive distribution is $p(\param|\Hist_k)$ by \cref{eq:proof_posterior_mean_identity}. Adding the deterministic cost terms gives
\begin{align}
\mathbb{E}_{Y_k|\Hist_k,\design_k}
\left[
A_{k+1}
\right]
=
A_k
+
\mathbb{E}_{Y_k|\Hist_k,\design_k}
\left[
\DKL\left(
p_{\Param|\Hist_{k+1}}\,||\,p_{\Param|\Hist_k}
\right)
+
c_k(\design_k)
\right].
\label{eq:proof_A_identity}
\end{align}

Let $V_k^{\mathrm{term}}(s_k)$ and $V_k^{\mathrm{incr}}(s_k)$ denote the value functions under the terminal and incremental reward formulations, respectively, for the same fixed policies $(\pi,\psi)$. We prove by backward induction that
\begin{align}
V_k^{\mathrm{term}}(s_k)
=
A_k
+
V_k^{\mathrm{incr}}(s_k),
\qquad k=0,\ldots,N.
\label{eq:proof_value_relation}
\end{align}

At $k=N$, the terminal formulation gives
\begin{align}
V_N^{\mathrm{term}}(s_N)=A_N,
\end{align}
whereas the incremental formulation gives
\begin{align}
V_N^{\mathrm{incr}}(s_N)=0.
\end{align}
Thus \cref{eq:proof_value_relation} holds at $k=N$.

Now suppose \cref{eq:proof_value_relation} holds at stage $k+1$. At stage $k<N$, there are two cases. If the fixed stopping policy stops, i.e., $\varphi_k(s_k)=1$, then the terminal formulation collects $A_k$, while the incremental formulation collects zero terminal reward. Hence
\begin{align}
V_k^{\mathrm{term}}(s_k)
=
A_k
=
A_k
+
V_k^{\mathrm{incr}}(s_k),
\end{align}
since $V_k^{\mathrm{incr}}(s_k)=0$ when stopping is selected.

If the fixed stopping policy continues, then $\design_k=\mu_k(s_k)$ and
\begin{align}
V_k^{\mathrm{term}}(s_k)
&=
\mathbb{E}_{Y_k|s_k,\design_k}
\left[
V_{k+1}^{\mathrm{term}}(s_{k+1})
\right]
\nonumber \\
&=
\mathbb{E}_{Y_k|s_k,\design_k}
\left[
A_{k+1}
+
V_{k+1}^{\mathrm{incr}}(s_{k+1})
\right]
\nonumber \\
&=
A_k
+
\mathbb{E}_{Y_k|s_k,\design_k}
\left[
\DKL\left(
p_{\Param|\Hist_{k+1}}\,||\,p_{\Param|\Hist_k}
\right)
+
c_k(\design_k)
+
V_{k+1}^{\mathrm{incr}}(s_{k+1})
\right]
\nonumber \\
&=
A_k
+
V_k^{\mathrm{incr}}(s_k),
\label{eq:proof_value_relation_induction}
\end{align}
where the third equality uses \cref{eq:proof_A_identity}, and the last equality follows from the recursive definition of $V_k^{\mathrm{incr}}(s_k)$ in \cref{eq:value_function_recursion} under the continuation decision. This completes the induction and proves \cref{eq:proof_value_relation} for all $k=0,\ldots,N$.

Finally, at the initial stage, $A_0=0$. Therefore
\begin{align}
U_T(\pi,\psi)
=
V_0^{\mathrm{term}}(s_0)
=
V_0^{\mathrm{incr}}(s_0)
=
U_I(\pi,\psi).
\end{align}
Since the policies $(\pi,\psi)$ were arbitrary, the terminal and incremental formulations yield the same expected utility for any design and stopping policies. Consequently, the two formulations have the same optimal value and the same set of optimal policy pairs.\end{proof}

\section{Proof of \cref{thm:policy_gradient}}
\label{app:policy_gradient_expression}

\begin{proof}
For notational simplicity, write
\begin{align}
\mu_{k,w}(s_k) := \mu_{k,w_k}(s_k),
\end{align}
with the understanding that only the component $w_k$ of $w$ enters the stage-$k$ design map. Let $\pi_w$ denote the corresponding design policy, let $\psi_w$ denote the stopping policy induced by the value-based stopping rule, and let
\begin{align}
\tau_w
=
\inf\{k \in \{0,\ldots,N-1\}:\varphi_{k,w}(s_k)=1\}
\wedge N
\end{align}
be the induced stopping time.

Since $U(w)=V_0^{\pi_w,\psi_w}(s_0)$, it suffices to compute the gradient of the value function. For $k<N$, the continuation value is
\begin{align}
Q_k^{\pi_w,\psi_w}(s_k,\design_k)
=
\mathbb{E}_{Y_k|s_k,\design_k}
\left[
r_k^C(s_k,\design_k,Y_k)
+
V_{k+1}^{\pi_w,\psi_w}(s_{k+1})
\right],
\label{eq:proof_pg_continuation_value}
\end{align}
where $s_{k+1}=\CF_k(s_k,\design_k,Y_k)$. In the partial derivative
$\nabla_{\design_k}Q_k^{\pi_w,\psi_w}(s_k,\design_k)$, the policies
$(\pi_w,\psi_w)$ are held fixed; the derivative is only with respect to the current design argument $\design_k$.

By \cref{thm:optimal_stopping_rule}, the value function can be written as
\begin{align}
V_k^{\pi_w,\psi_w}(s_k)
=
\max
\left\{
r_T^S(s_k),
Q_k^{\pi_w,\psi_w}
\left(
s_k,\mu_{k,w}(s_k)
\right)
\right\}.
\label{eq:proof_pg_value_max}
\end{align}
The assumed zero-probability stopping-boundary condition implies that, almost surely under the trajectory distribution induced by $(\pi_w,\psi_w)$, infinitesimal perturbations of $w$ do not switch the active branch in \cref{eq:proof_pg_value_max}. Thus the derivative of the active branch is valid almost surely, and boundary-motion terms do not contribute to the expectation.

If $s_k$ lies in the stopping region, then $V_k^{\pi_w,\psi_w}(s_k)=r_T^S(s_k)$ and no stage-$k$ design is selected. With the state $s_k$ fixed, this term has no explicit dependence on future design-policy parameters, so
\begin{align}
\nabla_w V_k^{\pi_w,\psi_w}(s_k)=0.
\label{eq:proof_pg_stop_gradient}
\end{align}
If $s_k$ lies in the continuation region, then
\begin{align}
V_k^{\pi_w,\psi_w}(s_k)
=
Q_k^{\pi_w,\psi_w}
\left(
s_k,\mu_{k,w}(s_k)
\right).
\end{align}
Taking the total derivative of this expression gives
\begin{align}
\nabla_w V_k^{\pi_w,\psi_w}(s_k)
&=
\nabla_w \mu_{k,w}(s_k)
\cdot
\nabla_{\design_k}
Q_k^{\pi_w,\psi_w}(s_k,\design_k)
\big|_{\design_k=\mu_{k,w}(s_k)}
\nonumber\\
&\quad+
\partial_w Q_k^{\pi_w,\psi_w}(s_k,\design_k)
\big|_{\design_k=\mu_{k,w}(s_k)}.
\label{eq:proof_pg_total_derivative}
\end{align}
The partial derivative in the second term holds the current design argument $\design_k$ fixed. Under the stated regularity assumptions, differentiation may be interchanged with the expectation in \cref{eq:proof_pg_continuation_value}. Since the immediate reward has no direct dependence on $w$ once $\design_k$ is fixed, the remaining dependence on $w$ enters through the future value function. Hence
\begin{align}
\partial_w Q_k^{\pi_w,\psi_w}(s_k,\design_k)
\big|_{\design_k=\mu_{k,w}(s_k)}
=
\mathbb{E}_{Y_k|s_k,\mu_{k,w}(s_k)}
\left[
\nabla_w V_{k+1}^{\pi_w,\psi_w}(s_{k+1})
\right].
\label{eq:proof_pg_future_derivative}
\end{align}
Combining \cref{eq:proof_pg_stop_gradient}--\cref{eq:proof_pg_future_derivative} gives the recursion
\begin{align}
\nabla_w V_k^{\pi_w,\psi_w}(s_k)
&=
\mathbf{1}_{s_k\notin \CT_{k,w}}
\Bigg[
\nabla_w \mu_{k,w}(s_k)
\cdot
\nabla_{\design_k}
Q_k^{\pi_w,\psi_w}(s_k,\design_k)
\big|_{\design_k=\mu_{k,w}(s_k)}
\nonumber\\
&\qquad\qquad+
\mathbb{E}_{Y_k|s_k,\mu_{k,w}(s_k)}
\left[
\nabla_w V_{k+1}^{\pi_w,\psi_w}(s_{k+1})
\right]
\Bigg],
\label{eq:proof_pg_gradient_recursion}
\end{align}
for $k=0,\ldots,N-1$, with terminal condition
\begin{align}
\nabla_w V_N^{\pi_w,\psi_w}(s_N)=0,
\label{eq:proof_pg_terminal_gradient}
\end{align}
because no future design decisions remain at stage $N$.

We now unroll \cref{eq:proof_pg_gradient_recursion}. Along any trajectory, the product of continuation indicators up to stage $k$ is equivalent to the event that the process has not stopped before stage $k$:
\begin{align}
\prod_{j=0}^{k}
\mathbf{1}_{s_j\notin \CT_{j,w}}
=
\mathbf{1}_{k<\tau_w}.
\label{eq:proof_pg_indicator_equivalence}
\end{align}
The one-step expectation over $Y_k$ induces the next-state distribution through the transition $s_{k+1}=\CF_k(s_k,\mu_{k,w}(s_k),Y_k)$. Repeatedly applying the tower property, i.e., the law of iterated expectations, converts the nested one-step conditional expectations into expectations over the marginal distribution of the visited states $s_k$ under $(\pi_w,\psi_w)$:
\begin{align}
\nabla_w U(w)
&=
\nabla_w V_0^{\pi_w,\psi_w}(s_0)
\nonumber\\
&=
\sum_{k=0}^{N-1}
\mathbb{E}_{s_k|\pi_w,\psi_w,s_0}
\left[
\mathbf{1}_{k<\tau_w}
\nabla_w \mu_{k,w}(s_k)
\cdot
\nabla_{\design_k}
Q_k^{\pi_w,\psi_w}(s_k,\design_k)
\big|_{\design_k=\mu_{k,w}(s_k)}
\right].
\label{eq:proof_pg_unrolled}
\end{align}
Replacing $\mu_{k,w}$ by $\mu_{k,w_k}$ gives
\begin{align}
\nabla_w U(w)
=
\sum_{k=0}^{N-1}
\mathbb{E}_{s_k|\pi_w,\psi_w,s_0}
\left[
\mathbf{1}_{k<\tau_w}
\nabla_w \mu_{k,w_k}(s_k)
\cdot
\nabla_{\design_k}
Q_k^{\pi_w,\psi_w}(s_k,\design_k)
\big|_{\design_k=\mu_{k,w_k}(s_k)}
\right],
\end{align}
which proves \cref{thm:policy_gradient}.
\end{proof}

\section{Analytical solution for the linear-Gaussian benchmark}
\label{app:linear_gaussian_analytic_solution}

This appendix derives the analytical solution for the linear-Gaussian benchmark in \cref{sec:case_linear_gaussian}. We use the terminal reward formulation and assume design-independent experimental costs $c_k$. The observation model is
\begin{align}
Y_k
=
\Param \design_k+\Epsilon_k,
\qquad
\Epsilon_k\sim \mathcal{N}(0,\sigma_{\epsilon}^2),
\label{eq:app_linear_gaussian_model}
\end{align}
with prior
\begin{align}
\Param \sim \mathcal{N}(m_0,\sigma_0^2).
\end{align}
In the numerical example, $m_0=0$, $\sigma_0=3$, $\sigma_{\epsilon}=1$, and $\design_k\in[0.1,3]$.

Because the model is conjugate, the posterior after $k$ experiments remains Gaussian:
\begin{align}
p(\param|\Hist_k)
=
\mathcal{N}(\param; m_k,\sigma_k^2).
\end{align}
The posterior precision and mean satisfy
\begin{align}
\frac{1}{\sigma_k^2}
&=
\frac{1}{\sigma_0^2}
+
\frac{1}{\sigma_{\epsilon}^2}
\sum_{i=0}^{k-1}\design_i^2,
\label{eq:app_lg_precision}\\
m_k
&=
\sigma_k^2
\left(
\frac{m_0}{\sigma_0^2}
+
\frac{1}{\sigma_{\epsilon}^2}
\sum_{i=0}^{k-1}y_i\design_i
\right).
\label{eq:app_lg_mean}
\end{align}
Equivalently, after performing experiment $k$, the variance update is
\begin{align}
\frac{1}{\sigma_{k+1}^2}
=
\frac{1}{\sigma_k^2}
+
\frac{\design_k^2}{\sigma_{\epsilon}^2}.
\label{eq:app_lg_variance_update}
\end{align}
Importantly, \cref{eq:app_lg_variance_update} is deterministic given the design and does not depend on the realized observation $y_k$. The posterior mean depends on $y_k$, but the posterior variance does not.

For Gaussian distributions, the KL divergence is
\begin{align}
\DKL\left(
\mathcal{N}(m_a,\sigma_a^2)
\,||\,
\mathcal{N}(m_b,\sigma_b^2)
\right)
=
\frac{1}{2}
\left[
\frac{\sigma_a^2}{\sigma_b^2}
+
\frac{(m_a-m_b)^2}{\sigma_b^2}
+
\log\frac{\sigma_b^2}{\sigma_a^2}
-
1
\right].
\label{eq:app_lg_gaussian_kl}
\end{align}
The EIG from experiment $k$ is the expected KL divergence from the updated posterior to the current prior:
\begin{align}
\mathbb{E}_{Y_k|s_k,\design_k}
\left[
\DKL\left(
p_{\Param|\Hist_{k+1}}
\,||\,
p_{\Param|\Hist_k}
\right)
\right].
\end{align}
Since the posterior variance after the experiment is deterministic given $\design_k$, this EIG equals the reduction in Gaussian differential entropy:
\begin{align}
\mathbb{E}_{Y_k|s_k,\design_k}
\left[
\DKL\left(
p_{\Param|\Hist_{k+1}}
\,||\,
p_{\Param|\Hist_k}
\right)
\right]
&=
H\left(p_{\Param|\Hist_k}\right)
-
\mathbb{E}_{Y_k|s_k,\design_k}
\left[
H\left(p_{\Param|\Hist_{k+1}}\right)
\right]
\nonumber\\
&=
\frac{1}{2}\log(2\pi e\sigma_k^2)
-
\frac{1}{2}\log(2\pi e\sigma_{k+1}^2)
\nonumber\\
&=
\frac{1}{2}\log\frac{\sigma_k^2}{\sigma_{k+1}^2}
\nonumber\\
&=
\frac{1}{2}
\log\left(
1+\frac{\sigma_k^2\design_k^2}{\sigma_{\epsilon}^2}
\right).
\label{eq:app_lg_information_gain}
\end{align}
Thus, for this benchmark, the EIG is deterministic given the current variance and design.

Because the experimental cost is independent of the design, the optimal design at any stage maximizes \cref{eq:app_lg_information_gain}. Since
\begin{align}
\frac{1}{2}
\log\left(
1+\frac{\sigma_k^2\design_k^2}{\sigma_{\epsilon}^2}
\right)
\end{align}
is monotone increasing in $|\design_k|$, and $\design_k\in[0.1,3]$, the optimal design is
\begin{align}
\design_k^{\ast}=3,
\qquad
k=0,\ldots,N-1.
\label{eq:app_lg_optimal_design}
\end{align}
Under this optimal design, the posterior variance after $K$ experiments is
\begin{align}
\sigma_K^{\ast 2}
=
\left(
\frac{1}{\sigma_0^2}
+
\frac{1}{\sigma_{\epsilon}^2}
\sum_{i=0}^{K-1}(\design_i^{\ast})^2
\right)^{-1}
=
\left(
\frac{1}{\sigma_0^2}
+
\frac{K(\design^{\ast})^2}{\sigma_{\epsilon}^2}
\right)^{-1},
\label{eq:app_lg_optimal_variance}
\end{align}
where $\design^{\ast}=3$.

The analytical utility obtained by stopping after exactly $N_{\mathrm{fixed}}$ experiments is therefore
\begin{align}
U_{N_{\mathrm{fixed}}}^{\mathrm{fixed}}
&=
\sum_{k=0}^{N_{\mathrm{fixed}}-1}
\left[
\frac{1}{2}
\log\frac{\sigma_k^{\ast 2}}{\sigma_{k+1}^{\ast 2}}
+
c_k
\right]
\nonumber\\
&=
\frac{1}{2}
\log\frac{\sigma_0^2}{\sigma_{N_{\mathrm{fixed}}}^{\ast 2}}
+
\sum_{k=0}^{N_{\mathrm{fixed}}-1} c_k
\nonumber\\
&=
\frac{1}{2}
\log\left(
1+
\frac{\sigma_0^2}{\sigma_{\epsilon}^2}
\sum_{k=0}^{N_{\mathrm{fixed}}-1}
(\design_k^{\ast})^2
\right)
+
\sum_{k=0}^{N_{\mathrm{fixed}}-1} c_k .
\label{eq:app_lg_fixed_utility}
\end{align}
For constant cost $c_k=c$, this reduces to
\begin{align}
U_{N_{\mathrm{fixed}}}^{\mathrm{fixed}}
=
\frac{1}{2}
\log\left(
1+
\frac{N_{\mathrm{fixed}}\sigma_0^2(\design^{\ast})^2}{\sigma_{\epsilon}^2}
\right)
+
N_{\mathrm{fixed}}c .
\label{eq:app_lg_fixed_utility_constant_cost}
\end{align}

The deterministic variance update also implies that the optimal stopping decision is observation-independent. In particular, after $k$ experiments, the value of conducting $J$ additional experiments under the optimal design is
\begin{align}
R_{k,J}
=
\frac{1}{2}
\log
\frac{\sigma_k^{\ast 2}}{\sigma_{k+J}^{\ast 2}}
+
\sum_{i=k}^{k+J-1}c_i,
\qquad
J=0,\ldots,N-k,
\label{eq:app_lg_remaining_utility}
\end{align}
with $R_{k,0}=0$. Therefore, the optimal continuation value from stage $k$ is determined only by the stage index and current variance:
\begin{align}
V_k^{\ast}(s_k)
=
\DKL\left(
p_{\Param|\Hist_k}
\,||\,
p_{\Param|\Hist_0}
\right)
+
\sum_{i=0}^{k-1}c_i
+
\max_{J\in\{0,\ldots,N-k\}} R_{k,J}.
\label{eq:app_lg_value_function}
\end{align}
Stopping is optimal at stage $k$ when the best remaining incremental utility is zero:
\begin{align}
0
=
R_{k,0}
\geq
\max_{J\in\{1,\ldots,N-k\}} R_{k,J}.
\label{eq:app_lg_stopping_condition}
\end{align}
Thus the adaptive stopping policy collapses to a deterministic stopping stage,
\begin{align}
N_{\mathrm{fixed}}^{\ast}
\in
\argmax_{K\in\{0,\ldots,N\}}
U_K^{\mathrm{fixed}},
\label{eq:app_lg_optimal_fixed_stage}
\end{align}
where $K=0$ corresponds to immediate stopping before any experiment and has $U_0^{\mathrm{fixed}}=0$. \Cref{tab:analytic_result} reports the nonzero fixed stages $N_{\mathrm{fixed}}=1,\ldots,4$. For $c=0$, the utility in \cref{eq:app_lg_fixed_utility_constant_cost} is monotone increasing in $N_{\mathrm{fixed}}$, so the optimal policy uses the maximum allowable number of experiments. For negative costs, the optimal stopping stage is determined by the tradeoff between the decreasing marginal information gain and the fixed cost per experiment.

\section{Hybrid-action reinforcement learning baseline}
\label{app:additional_results}

Direct baselines for learned stopping in sequential BED are limited. As a general-purpose comparator, we implement a hybrid-action reinforcement learning baseline in which stopping is treated as part of the action space. At each stage, the policy selects both a discrete stop--continue action and a continuous experimental design. The discrete action is represented by a stochastic stopping policy, while the continuous action is represented by a stochastic design policy. Both components are trained using a soft actor-critic objective \cite{Haarnoja2018}, yielding a generic hybrid-control method that does not use problem-specific stopping heuristics.
\begin{figure}[htbp]
  \centering
  \includegraphics[width=0.49\linewidth]{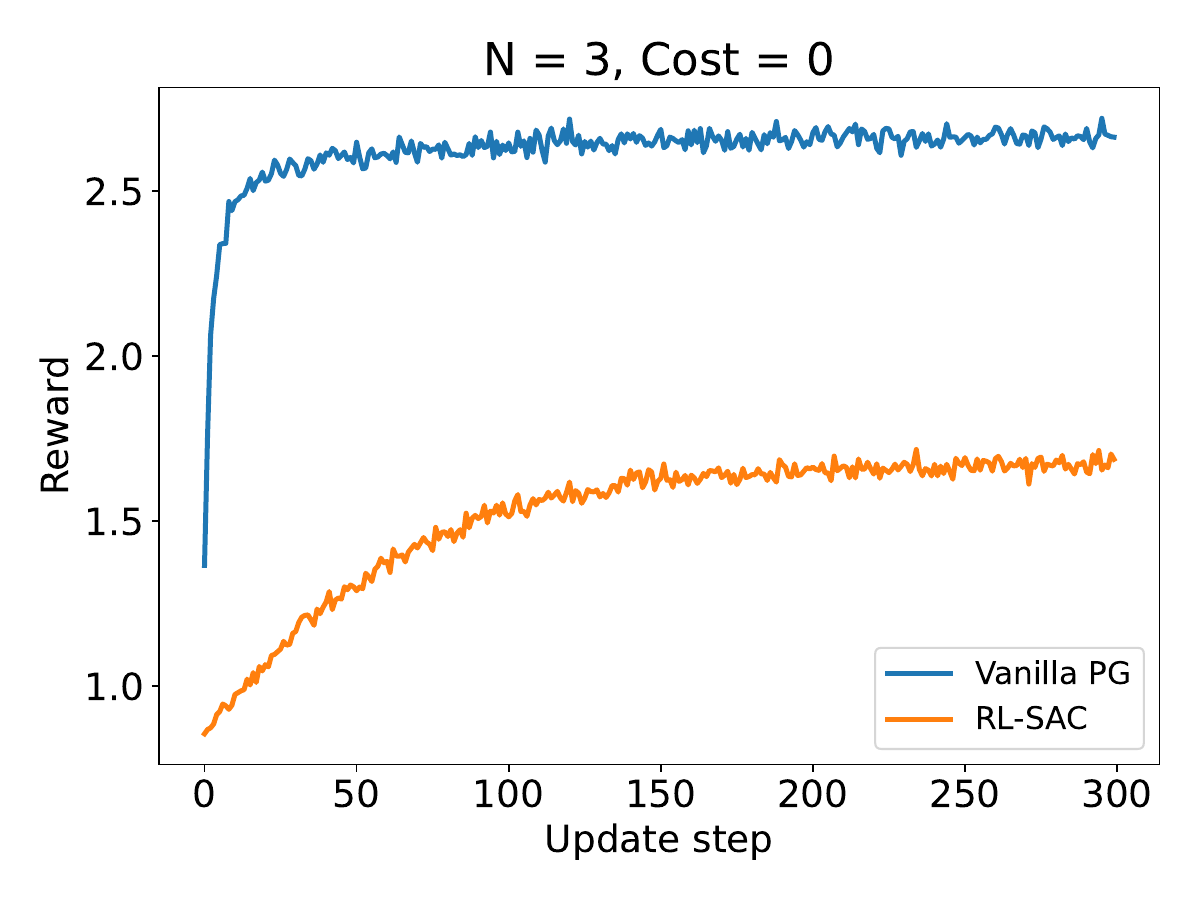}
  \includegraphics[width=0.49\linewidth]{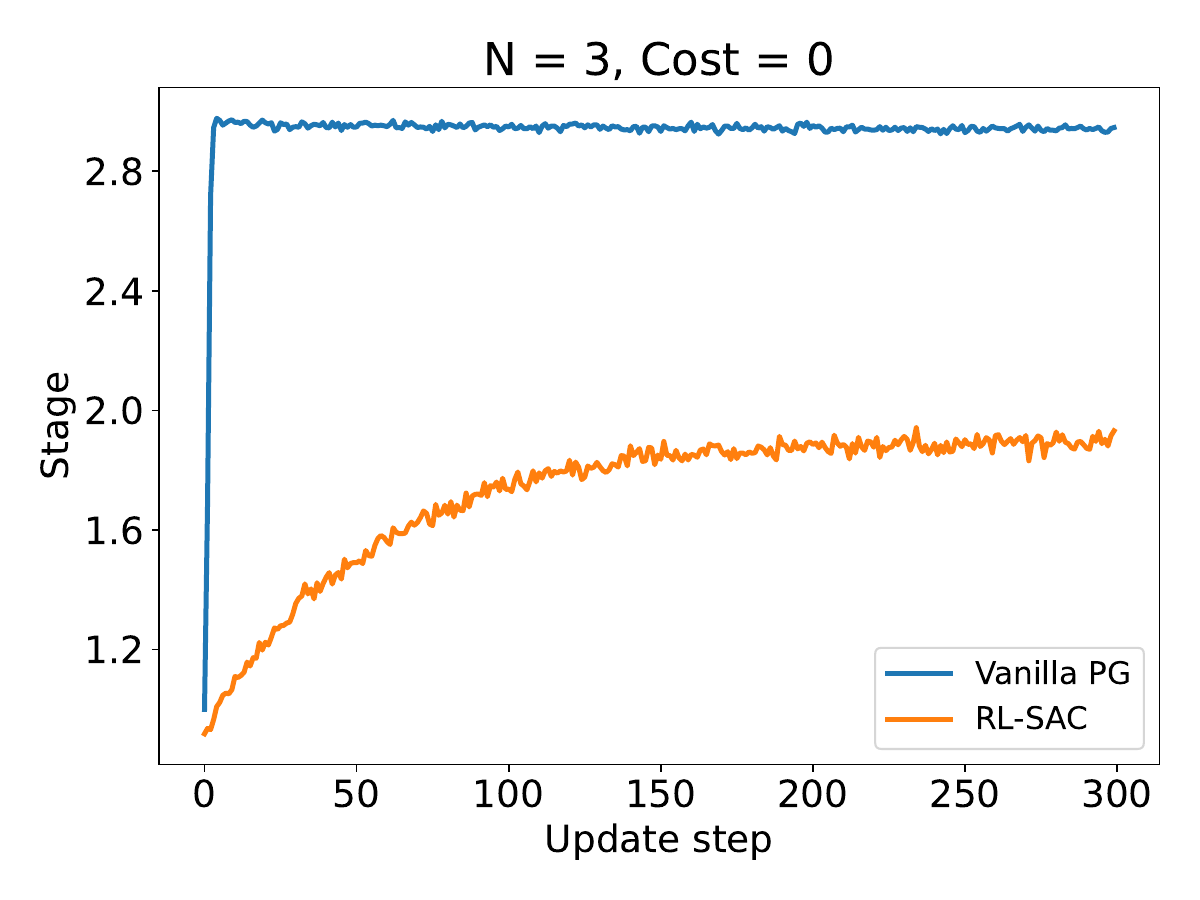}
  \caption{Comparison between the proposed vanilla policy-gradient method and a generic hybrid-action soft actor-critic baseline on the linear-Gaussian benchmark with zero experimental cost ($c_k=0$). Left: average reward. Right: average stopping stage.}
  \label{fig:linear_gaussian_sac_baseline}
\end{figure}

\Cref{fig:linear_gaussian_sac_baseline} compares this baseline with the proposed vanilla policy-gradient method on the linear-Gaussian benchmark with zero experimental cost. This setting is deliberately simple: because experiments are cost-free and the posterior variance decreases deterministically, the optimal policy continues to the maximum allowable stage. Nevertheless, in this experiment the hybrid-action baseline does not reliably identify this stopping behavior and obtains a lower reward than the proposed method.

This comparison illustrates the computational value of the optimal stopping characterization in \cref{thm:optimal_stopping_rule}. A generic hybrid-action formulation treats stopping as a discrete action that must be learned stochastically, because discrete stop--continue decisions are not directly amenable to pathwise differentiation. In contrast, \cref{thm:optimal_stopping_rule} shows that, for a fixed design policy, the optimal stopping decision is deterministic and is obtained by comparing the terminal reward with the continuation value. The proposed method exploits this structure: rather than optimizing over a stochastic stopping action, it learns a continuous design policy and evaluates stopping through the learned continuation value. This baseline result suggests that the reduction is not only theoretically convenient but also computationally useful, since a generic hybrid-action approach can underperform even in a simple benchmark where the optimal stopping behavior is analytically clear.

\section*{Acknowledgments}
During the preparation of this work, the authors used ChatGPT to assist with language editing, organization, clarity of presentation, and proof checking. The authors reviewed and revised all AI-assisted text and take full responsibility for the content of the manuscript.

\bibliographystyle{cas-model2-names}
\bibliography{references}

\end{document}